\documentclass[review,3p,times,sort&compress]{elsarticle}

\usepackage{lineno,hyperref}
\usepackage{amsmath,amsthm,amssymb}
\usepackage{mathrsfs}
\usepackage{subcaption} 
\usepackage{booktabs}
\usepackage{xcolor}
\usepackage{verbatim} 
\usepackage[export]{adjustbox}

\modulolinenumbers[5]

\journal{Computer Methods in Applied Mechanics and Engineering}

\begin{document}

\begin{frontmatter}

\title{Variational Multiscale Closures for Finite Element Discretizations Using the Mori-Zwanzig Approach}

%% Group authors per affiliation:
\author{Aniruddhe Pradhan}
\ead{anipra@umich.edu}
\author{Karthik Duraisamy}
\ead{kdur@umich.edu}
\address{Department of Aerospace Engineering, University of Michigan, Ann Arbor, MI 48109, USA}

\begin{abstract}
The simulation of multiscale problems remains a challenge due to the disparate range of spatial and temporal scales and the complex interaction between the resolved and unresolved scales.  This work develops a coarse-grained modeling approach for Galerkin discretizations by combining the Variational Multiscale decomposition and the Mori-Zwanzig (M-Z) formalism. An appeal of the M-Z formalism is that - akin to Greens functions for linear problems -  the impact of unresolved dynamics on resolved scales can be {\em formally} represented as a convolution (or memory) integral  in a {\em non-linear} setting. To ensure tractable and efficient models, Markovian closures are developed for the M-Z memory integral. The resulting sub-scale model has some similarities to adjoint stabilization and orthogonal sub-scale models. The model is made parameter free by adaptively determining the memory length during the simulation. To illustrate the generalizablity of this model, it is employed in coarse-grained simulations for the one-dimensional Burgers equation and in incompressible turbulence problems.    
\end{abstract}

\begin{keyword}
\texttt{Mori-Zwanzig, Continuous Galerkin, Variational Multiscale Method, Coarse-grained Modeling}
\end{keyword}

\end{frontmatter}

\section{Introduction}
Numerical simulation of multi-scale phenomena requires the development of coarse-grained models, which attempts to resolve a sub-set of the scales while providing a model for unresolved features. As an example, a popular approach employed  in the simulation of turbulent flows is large-eddy simulation~\cite{SMAG,VREMEN,WALE,DSM,DSM2,GDSM}, which filters the flow field to resolve the largest energy containing scales and providing a model for the scales smaller than the filter length referred to as the sub-grid scales (SGS).  The success of these  sub-scale  models largely depends on the validity of assumptions at simulated flow conditions. For example, the Smagorinsky model \cite{SMAG}, which is one of the most commonly employed SGS models, is based on the assumption that modeled rate for turbulence kinetic energy transfer from large to small scale balances dissipation \cite{pope}. This assumption is clearly not valid for all turbulent flows. An alternate approach to SGS modeling without employing phenomenological assumption, which we pursue, is to derive sub-grid models directly from the structure of the PDE and the numerical discretization.

{\color{black}In addition to physics-based sub-grid scale models \cite{SMAG,WALE,VREMEN,SIGMA} presented above, models based on the variational multiscale method have proven to be quite successful for both linear \cite{VMS,GLS,SUPG,ADJ,SUPG2} and non-linear problems \cite{OSS2,OSS,VMS3,VMSE,NLVMS,NLVMS2}}. The variational multiscale method is based on a similar idea of decomposing the flow field into resolved (coarse-scale) and un-resolved (fine-scale) variables. The fine-scales are then approximated using simple algebraic operators acting on the residual of the coarse scales which give rise to additional stabilization terms in the standard Galerkin procedure \cite{GLS,SUPG,ADJ}. A link between the stabilization terms and  implicit sub-grid models \cite{VMS} has been established. Different stabilization techniques such as the Galerkin least square (GLS) \cite{GLS}, the streamwise upwind Petrov Galerkin (SUPG) \cite{SUPG,SUPG2}, the  adjoint-stabilized methods \cite{ADJ,TAU,OSS}, and the orthogonal sub-scale stabilization \cite{OSS} can be derived based on the type of algebraic model for the fine-scales. {\color{black}In the early days of the development of the VMS techniques, these methods were  formulated for linear problems, and their application to the non-linear multiscale problems such as the Navier-Stokes equations though successful, depended on constructs such as transformations to linear problems such as the Oseen equations at every non-linear iteration. Recently, however, several attempts have been extended to develop non-linear VMS closure models by Codina et al. \cite{OSS2,OSS}, Bazilevs et al. \cite{VMS3} and many others \cite{VMSE,NLVMS,NLVMS2}. These methods utilize a stabilization parameter $\tau$ which is an approximation to the inverse of the differential operator of the governing equation. This model parameter $\tau$ is typically defined in terms of a local length-scale, elemental Reynolds and Courant numbers \cite{TAU} or derived from the Fourier analysis of the fine scale equation \cite{OSS}. Further, it is chosen to aid optimal convergence and stability of the method. %However, choosing $\tau$ in this manner might not lead to the best representation of the operator inverse. 

In this work, we aim to develop a general  coarse-graining approach in the context of the continuous Galerkin method, that is: (i) built using a  {\em non-linear} model reduction strategy akin to Greens functions for linear problems;} (ii) capable of generating a fine-scale description directly from the structure of the PDE and the underlying numerics; and (iii) model parameters are adaptive to the resolution, and are dynamically determined. 
     
 The VMS decomposition of a PDE leads to a set of coupled equations which govern the coarse-scales (resolved) and the fine-scales (un-resolved) respectively. However,  the fine-scale closure problem still persists. In our approach, the dependence of the fine-scale variables on the coarse-scale variable is removed by using the optimal prediction framework developed by Chorin \cite{CHORIN,CHORIN2}. This framework, originally developed in the context of non-equilibrium statistical mechanics, enables the higher dimensional non-linear Markovian dynamical system to be written into an exactly equivalent lower dimensional non-Markovian dynamical system \cite{MZ1,MZ2,MZ3}. The advantage is that the evolution of any observable in time can be represented solely in terms of the resolved variables. The cost of evaluating the resulting closure term, however, is enormous. The possible simplifications will be discussed later in the paper. Similar ideas have been put forward by Stinis \cite{STINIS}, Parish and Duraisamy \cite{MZ1,MZ2} in context of spectral methods and discontinuous Galerkin (DG) \cite{MZVMS} methods. {\color{black}The MZ-VMS philosophy presented herein closely follows the approach presented in \cite{MZVMS,ETHESIS}. However, the derivation of the closure terms in a continuous Galerkin setting requires specific approximations. As will be discussed  in this article, the contributing term to the final closure model in CG is different from that in DG. The main contribution of this work is to extend these dynamic closure models to the continuous Galerkin (CG) method previously not explored by Parish and Duraisamy \cite{MZVMS,ETHESIS} and testing them on canonical turbulent flow problems. }%In comparison to DG, the current approach will require lesser number of degrees of freedom per element for the same order of approximating polynomial.

The outline of the paper is as follows: We introduce the M-Z formalism in Section 2 and the VMS method in Section 3. In Section 4, we develop the VMS-MZ method in the context of a continuous Galerkin (CG) discretization. In Section 4, we derive a dynamic model for the estimation of the memory length of the convolution integral to provide a parameter free closure to the model. In the final part, we discuss results for canonical turbulence cases in Section 5. Finally, we conclude our work in Section 6.  

\begin{figure}
	\includegraphics[width=9cm]{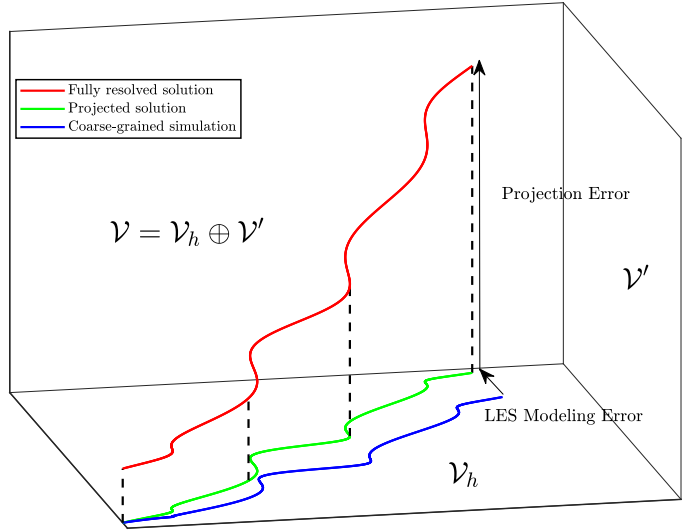}
	\centering
	\caption{Time evolution of the DNS, projected DNS and LES solutions.}
	\label{les_idea}
\end{figure}

\section{The Mori-Zwanzig (M-Z) formalism}
In this section, we introduce the general principles of the M-Z formalism \cite{CHORIN}. The concept of M-Z was first introduced in the context of statistical mechanics \cite{MORI,ZWANZIG} but was later extended by Chorin \cite{CHORIN} to more generalized systems. To demonstrate the basic idea of M-Z, we introduce it in a simple  linear dynamical system with two degrees of freedom. Following this, we present the generalization of this concept to non-linear systems via the Generalized Langevin Equations (GLEs). 

\subsection{Linear Dynamical System - An Example}
Consider a dynamical system with two degrees of freedom given by

\begin{equation}
{{dx}\over{dt}} = A_{11}x + A_{12}y
\label{toy1}
\end{equation}

\begin{equation}
{{dy}\over{dt}} = A_{21}x + A_{22}y,
\label{toy2}
\end{equation}
where $x\in \mathbb{R}$ and $y\in \mathbb{R}$ are the state space, and time $t \in (0,T\rbrack$ with  initial conditions: $x(0)$ and $y(0)$ provided. Our aim is to write an exact evolution equation for just one variables, say $x$, i.e.

\begin{equation}
{{dx}\over{dt}} = A_{11}x + F(x).
\end{equation}
Using the appropriate integration factor and integrating Equation \eqref{toy2} we have the following equation in $x$:

\begin{equation}
{{dx}\over{dt}} = A_{11}x + A_{12}\int_{0}^{t}e^{A_{22}s}A_{21}x(t-s)ds + A_{12}e^{A_{22}t}y(0)
\label{toyMZ1}.
\end{equation}
Equation \eqref{toyMZ1} has three terms: (i.) the first term represents the Markovian term containing the resolved variable; (ii.) the second term is the memory integral; and (iii.) the third term represents the dependence on the initial condition of $y$ on $x$. Although Equation \eqref{toyMZ1} represents the evolution of $x$ without any dependence on the second variable $y$, the flow of $x$ at any point of time not only depends on the current values of $x$ but also on its history weighted by an exponential factor. The two variable Markovian system is now converted to a one variable non-Markovian system without loss of accuracy. For coarse-grained model development for $x$, Equation \eqref{toyMZ1} requires the inclusion of closure for the memory integral term.   

\subsection{The Generalized Langevin Equation}
Although the discussion in the previous section was limited to a linear system, the M-Z formalism can be generalized to non-linear problems as well. To this end, consider the spatial discretization of a space-time problem leading to the following set of $N$ coupled ODEs in time:

\begin{equation}
{d{\phi} \over dt} = R(\phi),
\label{nonlinprob}
\end{equation}
where $\phi = \{\hat{\phi},\tilde{\phi}\}$, $\hat{\phi} \in \mathbb{R}^M$ and $\tilde{\phi}\in \mathbb{R}^{N-M}$ are the modes we want to resolve and model respectively. The choice of spatial discretization can be of non-tailored basis such as spectral methods \cite{MZ1,MZ2}, continuous, and the discontinuous Finite Element (FE) basis functions \cite{MZVMS} or tailored basis obtained from purely data driven techniques such as the proper orthogonal decomposition (POD) \cite{CHRIS}. By assuming the initial condition of the problem to be $\phi_0$, we aim to solve for  $\hat{\phi}$ without solving for the un-resolved modes $\tilde{\phi}$ to reduce the computation and cost. However, unlike the linear problem discussed in the previous section, non-linearity restricts us from using the integration factor approach. {\color{black}The Mori-Zwanzig approach \cite{CHORIN,CHORIN2,CHORIN3} allows us to cast the above non-linear problem (Equation \eqref{nonlinprob})  in the form of a linear PDE in the space of initial condition variables $\phi_0$ and time $t$ as follows
\begin{equation}
{\partial \over \partial t} u(\phi_0,t) = \mathcal{L}u(\phi_0,t),
\label{LOU}
\end{equation}
where the Liouville operator $\mathcal{L}$ corresponding to Equation \eqref{nonlinprob} is defined as
\begin{equation}
\mathcal{L} = \sum_{k=1}^{N} R_{k}({\phi_{0}}){\partial \over \partial \phi_{0k}},
\label{LV}
\end{equation}
and initial conditions $u(\phi_0,0)=g(\phi(\phi_0,0))$, where $g:\mathbb{R}^N\rightarrow\mathbb{R}$ is a scalar valued observable. {\color{black}Equation \eqref{LOU} can be shown to have the following solution \cite{CHORIN,CHORIN2}}:
\begin{equation}
u(\phi_0,t) = g(\phi(\phi_0,t)).
\end{equation}
Next, the semi-group notation is introduced:
\begin{equation}
u(\phi_0,t)=e^{t\mathcal{L}}g(\phi(\phi_0,0))=g(\phi(\phi_0,t)),
\label{koopman}
\end{equation}
where $e^{t\mathcal{L}}$ is called the Koopman operator, which is an infinite dimensional linear operator which when applied to an observable  $g(\phi_0,0)$ evolves it in time $t$. As a special case, the observables are chosen to be the same as the initial states $g(\phi_0)=\phi_{0j}$. As a consequence, Equation \eqref{LOU} along with Equation \eqref{koopman} results in the following
 \begin{equation}
 {\partial \over \partial t} e^{t\mathcal{L}}\phi_{0j} =  \mathcal{L} e^{t\mathcal{L}}\phi_{0j} = e^{t\mathcal{L}} \mathcal{L}\phi_{0j},
 \label{LOUV}
 \end{equation}
  where the last equality is a result of commutative property \cite{CHORIN2} between $\mathcal{L}$ and $e^{t\mathcal{L}}$. We decompose the right hand side of Equation \eqref{LOUV} into spaces of resolved initial conditions and un-resolved initial conditions as follows:
 \begin{equation}
 {\partial \over \partial t} e^{t\mathcal{L}}\phi_{0j} = e^{t\mathcal{L}} \mathcal{P}\mathcal{L}\phi_{0j} + e^{t\mathcal{L}} \mathcal{Q}\mathcal{L}\phi_{0j},
 \label{projs}
 \end{equation}
 where $\mathcal{P}:L^2 \rightarrow \hat{L}^2$ is the projection operator, where the spaces formed by all initial conditions and resolved initial conditions are denoted by $L^2$ and  $\hat{L}^2$ respectively and $\mathcal{Q}=I-\mathcal{P}$. Different forms of projectors $\mathcal{P}$ can be used \cite{MZ1,MZ2,MZ3}. In the present work, we use a truncation projector \cite{CHORIN2,MZVMS}, i.e. the application of the projector $\mathcal{P}$ to the function $f(\hat{\phi_0},\tilde{\phi_0})$ results in the truncation of the unresolved initial conditions $f(\hat{\phi_0},0)$. {\color{black}The projector $\mathcal{P}$ acts on the space formed by initial conditions and is different from the $L^2$-projectors commonly used to project onto finite dimensional spaces.} By applying Duhamel's formula \cite{CHORIN2},
 \begin{equation}
 e^{t\mathcal{L}}=e^{tQ\mathcal{L}}+\int_0^{t}e^{(t-s)\mathcal{L}}\mathcal{PL}e^{s\mathcal{QL}}ds,
 \label{dyson}
 \end{equation}
 in Equation \eqref{projs}, which is equivalent to the integration factor approach for linear systems, we obtain the generalized Langevin equation (GLE) \cite{CHORIN,CHORIN2,CHORIN3} also known as the Mori-Zwanzig identity:
 \begin{equation}
 {\partial \over \partial t} e^{t\mathcal{L}}\phi_{0j} = e^{t\mathcal{L}} \mathcal{P}\mathcal{L}\phi_{0j} + e^{t\mathcal{QL}} \mathcal{Q}\mathcal{L}\phi_{0j} + \int_0^{t}e^{(t-s)\mathcal{L}}\mathcal{PL}e^{s\mathcal{QL}}\mathcal{Q}\mathcal{L}\phi_{0j}ds.
 \label{mz_identity}
 \end{equation}
 An important observation is that Equation \eqref{mz_identity} has a similar structure to Equation \eqref{toyMZ1}. The first term is the Markovian term, the second term is the noise due to uncertainty in the initial condition and the last term is called the memory integral. The noise term given by $\mathcal{F}_j(\phi_0,t)=e^{t\mathcal{QL}} \mathcal{Q}\mathcal{L}\phi_{0j}$ is precisely the solution to the orthogonal dynamics \cite{CHORIN,FABER} equation given by  
  \begin{equation}
 {\partial \over \partial t} \mathcal{F}_j(\phi_0,t) =  \mathcal{QL}\mathcal{F}_j(\phi_0,t).
 \label{ODE}
 \end{equation}
  It can also be shown \cite{CHORIN2} that $\mathcal{F}_j(\phi_0,t)$ lies in the null space of the projector $\mathcal{P}$ i.e.  $\mathcal{PF}_j(\phi_0,t)=0$. As a result, application of $\mathcal{P}$ on Equation \eqref{mz_identity} results in the following simplification
 \begin{equation}
 {\partial \over \partial t} e^{t\mathcal{L}}\hat{\phi}_{0j} = e^{t\mathcal{L}} \mathcal{P}\mathcal{L}\hat{\phi}_{0j} + \int_0^{t}e^{(t-s)\mathcal{L}}\mathcal{PL}e^{s\mathcal{QL}}\mathcal{Q}\mathcal{L}\hat{\phi}_{0j}ds.
 \label{mz_identity2}
 \end{equation}
Equation \eqref{mz_identity2} is exact and governs the evolution of the resolved modes $\hat{\phi}$ without any dependence on the unresolved modes. However, it does not lead to reduction in the overall computational cost as it requires the solution of the orthogonal dynamics equation (Equation \eqref{ODE}) which is a high-dimensional PDE and solving it is intractable. However, this marks the starting point for deriving coarse-grained models based on different approximations \cite{CHORIN2,CHORIN3,STINIS2,STINIS3,MZ2,FABER} to the memory term. In this paper, we use the fixed memory type model which assumes that the memory integral is correlated to its integrand at $s=0$ and has finite support in time i.e.
 \begin{equation}
 \int_0^{t}e^{(t-s)\mathcal{L}}\mathcal{PL}e^{s\mathcal{QL}}\mathcal{Q}\mathcal{L}\hat{\phi}_{0j}ds 
 \approx \tau e^{t\mathcal{L}}\mathcal{PL}\mathcal{Q}\mathcal{L}\hat{\phi}_{0j} .
 \label{mz_identity3}
 \end{equation}
 where $\tau$ is called the memory length. Examples on application of this formalism for non-linear model reduction of toy problems can be found in \cite{CHORIN}.}
 
\section{The Variational Multiscale Method}
 {\color{black}We now present a brief overview of the variational multiscale (VMS) method, which was originally formalized by Hughes et al. \cite{VMS} . Consider the following PDE on an open and bounded domain $\Omega \subset \mathbb{R}^d$, where  $d\geq1$ is the dimension of the problem, with a smooth boundary $\Gamma = \partial \Omega$: 
\begin{equation}
{\partial u \over \partial t}+\mathcal{R}(u)-f=0, 
\end{equation}
 where the operator $\mathcal{R} : \mathbb{R}^d \rightarrow  \mathbb{R}^d$ can be both linear or non-linear, the function $f:\Omega\rightarrow\mathbb{R}$ and the time varying from $t \in ( 0,T \rbrack$. Let, $ \mathcal{V}\equiv\mathcal{H}^1(\Omega)$ denote the Sobolev space containing square integral functions with square integral derivatives. We define the variational problem as follows:
\begin{equation}
({\partial u \over \partial t},w)+(\mathcal{R}(u),w)=(f,w), 
\label{vprob}
\end{equation}
find $u \in \mathcal{V}$ for all $w \in \mathcal{V}$, where $(\cdot,\cdot)$ denotes the $L_2$ inner product. The solution and weighting space are decomposed as follows:
\begin{equation}
\mathcal{V} = \tilde{\mathcal{V}} \oplus \mathcal{V}',
\end{equation}
where $\oplus$ represents a direct sum of $\tilde{\mathcal{V}}$ and $\mathcal{V}'$. From the perspective of a numerical method, $\tilde{\mathcal{V}}$ is the resolved finite dimensional space  and $\mathcal{V}'$ represents the space of functions which is not resolved. This leads to a decomposition for $u$ and $w$:
\begin{equation}
u = \tilde{u} + u',
\label{vmsu}
\end{equation}
\begin{equation}
w = \tilde{w} + w',
\label{vmsw}
\end{equation}
where $\tilde{u}, \tilde{w} \in \tilde{\mathcal{V}}$ and $u', w' \in \mathcal{V}'$. By substituting  Equations \eqref{vmsu} and \eqref{vmsw} into the variational problem given by equation \eqref{vprob} the following is obtained, 
\begin{equation}
({\partial( \tilde{u}+u') \over \partial t},\tilde{w}+w')+(\mathcal{R}(\tilde{u}+u'),\tilde{w}+w')=(f,\tilde{w}+w').
\label{vmsdec}
\end{equation} 
Due to the linear independency of $w'$ and $\tilde{w}$, equation \eqref{vmsdec} is separated into the coarse scale and fine-scale equations, respectively:
\begin{equation}
({\partial( \tilde{u}+u') \over \partial t},\tilde{w})+(\mathcal{R}(\tilde{u}+u'),\tilde{w})=(f,\tilde{w}),
\label{vmsc}
\end{equation} 
\begin{equation}
({\partial( \tilde{u}+u') \over \partial t},w')+(\mathcal{R}(\tilde{u}+u'),w')=(f,w').
\label{vmsf}
\end{equation}  
When the coarse-scale Equation \eqref{vmsc} is rearranged, the following is obtained,
\begin{equation}
({\partial \tilde{u} \over \partial t},\tilde{w})+(\mathcal{R}(\tilde{u}),\tilde{w})-(f,\tilde{w})=-({\partial u'\over \partial t},\tilde{w})-(\mathcal{R}(\tilde{u}+u')-\mathcal{R}(\tilde{u}),\tilde{w}).
\label{vmsc2}
\end{equation} 
The LHS of Equation \eqref{vmsc2} contains terms present in the standard Galerkin procedure. However, it is also depends on the solution to the fine-scale equation which can be considered as the error in the coarse scale approximation. The goal of VMS sub-grid modelling is to approximate the fine-scale solution using Equation \eqref{vmsf} and substitute it in Equation \eqref{vmsc}. Different closures can be obtained for the fine-scales depending on the type of approximation, especially when the problem is non-linear \cite{OSS2,OSS,VMS3,VMSE,NLVMS,NLVMS2}. However, for linear problems  Hughes et al. \cite{VMS} demonstrated that the fine scale solution $u'$ is related to the Green's function $g'(x,y)$ of the adjoint operator $\mathcal{R}^*$ and the coarse scale residual as follows:
\begin{equation}
u'(y) = - \int_{\Omega}g'(x,y)(\mathcal{R}(\tilde{u})-f)(x)d\Omega_x.
\label{greeni}
\end{equation}
The simplest approximation to which is given by
\begin{equation}
u'=-\tau(\mathcal{R}(\tilde{u})-f).
\label{model_vms}
\end{equation}
Although, Equation \eqref{model_vms} and Equation \eqref{greeni} have been derived for linear problems, the idea that the coarse-scale residual can be linked to the fine-scale solution, remains the basis for developing non-linear VMS models \cite{OSS2,OSS,VMS3,VMSE,NLVMS,NLVMS2} as well.}

\section{The CG-MZ-VMS Framework}
In this section, we combine methods from sections 2 and 3 to derive a coarse grained model. Figure \ref{les_idea} shows two kinds of errors in coarse grain modeling: (i) projection; and (ii) model error. A perfect model will give us the exact projection (green line), based on some optimality condition, of the full order solution on the subspace we are approximating our solution in. For a given set of FE basis functions, the projection error can never be reduced i.e. the high-dimensional full order solution cannot be represented using a small number of FE basis functions. In the present work, we seek to develop a model to accurately predict a low-dimensional projected solution rather than the high-dimensional solution itself. We begin with the governing equation in the domain $\Omega \subset \mathbb{R}^d$ with the boundary $\Gamma = \partial \Omega$, where $d\geq1$ is the dimension of the problem as follows,

\begin{equation}
\frac{\partial u}{\partial t} + \mathcal{R}(u)-f=0,
\end{equation}
where $u=g$ at the boundary $\Gamma$ and time $t \in ( 0,T \rbrack$. The weak form of the above PDE, obtained after integration by parts, can be written as follows,

\begin{equation}
\bigg(\frac{\partial u}{\partial t},w\bigg)_{\Omega} + (R(u),w)_{\Omega} +(b(u),w)_{\Gamma} = (f,w)_{\Omega} \quad \forall {w} \in {\mathcal{V}},
\label{wform}
\end{equation}
where $u \in \mathcal{V}$. The Sobolev space of functions $\mathcal{V}\equiv\mathcal{H}^1(\Omega)$ and first derivatives are square integrable. The functional space $\mathcal{V}$ is an infinite dimensional and must be approximated by a finite dimensional approximation $\tilde{\mathcal{V}}$. We consider the tessellation of $\Omega$ into non-overlapping finite elements. The domain and boundary of an element marked by $\Omega_e$ and $\Gamma_e$ respectively. Also consider the following notations:

\begin{equation}
\Omega' = \bigcup_{i=1}^{n_{el}} \Omega^{e},
\end{equation}

\begin{equation}
\Gamma' = \bigcup_{i=1}^{n_{el}} \Gamma^{e}.
\end{equation}
where $\Omega'$, $\Gamma'$ denote the interior and boundaries of all the elements respectively. Let $\tilde{\mathcal{V}} \subset C^{0} \cap \mathcal{H}^1(\Omega)$ denote our finite dimensional FE approximation space containing basis functions having $C^{0}$ continuity everywhere including element boundaries. Approximating $w$ by $\tilde{w}$ and $u$ by $\tilde{u}$ in Equation \eqref{wform}, leads to the standard Galerkin procedure given by

\begin{equation}
\bigg(\frac{\partial \tilde{u}}{\partial t},\tilde{w}\bigg)_{\Omega'} + (R(\tilde{u}),\tilde{w})_{\Omega'} +(b(\tilde{u}),\tilde{w})_{\Gamma'} = (f,\tilde{w})_{\Omega'} \quad \forall {\tilde{w}} \in {\tilde{\mathcal{V}}},
\label{weakpde}
\end{equation}
where $\tilde{u} \in \tilde{\mathcal{V}}$. The above method, although directly applicable to diffusion dominated problems, encounters stability issues when applied to convection dominated problems. The VMS procedure provides a solution to this problem by elegantly accounting for the sub-grid scale effects. By splitting the space of the solution $u=\tilde{u}+u'$ and the weighting function $w=\tilde{w}+w'$, and substituting it into Equation \eqref{weakpde}, we obtain the following integral equations for the coarse and fine scales, respectively,

\begin{equation}
\bigg(\frac{\partial \tilde{u}}{\partial t}, \mathbf{\tilde{w}}\bigg)_{\Omega'}  + (R(\tilde{u}),\mathbf{\tilde{w}})_{\Omega'}  + (R(u)-R(\tilde{u}),\mathbf{\tilde{w}})_{\Omega'} + (b(\tilde{u}),\mathbf{\tilde{w}})_{\Gamma'} + (b(u)-b(\tilde{u}),\mathbf{\tilde{w}})_{\Gamma'}= (f,\mathbf{\tilde{w}})_{\Omega'},
\label{VMS_CS}
\end{equation}

\begin{equation}
\bigg(\frac{\partial {u'}}{\partial t}, \mathbf{w'}\bigg)_{\Omega'} + (R(\tilde{u}),\mathbf{{w'}})_{\Omega'} + (R(u)-R(\tilde{u}), \mathbf{{w'}}))_{\Omega'} + (b(\tilde{u}),\mathbf{{w'}})_{\Gamma'} + (b(u)-b(\tilde{u}),\mathbf{{w'}})_{\Gamma'} = 0,
\label{finescaleeq}
\end{equation}
where $u'$ and $w'$ lie in a space orthogonal to $\tilde{u}$ and $\tilde{w}$ i.e. ${u'},{w'} \in {\mathcal{V'}}$ and $f\in\tilde{\mathcal{V}}$. This idea of decomposing the full space into orthogonal spaces has previously been pursued, for instance the Orthogonal Sub-Scale (OSS) method by Codina \cite{OSS} and Parish and Duraisamy \cite{ETHESIS,MZ2}. By substituting the resolved and un-resolved variables in terms of their modal coefficients and their basis function as follows,
\begin{equation}
\tilde{u} = \mathbf{\tilde{a}{\tilde{w}}^{T}},
\end{equation}

\begin{equation}
{\color{black} {u'} = \mathbf{{a'}{w'}^{T}}.}
\end{equation}
we get the following ODE systems for modal coefficients of the coarse and fine scales,

\begin{equation}
\frac{d\mathbf{\tilde{a}}}{dt} = \mathbf{\tilde{M}^{-1}}(-(R(\tilde{u}),\mathbf{\tilde{w}})_{\Omega'} - (R(u)-R(\tilde{u}),\mathbf{\tilde{w}})_{\Omega'} - (b(\tilde{u}),\mathbf{\tilde{w}})_{\Gamma'} - (b(u)-b(\tilde{u}),\mathbf{\tilde{w}})_{\Gamma'}+(f,\mathbf{\tilde{w}})_{\Omega'}),
\label{larges}
\end{equation}
\begin{equation}
\frac{d\mathbf{{a'}}}{dt} = \mathbf{{M'}^{-1}}(-(R(\tilde{u}),\mathbf{{w'}})_{\Omega'} - (R(u)-R(\tilde{u}), \mathbf{{w'}}))_{\Omega'} - (b(\tilde{u}),\mathbf{{w'}})_{\Gamma'} - (b(u)-b(\tilde{u}),\mathbf{{w'}})_{\Gamma'}),
\label{smalls}
\end{equation}
where the mass matrices for resolved scales and un-resolved orthogonal scales can be written as
\begin{equation}
\mathbf{\tilde{M}} = \mathbf{(\tilde{w}^{T},\tilde{w})},
\end{equation}
\begin{equation}
\mathbf{{M}'} = \mathbf{({w'}^{T},{w'})}.
\end{equation}
The RHS of equation By utilizing the Mori-Zwanzig procedure to integrate out variables in Equation \eqref{smalls} from \eqref{larges}, we get the following system:
\begin{equation}
\bigg(\frac{\partial \tilde{u}}{\partial t}, \mathbf{\tilde{w}}\bigg)_{\Omega'} + (R(\tilde{u}),\mathbf{\tilde{w}})_{\Omega'} = \mathbf{\tilde{M}} \int_{0}^{t} K(\mathbf{\tilde{a}}(t-s),s)ds,
\end{equation}
where the additional term to the RHS is due to the memory effects. {\color{black} Different ways to model the memory term have been explored in the literature \cite{CHORIN2,CHORIN3,STINIS2,STINIS3,MZ2,FABER}. In the present formulation, we will use the fixed memory model $\int_{0}^{t} K(\mathbf{\tilde{a}}(t-s),s)ds \approx \tau K(\mathbf{\tilde{a}}(t),0)$ which results in the following simplification:}
\begin{equation}
\bigg(\frac{\partial \tilde{u}}{\partial t}, \mathbf{\tilde{w}}\bigg)_{\Omega'} + (R(\tilde{u}),\mathbf{\tilde{w}})_{\Omega'} = \tau \mathbf{\tilde{M}}K(\mathbf{\tilde{a}}(t),0),
\end{equation}
where $\tau$ is the memory length. The memory kernel at $s=0$ is given by
\begin{equation}
K(\mathbf{\tilde{a}}(t),0) = e^{\mathcal{L}t}\mathcal{PLQL}\mathbf{\mathbf{\tilde{a}}_0}.
\end{equation}
First, we apply $\mathcal{L}$ on $\mathbf{\mathbf{\tilde{a}}_0}$, resulting in the RHS of Equation \eqref{larges} given by
\begin{equation}
e^{\mathcal{L}t}\mathcal{L}\mathbf{\mathbf{\tilde{a}}_0} = \mathbf{\tilde{M}^{-1}}(-(R(\tilde{u}),\mathbf{\tilde{w}})_{\Omega'} - (R(u)-R(\tilde{u}),\mathbf{\tilde{w}})_{\Omega'} - (b(\tilde{u}),\mathbf{\tilde{w}})_{\Gamma'} - (b(u)-b(\tilde{u}),\mathbf{\tilde{w}})_{\Gamma'}+(f,\mathbf{\tilde{w}})_{\Omega'}),
\label{MZ_L}
\end{equation}
Second, we apply the projection $\mathcal{Q=I-P}$ to Equation \eqref{MZ_L} which results in the following expression,
\begin{equation}
e^{\mathcal{L}t}\mathcal{QL}\mathbf{\mathbf{\tilde{a}}_0} = \mathbf{\tilde{M}^{-1}}(- (R(u)-R(\tilde{u}),\mathbf{\tilde{w}})_{\Omega'} - (b(u)-b(\tilde{u}),\mathbf{\tilde{w}})_{\Gamma'}).
\label{MZ_QL}
\end{equation}
{\color{black}Third, we apply the Liouville operator $\mathcal{L}$ to obtain $e^{\mathcal{L}t}\mathcal{LQL}\mathbf{\mathbf{\tilde{a}}_0}$. The effect of application of the Liouville operator $\mathcal{L}$ to any scalar function results in the Frechet derivative evaluated in the direction of the RHS i.e. $\mathcal{L}\mathbf{a}_0$. For example, for a scalar function g we have the following:  
\begin{equation}
{\mathcal{L}}g(u(\mathbf{a}_0)) = {\partial g \over \partial \mathbf{a}_0}\mathcal{L}\mathbf{a}_0
\end{equation}
Recognising that $u=\mathbf{w}^T {\mathbf{a}}$ and applying chain rules we get
\begin{equation}
{\mathcal{L}}g(u(\mathbf{a}_0)) = {\partial g \over \partial {u}_0}\mathbf{w}^T\mathcal{L}\mathbf{a}_0
\end{equation}}
Finally $e^{\mathcal{L}t}\mathcal{LQL}\mathbf{\mathbf{\tilde{a}}_0}$ is obtained by linearising w.r.t to $u$ and evaluating the result model at RHS of Equation \eqref{larges} and \eqref{smalls} as follows:
\begin{align}
\begin{split}
e^{\mathcal{L}t}\mathcal{LQL}\mathbf{\tilde{a}_0} = -\mathbf{\tilde{M}^{-1}} (R'(\mathbf{{w}^{T}}[\mathbf{{M}^{-1}}(-(R(\tilde{u})-f,\mathbf{{w}})_{\Omega'} - (R(u)-R(\tilde{u}),\mathbf{{w}})_{\Omega'} - (b(\tilde{u}),\mathbf{{w}})_{\Gamma'} - (b(u)-b(\tilde{u}),\mathbf{{w}})_{\Gamma'})]) \\
-R'(\mathbf{\tilde{w}^{T}}[\mathbf{\tilde{M}^{-1}}(-(R(\tilde{u})-f,\mathbf{\tilde{w}})_{\Omega'} - (R(u)-R(\tilde{u}),\mathbf{\tilde{w}})_{\Omega'} - (b(\tilde{u}),\mathbf{\tilde{w}})_{\Gamma'} - (b(u)-b(\tilde{u}),\mathbf{\tilde{w}})_{\Gamma'})]),\mathbf{\tilde{w}})_{\Omega'}\\ -\mathbf{\tilde{M}^{-1}}(b'(\mathbf{{w}^{T}}[\mathbf{\tilde{M}^{-1}}(-(R(\tilde{u})-f,\mathbf{{w}})_{\Omega'} - (R(u)-R(\tilde{u}),\mathbf{{w}})_{\Omega'} - (b(\tilde{u}),\mathbf{{w}})_{\Gamma'} - (b(u)-b(\tilde{u}),\mathbf{{w}})_{\Gamma'}])-\\
b'(\mathbf{\tilde{w}^{T}}[\mathbf{\tilde{M}^{-1}}(-(R(\tilde{u})-f,\mathbf{\tilde{w}})_{\Omega'} - (R(u)-R(\tilde{u}),\mathbf{\tilde{w}})_{\Omega'} - (b(\tilde{u}),\mathbf{\tilde{w}})_{\Gamma'} - (b(u)-b(\tilde{u}),\mathbf{\tilde{w}})_{\Gamma'})]),\mathbf{\tilde{w}})_{\Gamma'}.
\end{split}
\end{align}
Finally, we apply the projector $\mathcal{P}$ which removes the dependence on un-resolved variables $\mathbf{a}'$ and results in
\begin{align}
\begin{split}
\mathbf{\tilde{M}}e^{\mathcal{L}t}\mathcal{PLQL}\mathbf{\mathbf{\tilde{a}}_0} = - (R'(\mathbf{{w}^{T}}[\mathbf{{M}^{-1}}(-(R(\tilde{u})-f,\mathbf{{w}})_{\Omega'} -  (b(\tilde{u}),\mathbf{{w}})_{\Gamma'})]) 
-R'(\mathbf{\tilde{w}^{T}}[\mathbf{\tilde{M}^{-1}}(-(R(\tilde{u})-f,\mathbf{\tilde{w}})_{\Omega'} - (b(\tilde{u}),\mathbf{\tilde{w}})_{\Gamma'})]),\mathbf{\tilde{w}})_{\Omega'} -\\ (b'(\mathbf{{w}^{T}}[\mathbf{{M}^{-1}}(-(R(\tilde{u})-f,\mathbf{{w}})_{\Omega'} - (b(\tilde{u}),\mathbf{{w}})_{\Gamma'} ])-
b'(\mathbf{\tilde{w}^{T}}[\mathbf{\tilde{M}^{-1}}(-(R(\tilde{u})-f,\mathbf{\tilde{w}})_{\Omega'}- (b(\tilde{u}),\mathbf{\tilde{w}})_{\Gamma'})]),\mathbf{\tilde{w}})_{\Gamma'},
\end{split}
\label{MZ_PLQL}
\end{align}
Equation \eqref{MZ_PLQL} can be compactly written as,
\begin{align}
\begin{split}
\mathbf{\tilde{M}}e^{\mathcal{L}t}\mathcal{PLQL}\tilde{\mathbf{a_0}} = \int_{\Omega'}\int_{\Omega'} \mathbf{\tilde{w}}R'(\Pi'(x,y)(R(\tilde{u})-f))d{\Omega'_y}d{\Omega'_x} 
+ \int_{\Omega'}\int_{\Gamma'} \mathbf{\tilde{w}}R'(\Pi'(x,y)(b(\tilde{u})))d{\Gamma'_y}d{{\Omega'}_x} \\
+\int_{\Gamma'}\int_{\Omega'}\mathbf{\tilde{w}} b'(\Pi'(x,y)(R(\tilde{u})-f))d{\Omega'_y}d{\Gamma'_x} + \int_{\Gamma'}\int_{\Gamma'}\mathbf{\tilde{w}} b'(\Pi'(x,y)(b(\tilde{u})))d{\Gamma'_y}d{\Gamma'_x},
\end{split}
\label{MZFE}
\end{align}
where $\Pi'$ is the orthogonal projector onto the space of the the fine scales i.e,
\begin{equation}
\Pi'(x,y) = \mathbf{w'}^{T}(x)\mathbf{M'^{-1}}\mathbf{w'}(y).
\end{equation}
{\color{black} An important aspect of the sub-scales $u'$ is that it is dynamic \cite{OSS,OSS2} in nature. When these sub-scales are approximated with the inverse of the spatial operator, the resulting  VMS models are non-Markovian \cite{OSS,OSS2} and the sub-scales $u'$ have to be tracked in time. The M-Z formalism on the other hand, precisely integrates out the time dependency of these sub-scales, making the final formulation Markovian on the resolved variables only. To make this formulation computationally tractable, we assume that the memory integral is correlated to its integrand at $s=0$ and has finite support in time. This is the main reason an approximation to the inverse of the differential operator - as is popularly used to derive VMS models - is not required here.}

{\color{black} The final form of closure given by Equation \eqref{MZFE} is valid for both CG and DG methods. For a simpler derivation in case of smooth orthogonal basis, readers are encouraged to read  Appendix A of \cite{MZVMS}. As shown by Parish and Duraisamy \cite{MZVMS,ETHESIS}, the main contributing term in Equation \eqref{MZFE} to the final closure model in DG is Term 4. However, in this formulation, we assume that the fine-scales vanish at the element boundaries analogous to the concept of bubble functions \cite{BUBBLE,BUBBLE1,BUBBLE2,BUBBLE3}.} This approximation has also been used in the OSS model \cite{OSS,OSS2}. By using this approximation, Term 2 and Term 4 in Equation \eqref{MZFE} are neglected, resulting in the following equation: 

\begin{align}
\begin{split}
\mathbf{\tilde{M}}e^{\mathcal{L}t}\mathcal{PLQL}\tilde{\mathbf{a_0}} = \int_{\Omega'}\int_{\Omega'} \mathbf{\tilde{w}}R'(\Pi'(x,y)(R(\tilde{u})-f))d{\Omega'_y}d{\Omega'_x} + \int_{\Gamma'}\int_{\Omega'}\mathbf{\tilde{w}} b'(\Pi'(x,y)(R(\tilde{u})-f))d{\Omega'_y}d{\Gamma'_x}. 
\end{split}
\end{align}
The scale separation by projection of the residual on the fine scale can be computed as follows,
\begin{equation}
\int_{\Omega'_y}\Pi'(x,y)(R(\tilde{u}(y))-f)d{\Omega'_y} = (R(\tilde{u}(x))-f) - \tilde{\Pi}(R(\tilde{u}(x))-f),
\end{equation}
where $\tilde{\Pi}$ is again the $L_2$ projector on the finite dimensional space spanned by $\mathbf{\tilde{w}}$. {\color{black}This concludes the derivation of CG-MZ-VMS framework for the fixed memory type model}.

\section{Dynamic Memory Estimation}
While the constant memory length model provides a closure to the memory term in the M-Z expression, the parameter $\tau$ should adapt to the evolving resolution and not necessarily remain constant. Another approach is to allow the parameter $\tau$ to dynamically vary in time to attempt to represent the variations of the effects of the fine-scale quantities on the coarse scales. {\color{black} To facilitate the stabilization of our method with fewer parameters and account for the temporal variations of the memory length, we seek  a dynamic memory length model using the variational counterpart of the Germano's identity \cite{DSM,GDSM2,OGERMANO,OGERMANO2}. A similar dynamic procedure has been previously used by Oberai et al. \cite{OGERMANO} and Akkerman et al. \cite{OGERMANO2} to estimate model coefficients.} We begin by applying a zero-variance phase space projector with a fully resolved initial condition with the large-scale equation (Eqn \eqref{VMS_CS}) to obtain an exact solution to the closure problem as following: 
\begin{equation}
\mathbf{\tilde{M}} \int_{0}^{t} K(\mathbf{\tilde{a}}(t-s),s)ds = (R(\tilde{u})-R(u),\mathbf{\tilde{w}})_{\Omega'} +  (b(\tilde{u})-b(u),\mathbf{\tilde{w}})_{\Gamma'}.
\end{equation}
By assuming that the memory term has a finite support we obtain
\begin{equation}
\tau_1 \mathbf{\tilde{M}} K(\mathbf{\tilde{a}}(t),0) = (R(\tilde{u})-R(u),\mathbf{\tilde{w}})_{\Omega'} +  (b(\tilde{u})-b(u),\mathbf{\tilde{w}})_{\Gamma'}.
\end{equation}
Similarly, for a separate coarser mesh with weighting function $\mathbf{\hat{w}} \in \hat{\mathcal{V}}$, where $\hat{\cdot}$ signifies a coarser mesh than $\tilde{\cdot}$, the memory terms can be written as
\begin{equation}
\tau_2 \mathbf{\hat{M}} K(\mathbf{\hat{a}}(t),0) = (R(\hat{u})-R(u),\mathbf{\hat{w}})_{\Omega'} +  (b(\hat{u})-b(u),\mathbf{\hat{w}})_{\Gamma'}.
\label{closure_1}
\end{equation}
 We choose $\mathbf{\tilde{w}}$ such that it spans the weighting function on the coarser mesh $\mathbf{\hat{w}}$ i.e. $\hat{\mathcal{V}} \subset \tilde{\mathcal{V}}$, which results in the following equation:
\begin{equation}
\tau_1 \mathbf{G\tilde{M}} K(\mathbf{\tilde{a}}(t),0) = (R(\tilde{u})-R(u),\mathbf{\hat{w}})_{\Omega'} +  (b(\tilde{u})-b(u),\mathbf{\hat{w}})_{\Gamma'}
\label{closure_2}
\end{equation}
where $\mathbf{G}$ is a matrix which transforms $\mathbf{\tilde{w}}$ to $\mathbf{\hat{w}}$ given by
\begin{equation}
\mathbf{G} \mathbf{\tilde{w}} = \mathbf{\hat{w}}
\end{equation}

\begin{comment}
\begin{figure}
	\includegraphics[width=10cm]{./dynamic_basis.eps}
	\centering
	\caption{Basis Functions at two different levels of coarse graining. Black line and red line (dashed) represent linear $C_0$ basis functions for the finer and coarser elements with sizes h and 2h respectively.}
	\label{dy_grid}
\end{figure}
\end{comment}

\begin{figure}
	\includegraphics[width=10cm,read=right]{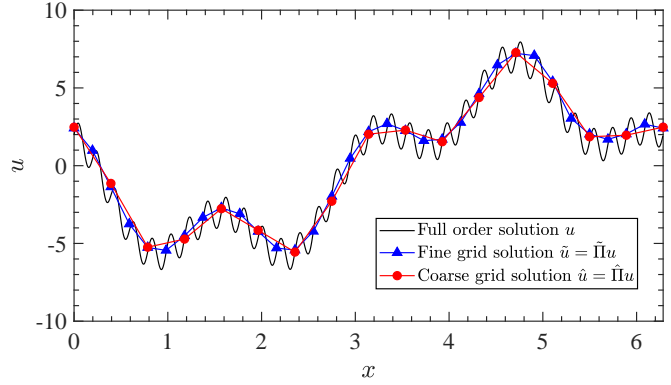}
	\centering
	\caption{$L_2$ projection of an example full order solution on two meshes with element sizes h and 2h respectively i.e $\tilde{u}$ and $\hat{u}$. }
	\label{dy_example}
\end{figure}
If the finer grid is obtained by element wise refinement of the coarse grid, the fine grid $\mathbf{\tilde{w}}$ basis functions span all the weighting functions on the coarser mesh $\mathbf{\hat{w}}$. By subtracting Equations \eqref{closure_1} and \eqref{closure_2} we obtain
\begin{equation}
\tau_1 \mathbf{G\tilde{M}} K(\mathbf{\tilde{a}}(t),0) -  \tau_2 \mathbf{\hat{M}} K(\hat{a}(t),0) = (R(\tilde{u})-R(\hat{u}),\mathbf{\hat{w}})_{\Omega'} +  (b(\tilde{u})-b(\hat{u}),\mathbf{\hat{w}})_{\Gamma'}.
\label{MZDY}
\end{equation}
To obtain $\hat{u}$ project the $\tilde{u}$ 
\begin{equation}
\hat{u} = \hat{\Pi} \tilde{u},
\end{equation}
where $\hat{\Pi}:\tilde{L}_2 \rightarrow \hat{L}_2$ is the $L_2$ projector on the coarse grid. Figure \ref{dy_example} shows the projection of a fully-resolved simulation onto $\tilde{u}$ and $\hat{u}$. This is similar to test filtering in the dynamic Smagorinsky model \cite{DSM} employed in LES. Here, we assume a scaling law similar to the one proposed by Parish and Duraisamy \cite{MZ2} relating the memory lengths $\tau$ at two different levels of coarsening as following 
\begin{equation}
{\tau_1 \over \tau_2} = \bigg[{\Delta_1 \over \Delta_2}\bigg]^{1.5},
\end{equation}
 where $\Delta_1$ and $\Delta_1$ denote the element sizes at the fine and coarse mesh. An important observation is Equation \eqref{MZFE} cannot be satisfied for all $\mathbf{\hat{w}}$ with a single value of $\tau$, but is true only in the average sense. To satisfy this condition, three different possibilities are considered here:

\begin{enumerate}
	\item Dynamic-$\tau$-AVG: Scale the modes with their respective modal values 
	
	\begin{equation}
	\tau_1 \mathbf{\hat{a}^T}\mathbf{G\tilde{M}} K(\mathbf{\tilde{a}}(t),0) -  \tau_2 \mathbf{\hat{a}^T}\mathbf{\hat{M}} K(\mathbf{\hat{a}}(t),0) = (R(\tilde{u})-R(\hat{u}),\hat{u})_{\Omega'} +  (b(\tilde{u})-b(\hat{u}),\hat{u})_{\Gamma'},
	\end{equation}
	
    which gives the following final form for the dynamic memory length, 
	\begin{equation}
	\tau = {{(R(\tilde{u})-R(\hat{u}),\hat{u})_{\Omega'} +  (b(\tilde{u})-b(\hat{u}),\hat{u})_{\Gamma'}}\over{\mathbf{\hat{a}^T}\mathbf{G\tilde{M}} K(\mathbf{\tilde{a}}(t),0) - \big({\Delta_2 \over \Delta_1}\big)^{1.5} \mathbf{\hat{a}^T}\mathbf{\hat{M}} K(\mathbf{\hat{a}}(t),0)}},
	\label{tau1}
	\end{equation}
	
	\item Dynamic-$\tau$-LS: Solve the overdetermined system based on some optimality condition,
	\begin{equation}
	\mathbf{L}=\mathbf{R}\tau,
	\end{equation}
	where \textbf{L} and \textbf{R} are given by
	\begin{equation}
	\mathbf{L}=(R(\tilde{u})-R(\hat{u}),\mathbf{\hat{w}})_{\Omega'} +  (b(\tilde{u})-b(\hat{u}),\mathbf{\hat{w}})_{\Gamma'},
	\end{equation}
	\begin{equation}
	\mathbf{R}=\mathbf{G\tilde{M}} K(\mathbf{\tilde{a}}(t),0) -  \bigg[{\Delta_2 \over \Delta_1}\bigg]^{1.5} \mathbf{\hat{M}} K(\mathbf{\hat{a}}(t),0).
	\end{equation}
	
	The above system can be solved using the least-squares approach which is commonly used with the DSM \cite{DSM} LES model, resulting in the following expression:
	
	\begin{equation}
	\tau={\mathbf{L^{T}R} \over \mathbf{R^{T}R}}.
	\end{equation}
	
	\item Dynamic-$\tau$-$l_2$: Approximate $\tau$ based on the following equation:
	
	\begin{equation}
	\tau={\mathbf{||L||} \over \mathbf{||R||}},
	\end{equation}
	
	where $||\cdot||$ denotes any kind of norm. In the present work, we have used $l_2$ or Euclidean norm for all our calculations. By using this averaging procedure, we obtain a value of $\tau$ that is (i) always positive; and (ii) free from division errors.
	
\end{enumerate}
Although the steps involved in derivation of the above formulation closely follow that of the DSM \cite{DSM,DSM2}, our approach is valid for general PDEs in that the functional form of the model is not chosen based on the underlying physical phenomena. Unlike other traditional LES SGS models such as DSM \cite{DSM,GDSM,DSM2}, WALE \cite{WALE}, VREMEN \cite{VREMEN}, and Sigma \cite{SIGMA} which are derived exclusively for the Navier-Stokes equation or other scalar transport equations, our model is not equation specific. 

\section{One-dimensional viscous Burgers equation}
As a first step towards deriving coarse-grained models for the Navier-Stokes equation, we apply our framework to a 1-D non-linear PDE exhibiting multiscale features. To this end, let $\mathcal{V}\equiv\mathcal{H}^1(\Omega)$ denote the Sobolev space where our solution $u$ and weighting functions $w$ exist. The viscous Burgers equation in the domain $\Omega \subset \mathbb{R}$ is given by the following equation:
\begin{equation}
\frac{\partial u}{\partial t} + u\frac{\partial u}{\partial x}= \nu  \frac{\partial^2 u}{\partial {x^2}},
\label{VBEeq}
\end{equation}
with periodic boundary conditions and the time varying from $t \in ( 0,T \rbrack$. The weak form of Equation \eqref{VBEeq} translates into a problem of finding ${u} \in {\mathcal{V}}$ such that
\begin{equation}
\bigg(\frac{\partial u}{\partial t},w\bigg)_{\Omega} + \bigg(u\frac{\partial {u}}{\partial x},w\bigg)_{\Omega} + \nu\bigg(\frac{\partial u}{\partial x},\frac{\partial w}{\partial x}\bigg)_{\Omega}= 0 \quad \forall {w} \in {\mathcal{V}}.
\label{wVBEeq}
\end{equation}
Using integration by parts we obtain \cite{JUMP},
\begin{equation}
\bigg(\frac{\partial u}{\partial t}+u\frac{\partial {u}}{\partial x}-\nu\frac{\partial^2 {u}}{\partial x^2},w\bigg)_{\Omega} + (J(u),w)_{\Gamma}= 0,
\end{equation}
where, $(a,b)_{\Gamma'} = \sum_{k}\int_{\Gamma'_k}ab \, d\Gamma'$ and $J(u) = \nu n_1.\nabla u_1 + \nu n_2.\nabla u_2$ (where subscripts 1 and 2 denote adjacent elements sharing a boundary). By utilizing the present coarse graining procedure to Equation \ref{wVBEeq} we get
\begin{equation}
\bigg(\frac{\partial \tilde{u}}{\partial t}+\tilde{u}\frac{\partial {\tilde{u}}}{\partial x}-\nu\frac{\partial^2 {\tilde{u}}}{\partial x^2},\tilde{w}\bigg)_{\Omega'} + (J(\tilde{u}),\tilde{w})_{\Gamma'}= \tau \mathbf{\tilde{M}} K(\mathbf{\tilde{a}}(t),0),
\label{burgermz}
\end{equation}
where the memory term $\mathbf{\tilde{M}} K(\mathbf{\tilde{a}}(t),0)$ is given by, 

\begin{equation}
\mathbf{\tilde{M}} K(\mathbf{\tilde{a}}(t),0) = \int_{\Omega'}\int_{\Omega'} \mathbf{\tilde{w}}R'(\Pi'(x,y)(R(\tilde{u})))d{\Omega'_y}d{\Omega'_x} + \int_{\Gamma'}\int_{\Omega'}\mathbf{\tilde{w}} J(\Pi'(x,y)(R(\tilde{u})))d{\Omega'_y}d{\Gamma'_x}
\label{burgermem} 
\end{equation}
and $R'$ denotes the linearization of the non-linear operator about $\tilde{u}$. Using integration by parts and neglecting the sub-scale contributions at the elemental boundaries we have    

\begin{equation}
\mathbf{\tilde{M}} K(\mathbf{\tilde{a}}(t),0) = \int_{\Omega'}R^*(\mathbf{\tilde{w}}(x))[\int_{\Omega'}\Pi'(x,y)R(\tilde{u})d{\Omega'_y}]d{\Omega'_x},
\label{memBURG}
\end{equation}
where $R^*$ is the adjoint of the linearized operator $R'$. The integrand  is computed as follows:
\begin{equation}
\int_{\Omega'_y}\Pi'(x,y)R(\tilde{u}(y))d{\Omega'_y} = R(\tilde{u}(x)) - \tilde{\Pi}(R(\tilde{u}(x))).
\end{equation}
{\color{black} The resulting closure is very similar to the adjoint stabilization method \cite{ADJ,VMS,TAU} except $R(\tilde{u}(x)) - \tilde{\Pi}(R(\tilde{u}(x)))$ is present instead of $R(\tilde{u}(x))$. Here, $R^*$ denotes the adjoint of the linearized operator $R'$ and not the operator $R$ itself.} The adjoint operator $R^*$ is given by

\begin{equation}
R^*({\tilde{w}}) = -\tilde{u}\frac{\partial \tilde{w}}{\partial x}-\nu\frac{\partial ^2\tilde{w}}{\partial x^2} 
\end{equation}
Substitution of Equation \eqref{memBURG} into Equation \eqref{burgermz} results in the following problem for the coarse scales ${\tilde{u}} \in {\tilde{\mathcal{V}}}$:

\begin{equation}
\bigg(\frac{\partial \tilde{u}}{\partial t}+\tilde{u}\frac{\partial {\tilde{u}}}{\partial x},\tilde{w}\bigg)_{\Omega'} + \nu\bigg(\frac{\partial \tilde{u}}{\partial x},\frac{\partial \tilde{w}}{\partial x}\bigg)_{\Omega'}= \tau \sum_{K}\int_{K}R^*(\tilde{w})[R(\tilde{u}(x)) - \tilde{\Pi}(R(\tilde{u}(x)))]d{\Omega'} \quad \forall {\tilde{w}} \in {\mathcal{\tilde{V}}}.
\label{mz_burg}
\end{equation}
Similarly the following coarse grained model can be derived for the linear advection-diffusion equation:

\begin{equation}
\bigg(\frac{\partial \tilde{u}}{\partial t}+a\frac{\partial {\tilde{u}}}{\partial x},\tilde{w}\bigg)_{\Omega'} + \nu\bigg(\frac{\partial \tilde{u}}{\partial x},\frac{\partial \tilde{w}}{\partial x}\bigg)_{\Omega'}= \tau \sum_{K}\int_{K}\bigg(-a\frac{\partial \tilde{w}}{\partial x}-\nu\frac{\partial ^2\tilde{w}}{\partial x^2}\bigg)[R(\tilde{u}(x)) - \tilde{\Pi}(R(\tilde{u}(x)))]d{\Omega'} \quad \forall {\tilde{w}} \in {\mathcal{\tilde{V}}}.
\label{mz_adv}
\end{equation}
Equations \eqref{mz_burg} and \eqref{mz_adv} are first discretized in time using the $\theta$ family of methods \cite{DONEA}. Equation \eqref{mz_burg} is then linearized using the standard Picard algorithm.

\subsection{Steepening of sine wave}
To benchmark our coarse-grained model, the solution to the viscous Burgers equation is computed at $T=3.0$ for an initial sine profile \cite{VMSE,ETHESIS} on a periodic domain of length $2\pi$. To discretize in time, we use the $\theta$ family of methods \cite{DONEA} with $\theta=0.5$ (Crank-Nicolson). The simulation parameters for DNS and coarse grained simulations are summarized in Table \ref{table_sine}. When sufficient resolution is available (i.e. DNS limit), the viscosity is responsible for dissipating energy at the shock. However, when the resolution is insufficient, sub-grid models are responsible for dissipating the energy. To ensure that the viscous dissipation due to large scales is negligible, viscosity $\nu$ has been set to a small value of $10^{-4}$. As a consequence, the primary contribution to the total dissipation comes from the sub-grid model.  

\begin{figure}
	\includegraphics[width=10cm]{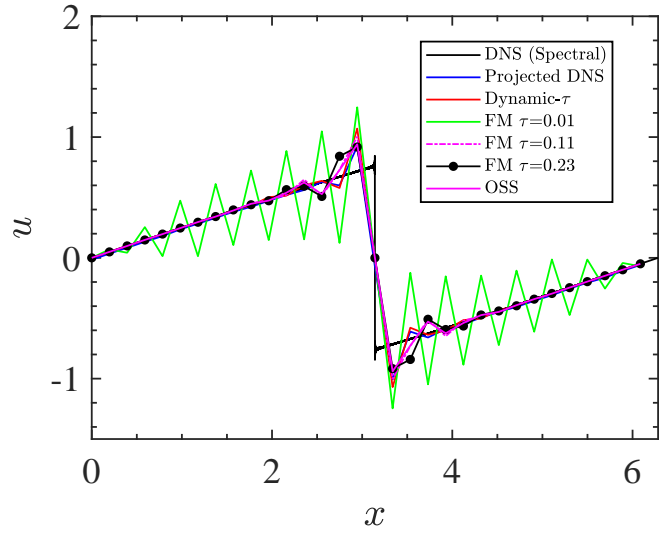}
	\centering
	\caption{Solution to the Burgers Equation at $T=3.0$ computed using different methods compared to projected DNS for the sine wave problem. }
	\label{burger_sine3}
\end{figure}

\begin{figure}
	\includegraphics[width=10cm]{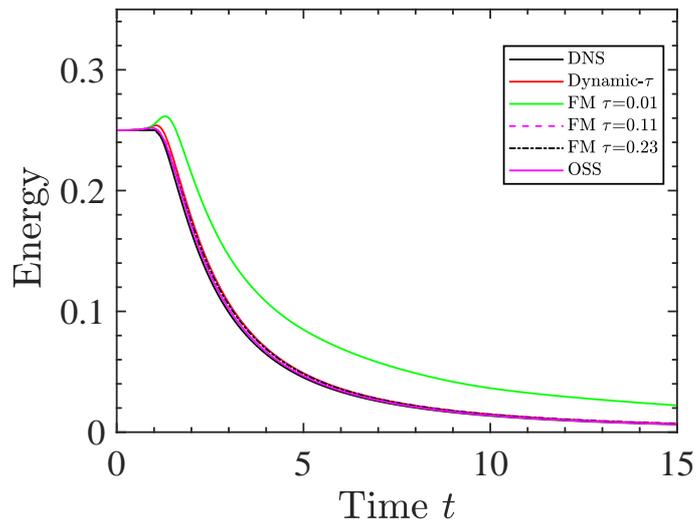}
	\centering
	\caption{Evolution of resolved KE compared to DNS for the sine wave problem using different coarse-graining methods. }
	\label{burger_energy}
\end{figure}

\begin{figure}
	\includegraphics[width=10cm]{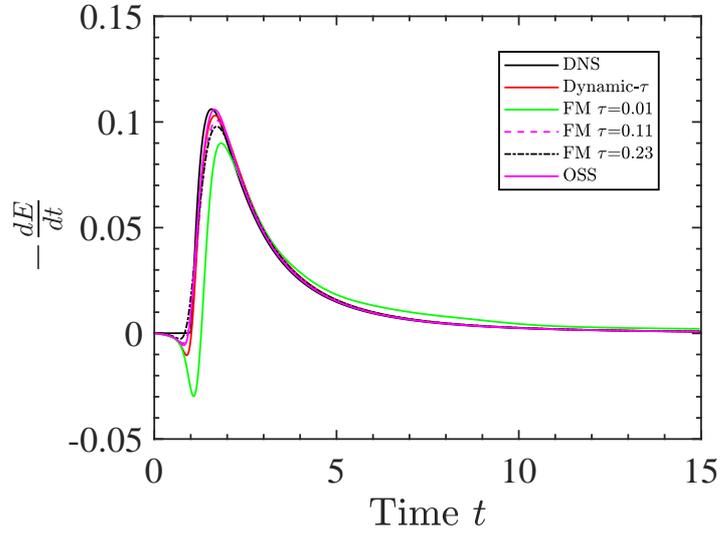}
	\centering
	\caption{Rate of energy decay compared to DNS for the sine wave problem using different coarse-graining methods. }
	\label{burger_diss}
\end{figure}

\begin{figure}
	\includegraphics[width=10cm]{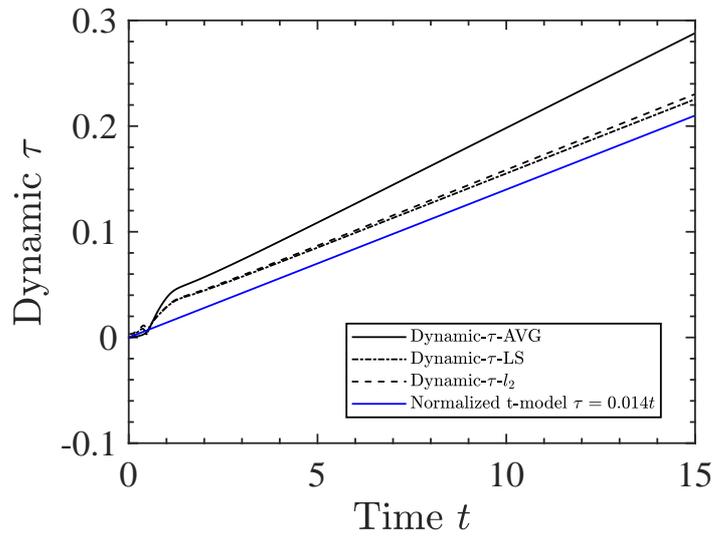}
	\centering
	\caption{Evolution of memory length $\tau$ predicted using different dynamic models for the sine wave problem. }
	\label{burger_tau}
\end{figure}

\begin{figure}
	\begin{subfigure}[b]{0.5\textwidth}
		\centering
		\includegraphics[width=\textwidth]{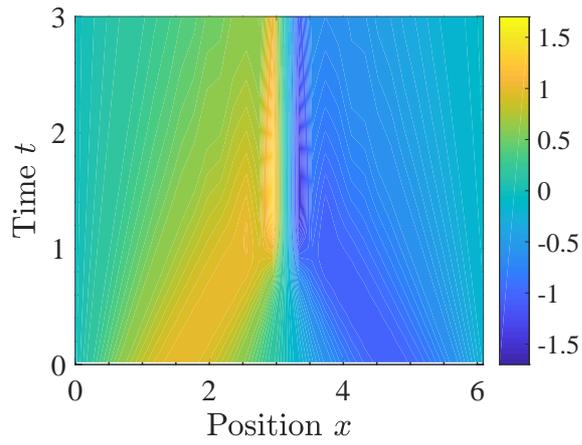}
		\caption{Dynamic-$\tau$ model.}
		\label{burger_xt1}
	\end{subfigure}
	\begin{subfigure}[b]{0.5\textwidth}
		\centering
		\includegraphics[width=\textwidth]{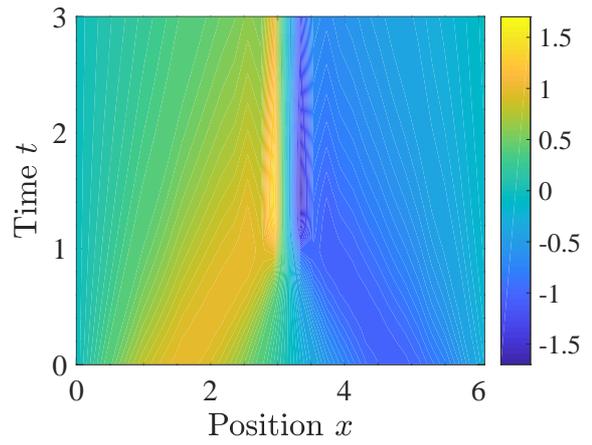}
		\caption{Projected DNS.}
		\label{burger_xt2}
	\end{subfigure}
    \begin{subfigure}[b]{0.5\textwidth}
    	\centering
    	\includegraphics[width=\textwidth]{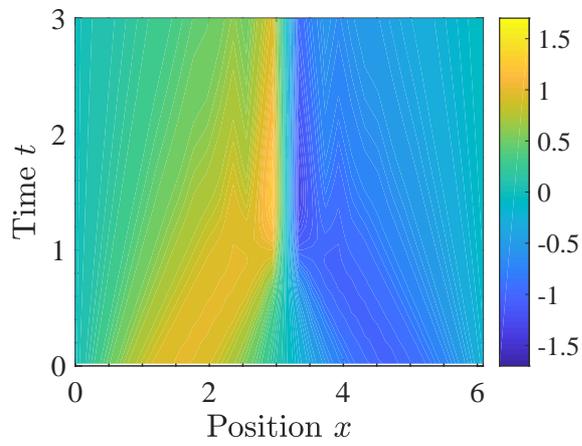}
    	\caption{OSS model \cite{OSS}.}
    	\label{burger_xt3}
    \end{subfigure}
    \begin{subfigure}[b]{0.5\textwidth}
    	\centering
    	\includegraphics[width=\textwidth]{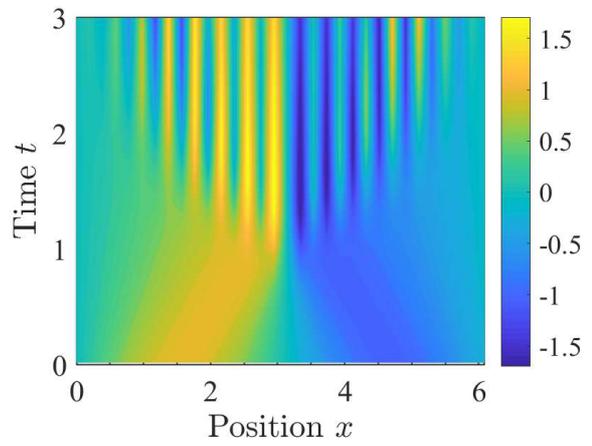}
    	\caption{No-Model.}
    	\label{burger_xt4}
    \end{subfigure}
\caption{Comparison of the wave system obtained using the dynamic-$\tau$ model, projected DNS, OSS and  No-model on the x-t diagram.}
\label{burger_xt}
\end{figure}

It can be observed in Figure \ref{burger_sine3}, the coarse-grained model solution approaches the projected DNS.  Figures \ref{burger_energy} \thinspace and \thinspace \ref{burger_diss} \thinspace show the time evolution of resolved KE and its rate of dissipation. These results indicate that the performance of all the models are comparable except the case when no sub-grid model was used or a fixed $\tau=0.01$ was used. A comparison between the solutions on the space-time diagram obtained using our dynamic-$\tau$ model, OSS, no-model and projected DNS has been presented in Figure \ref{burger_xt}. Among our Finite Memory (FM) models, the case with $\tau=0.01$ performs the worst, as can be seen in Figures \ref{burger_sine3}, \ref{burger_energy} and \ref{burger_diss}. The solution at $T=3$ improves when $\tau$ is increased to $0.11$ and becomes worse when further increased to $\tau=0.23$, which suggests the existence of an optimum $\tau$ value. The uncertainty in choosing the value of $\tau$ close to its optimum value can be reduced with a dynamic model. To this end, methods described in Section 5 are used to compute $\tau$ dynamically in Figure \ref{burger_tau}. Results indicate that the Dynamic-$\tau$-AVG, the Dynamic-$\tau$-LS and the Dynamic-$\tau$-$l_2$ models predict a similar magnitude of $\tau$ for the period of time considered. However, the Dynamic-$\tau$-$l_2$ model, which ensures positivity of the $\tau$, was found to be most stable and was used for all the following calculations in the paper. Although it is possible to use Method 1 and Method 2 by clipping $\tau$ above zero, they were found to be unstable for the TGV problem which will be discussed later in Section 7. 

At $T=3$, the Dynamic-$\tau$-$l_2$ predicts  $\tau \approx 0.05$, which supports our argument that an optimum $\tau$ exists in the range of $0.01$ and $0.23$. The $t$-model which assumes $\tau=t$ predicts a $\tau$ which does not perform well in this case and becomes unstable. However, as noted by Stinis \cite{STINIST}, the t-model needs to be re-normalized with a coefficient for the correct prediction of the memory length i.e. $\tau = C_N t$. When renormalization is used, $\tau=0.014t$ is the correct representative of the memory length with $C_N=0.014$ as shown in Figure \ref{burger_tau}.     

\begin{table}[h!]
	\centering
\begin{tabular}{ccccccc}
	\toprule
	Case            &  Domain Size  $L$ & Degrees of Freedom $N$ & Grid Size $dx$ &Time Step $dt$ & Viscosity $\nu$ & Memory Length $\tau$\\ \midrule
	DNS (Spectral)  & $2\pi$                & 4096 modes             & $7.67\times10^{-4}$  & $3.83\times10^{-4}$ & $10^{-4}$&-\\ 
	Dynamic $\tau$  & $2\pi$                & 32 elements            & $1.96\times10^{-1}$  & $1.96\times10^{-2}$& $10^{-4}$&Dynamic\\ 
    FM $\tau$=0.01  & $2\pi$                & 32 elements            & $1.96\times10^{-1}$  & $1.96\times10^{-2}$& $10^{-4}$&0.01\\ 
    FM $\tau$=0.11  & $2\pi$                & 32 elements            & $1.96\times10^{-1}$  & $1.96\times10^{-2}$& $10^{-4}$&0.11\\ 
    FM $\tau$=0.23  & $2\pi$                & 32 elements            & $1.96\times10^{-1}$  & $1.96\times10^{-2}$& $10^{-4}$&0.23\\ 
    OSS \cite{OSS}   & $2\pi$                & 32 elements            & $1.96\times10^{-1}$  & $1.96\times10^{-2}$& $10^{-4}$& - \\ 
    No-Model        & $2\pi$                & 32 elements            & $1.96\times10^{-1}$  & $1.96\times10^{-2}$& $10^{-4}$&0\\ \bottomrule
\end{tabular}
\caption{Simulation parameters for DNS and LES of the Burgers Equation for an initial sine profile.}
\label{table_sine}
\end{table}

\subsection{Burgers turbulence}
To further assess the performance of our coarse-grained model for turbulence, we use it to study the Burgers turbulence problem \cite{MZ1,MZ3}. The solution to the Burgers equations exhibits some similarities to realistic turbulence, both having an inertial and a dissipation range \cite{VMSE}. However, the solution to  Burgers turbulence is non-chaotic, unlike physically realistic turbulence obtained though the Navier-Stokes equations. To obtain the initial flow field satisfying a given energy spectrum, the following initial condition has been used
\begin{equation}
U(x,0) = \sum_{k=1}^{K_c} U^{*}\sqrt{2E(k)}sin(kx+\beta),
\end{equation}
where, the phase $\beta$ is randomly set from $[-\pi,\pi]$ and the energy spectra $E(k)$ is set to $5^{-5/3}$ for $k$ = 1 to 5 and $k^{-5/3}$ thereafter. Two different test cases are considered here: (i.) a high viscosity case A with $U^*=1$, $K_c = 8$  and $\nu=0.01$, and (ii.) a low viscosity case B with $U^*=10$, $K_c = 32$  and $\nu=0.0005$. Simulation parameters  are summarized in Table \ref{table_BT}. The two cases are considered to demonstrate the effect of the sub-grid model on moderately and highly under-resolved simulations respectively. 

\begin{figure}
	\begin{subfigure}[b]{0.5\textwidth}
		\centering
		\includegraphics[width=\textwidth]{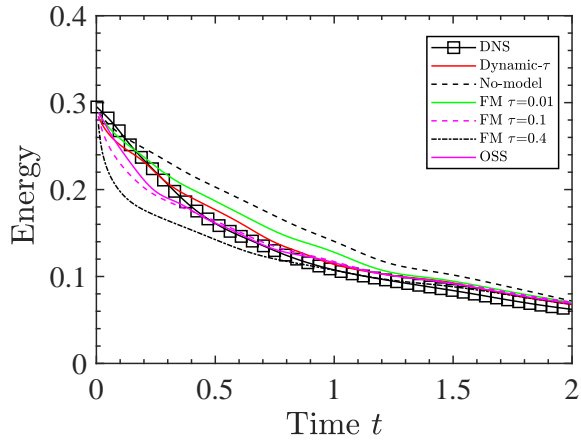}
		\caption{Case A}
		\label{BT_energy_1}
	\end{subfigure}
	\begin{subfigure}[b]{0.5\textwidth}
		\centering
		\includegraphics[width=\textwidth]{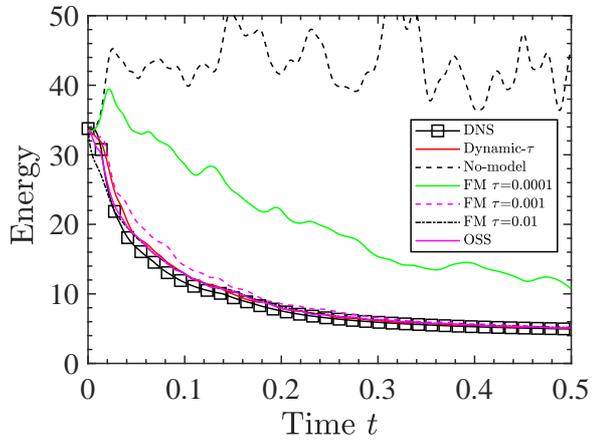}
		\caption{Case B}
		\label{BT_energy_2}
	\end{subfigure}
	\caption{Evolution of resolved KE using different methods compared to DNS for the Burgers Turbulence problem.}
	\label{BT_energy}
\end{figure}

\begin{figure}
	\begin{subfigure}[b]{0.5\textwidth}
		\centering
		\includegraphics[width=\textwidth]{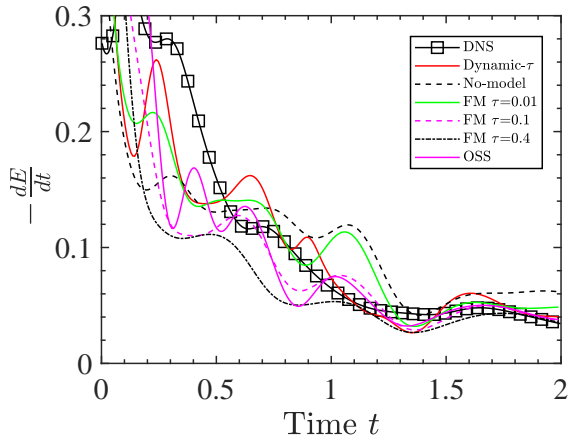}
		\caption{Case A}
		\label{BT_diss_1}
	\end{subfigure}
	\begin{subfigure}[b]{0.5\textwidth}
		\centering
		\includegraphics[width=\textwidth]{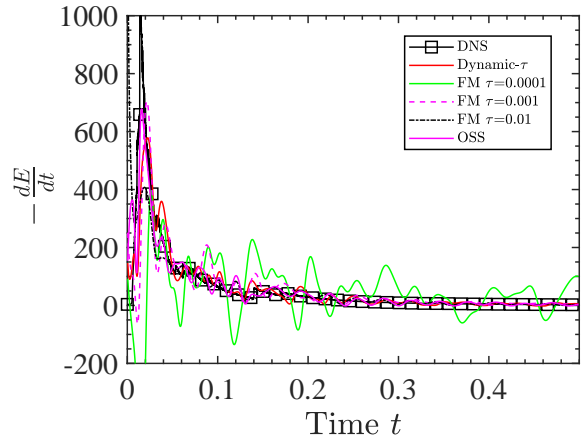}
		\caption{Case B}
		\label{BT_diss_2}
	\end{subfigure}
	\caption{Rate of energy decay due to dissipation by the sub-grid model and viscous dissipation by large scales using different methods compared to DNS for the Burgers Turbulence problem.}
	\label{BT_diss}
\end{figure}

\begin{figure}
	\begin{subfigure}[b]{0.5\textwidth}
		\centering
		\includegraphics[width=\textwidth]{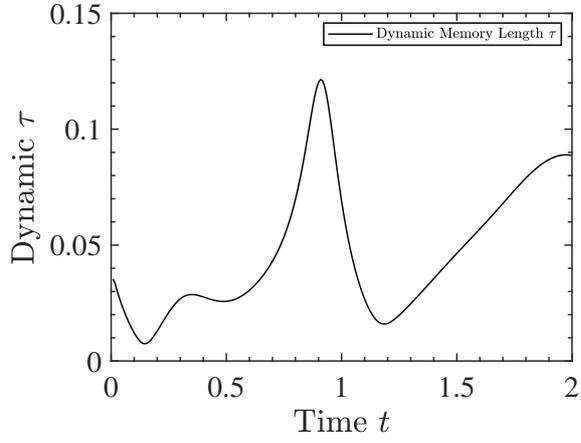}
		\caption{Case A}
		\label{BT_tau_1}
	\end{subfigure}
	\begin{subfigure}[b]{0.5\textwidth}
		\centering
		\includegraphics[width=\textwidth]{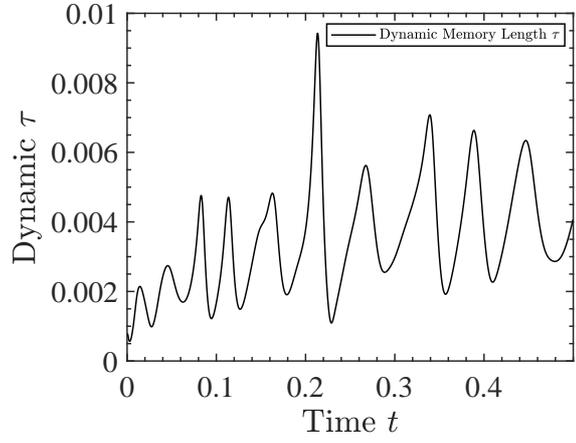}
		\caption{Case B}
		\label{BT_tau_2}
	\end{subfigure}
	\caption{Evolution of memory length $\tau$ predicted using our dynamic model for the Burgers Turbulence problem.}
	\label{BT_tau}
\end{figure}

\begin{figure}
	\begin{subfigure}[b]{0.5\textwidth}
		\centering
		\includegraphics[width=\textwidth]{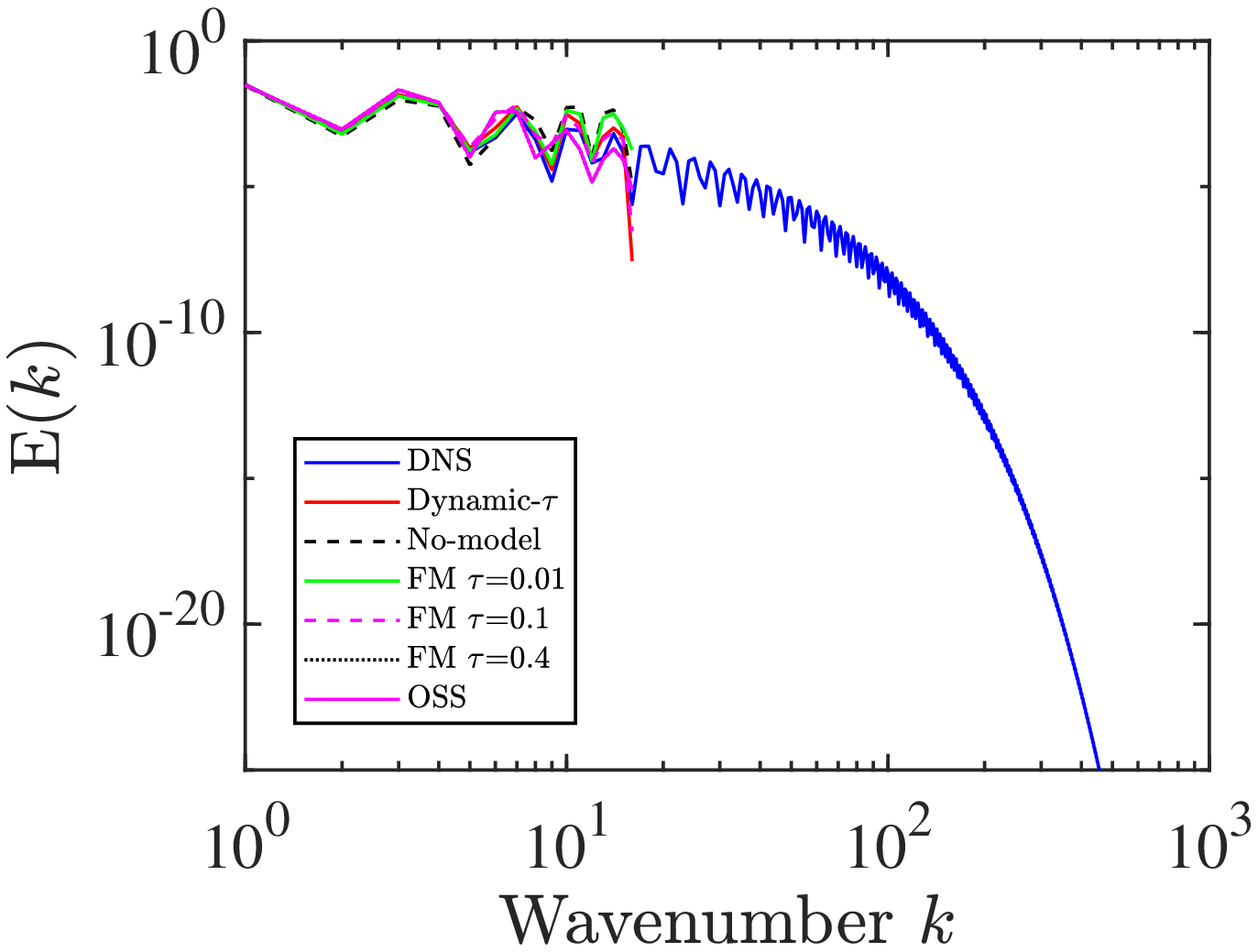}
		\caption{Case A at time $T = 2.0$}
		\label{BT_spectra_1}
	\end{subfigure}
	\begin{subfigure}[b]{0.5\textwidth}
		\centering
		\includegraphics[width=\textwidth]{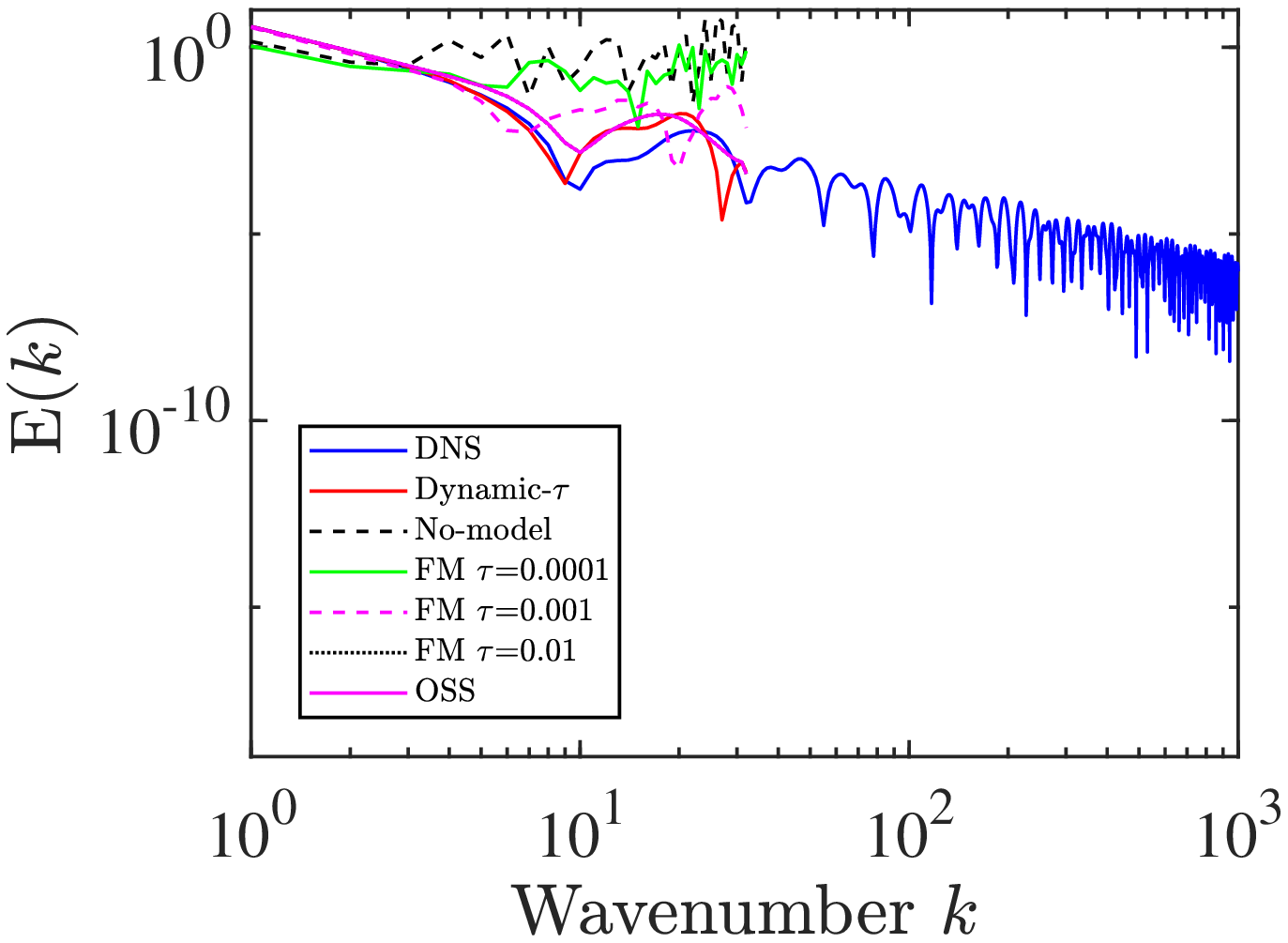}
		\caption{Case B at time $T = 0.5$}
		\label{BT_spectra_2}
	\end{subfigure}
	\caption{Energy spectra obtained using different methods compared to DNS for the Burgers Turbulence problem.}
	\label{BT_spectra}
\end{figure}

In the first case, when no sub-grid model is employed, the time variation of resolved kinetic energy is close to the DNS solution. However, for the low viscosity case, a sub-grid model becomes necessary. For comparison, DNS using the Fourier-Galerkin method is performed using 1024 and 4096 modes for case A and case B, respectively. The de-aliasing of the non-linear terms for the Fourier-Galerkin method is conducted by zero-padding (3/2-rule). The LES is conducted using the present coarse grained model with just 32 and 64 linear elements, respectively.  For each case, results from the FM model and dynamic-$\tau$ are compared to results obtained without using a sub-grid model, the OSS model and the DNS. For the FM model, different values of $\tau$ are considered in the range where our simulations are stable. Figure \ref{BT_energy} \thinspace and \thinspace \ref{BT_diss} \thinspace show the time evolution of resolved KE and its rate of dissipation for all these cases. Both these figures indicate that both the dynamic-$\tau$ model and OSS model accurately predict the time-evolution of the resolved kinetic energy in comparison DNS. Figure \ref{BT_tau} shows the variation of $\tau$ obtained from our dynamic model which for both case A and B, predict a large variation of $\tau$ in time, suggesting the importance of the adaptive selection of $\tau$. Figure \ref{BT_spectra} also shows the energy spectra at the final time. It can be observed for Case A, that all the models perform similarly and the resolved modes are able to capture most of energy. For Case B, the present dynamic-$\tau$ model performs better than other models especially at lower wavenumbers (large scales). 

\begin{table}[h!]
	\centering
	\begin{tabular}{ccccccccc}
		\toprule
		Case          & $L$     & $N$         &  $dx$    & $dt$     & $\nu$ & $U^{*}$& $K_c$ & FM $\tau$'s\\ \midrule
		Case A (DNS)  & $2\pi$  & 1024 modes  & $3.06\times10^{-3}$ & $2.33\times10^{-4}$ & $10^{-2}$  &       1& 8     &-                     \\ 
		Case A (LES)  & $2\pi$  & 32 elements & $1.96\times10^{-1}$ & $8.5\times10^{-3}$ &  $10^{-2}$  &       1& 8     &0.01, 0.1 and 0.4     \\ 
		Case B (DNS)  & $2\pi$  & 4096 modes  & $7.66\times10^{-4}$ & $3.41\times10^{-6}$ &  $5\times10^{-4}$  &      10& 32    &-                     \\ 
		Case B (LES)  & $2\pi$  & 64 elements & $9.81\times10^{-2}$ & $4.67\times10^{-4}$ & $5\times10^{-4}$ &      10& 32    &0.0001, 0.001 and 0.01\\ 
	\bottomrule
	\end{tabular}
	\caption{Simulation parameters for DNS and LES of the Burgers Equation for the Burger turbulence problem.}
	\label{table_BT}
\end{table}   

\section{Application to the Navier-Stokes Equations}
In this section, the coarse grained model is extended to the incompressible Navier-Stokes equations. Let $\mathcal{V}_d\equiv(\mathcal{H}^1(\Omega))^d$ and $\mathcal{K}\equiv{L}^2(\Omega)$ denote the Sobolev and Lebesque spaces where our solution and weighting functions exist and $d \geq 2$ in general. The weak form of the Navier-Stokes equations consists of finding $\mathbf{u}:\lbrack0,T\rbrack \rightarrow \mathcal{V}_d$, $p:\lbrack0,T\rbrack \rightarrow \mathcal{K}$  such that

\begin{equation}
(\partial_{t}\mathbf{u},\mathbf{w}) + \nu(\nabla\mathbf{u},\nabla\mathbf{w}) + (\mathbf{u}\cdot\mathbf{\nabla \mathbf{u}},\mathbf{w})-(p,\nabla\cdot\mathbf{w}) = (\mathbf{f},\mathbf{w}),
\end{equation}

\begin{equation}
(k,\nabla.\mathbf{u}) = 0,
\end{equation}
for all $[\mathbf{w},k] \in \mathcal{V}_d \times \mathcal{K}$.  

We start by assessing how our coarse-grained model stabilizes the forced viscous Burgers equation in higher dimensions. The weak form for the Burgers equations in higher dimensions can be written as

\begin{equation}
\bigg(\frac{\partial \mathbf{u}}{\partial t},\mathbf{w}\bigg)_{\Omega} + (\mathbf{u}\cdot\nabla \mathbf{u},\mathbf{w})_{\Omega} + \nu(\nabla \mathbf{u},\nabla \mathbf{w})_{\Omega}= (\mathbf{f},\mathbf{w}) \quad \forall {\mathbf{w}} \in {\mathcal{V}_d}.
\label{NS_der1}
\end{equation}
By applying integration by parts to the viscous term we get

\begin{equation}
\bigg(\frac{\partial \mathbf{u}}{\partial t},\mathbf{w}\bigg)_{\Omega} + (\mathbf{u}\cdot\nabla \mathbf{u},\mathbf{w})_{\Omega} - \nu(\nabla^2 \mathbf{u},\mathbf{w})_{\Omega} + (\mathbf{J}(\mathbf{u}),\mathbf{w})_{\Gamma} = (\mathbf{f},\mathbf{w}) \quad \forall {\mathbf{w}} \in {\mathcal{V}_d},
\label{NS_der2}
\end{equation}
where $\mathbf{J}(\mathbf{u}) = \nu\mathbf{n}_1\cdot\nabla \mathbf{u}_1 + \nu \mathbf{n}_2\cdot\nabla \mathbf{u}_2$ is the diffusive flux from adjacent elements sharing a boundary. Equations \eqref{NS_der1} and \eqref{NS_der2} can be equivalently written as,

\begin{equation}
\bigg(\frac{\partial \mathbf{u}}{\partial t},\mathbf{w}\bigg)_{\Omega} + (\mathbf{R}({\mathbf{u})},\mathbf{w})_{\Omega} + (\mathbf{J}(\mathbf{u}),\mathbf{w})_{\Gamma} = (\mathbf{f},\mathbf{w}) \quad \forall {\mathbf{w}} \in {\mathcal{V}_d}.
\end{equation}
By decomposing the spaces into $\mathcal{V}_d=\mathcal{\tilde{V}}_d \oplus \mathcal{V}_d'$ and $\mathcal{K}=\mathcal{\tilde{K}} \oplus \mathcal{K}'$, and applying our finite memory based framework leads to  the following formulation for the coarse scales $\tilde{\mathbf{u}} \in \tilde{\mathcal{V}_d}$:

\begin{equation}
\bigg(\frac{\partial \mathbf{\tilde{u}}}{\partial t},\mathbf{\tilde{w}}\bigg)_{\Omega'} + (\mathbf{R}({\mathbf{\tilde{u}})},\mathbf{\tilde{w}})_{\Omega'} + (\mathbf{J}(\mathbf{\tilde{u}}),\mathbf{\tilde{w}})_{\Gamma'} = (\mathbf{f},\mathbf{\tilde{w}}) + \tau (\mathbf{R'}({\mathbf{q})},\mathbf{\tilde{w}})_{\Omega'} + \tau (\mathbf{J'}({\mathbf{q})},\mathbf{\tilde{w}})_{\Gamma'} \quad \forall \tilde{\mathbf{w}} \in \tilde{\mathcal{V}_d},
\label{subscale_find}
\end{equation}
where $\tilde{\mathcal{V}_d}$ is our FE approximation space, and $\mathbf{R'}$ and  $\mathbf{b'}$ represent the linearizations of  $\mathbf{R}$ and $\mathbf{b}$ with respect to $\mathbf{\tilde{u}}$. The fine-scale variable $ \mathbf{q} \in {\mathcal{V}_d}'$ involving projection of the residuals on the fine-scales is given by

\begin{equation}
(\mathbf{q},\mathbf{w'})_{\Omega'} =  (\mathbf{R}({\mathbf{\tilde{u}})}-\mathbf{f},\mathbf{w'})_{\Omega'} + (\mathbf{J}(\mathbf{\tilde{u}}),\mathbf{w'})_{\Gamma'} \quad \forall {\mathbf{w}}' \in {\mathcal{V}_d}'.
\end{equation}
In the above equation, we assume that the fine-scales vanish at elemental boundaries \cite{BUBBLE,BUBBLE1,BUBBLE2,BUBBLE3}, a thus neglect the second term. The quantity $\mathbf{q}$ is approximated as follows
\begin{equation}
\mathbf{q} = \Pi'(\mathbf{R}({\mathbf{\tilde{u}})}-\mathbf{f}) = (\mathbf{R}({\mathbf{\tilde{u}})}-\mathbf{f}) - \tilde{\Pi}((\mathbf{R}({\mathbf{\tilde{u}})}-\mathbf{f})).
\end{equation}
By further simplifying the memory term we obtain the following:
\begin{equation}
\tau (\mathbf{R'}({\mathbf{q})},\mathbf{\tilde{w}})_{\Omega'} + \tau(\mathbf{J'}({\mathbf{q})},\mathbf{\tilde{w}})_{\Gamma'} = \tau[(\mathbf{q}\cdot\nabla \mathbf{\tilde{u}},\mathbf{\tilde{w}})_{\Omega'} + (\mathbf{\tilde{u}}\cdot\nabla \mathbf{q},\mathbf{\tilde{w}})_{\Omega'} - \nu(\mathbf{q},\nabla^2{\mathbf{\tilde{w}}})_{\Omega'}].
\end{equation}
Where the second term is simplified using the Green's identity and calculated as follows
\begin{equation}
(\mathbf{\tilde{u}}{\color{black}.}\nabla \mathbf{q},\mathbf{\tilde{w}})_{\Omega'} = - (\mathbf{q},\mathbf{\tilde{u}}\cdot\nabla{\mathbf{\tilde{w}}}) - (\nabla\cdot\mathbf{\tilde{u}},\mathbf{q}\cdot\mathbf{\tilde{w}}).
\end{equation}
From an implementation perspective, all the above terms are computed using numerical integration at the quadrature points. This results in the following problem for the coarse scales $\tilde{\mathbf{u}}\in \tilde{\mathcal{V}_d}$:
\begin{equation}
\bigg(\frac{\partial \mathbf{\tilde{u}}}{\partial t},\mathbf{\tilde{w}}\bigg)_{\Omega'} + (\mathbf{\tilde{u}}\cdot\nabla \mathbf{\tilde{u}},\mathbf{\tilde{w}})_{\Omega'} + \nu(\nabla \mathbf{\tilde{u}},\nabla \mathbf{\tilde{w}})_{\Omega'}= (\mathbf{f},\mathbf{\tilde{w}}) + \tau[(\mathbf{q}\cdot\nabla \mathbf{\tilde{u}},\mathbf{\tilde{w}})_{\Omega'} + (\mathbf{\tilde{u}}\cdot\nabla \mathbf{q},\mathbf{\tilde{w}})_{\Omega'} - \nu(\mathbf{q},\nabla^2{\mathbf{\tilde{w}}})_{\Omega'}] 
\quad \forall \tilde{\mathbf{w}} \in \tilde{\mathcal{V}_d},
\end{equation}

\begin{equation}
\mathbf{q} = (\mathbf{R}({\mathbf{\tilde{u}})}-\mathbf{f}) - \tilde{\Pi}((\mathbf{R}({\mathbf{\tilde{u}})}-\mathbf{f})). 
\end{equation}
The role of pressure herein is to impose the divergence free condition on the velocity field. In this formulation, only the velocity sub-scales have been accounted for, and the pressure terms arising from standard Galerkin procedure are retained and treated like a forcing function. This leads to additional stabilization terms to the standard Galerkin procedure given by,

\begin{equation}
\bigg(\frac{\partial \mathbf{\tilde{u}}}{\partial t},\mathbf{\tilde{w}}\bigg)_{\Omega'} + (\mathbf{\tilde{u}}\cdot\nabla \mathbf{\tilde{u}},\mathbf{\tilde{w}})_{\Omega'} + \nu(\nabla \mathbf{\tilde{u}},\nabla \mathbf{\tilde{w}})_{\Omega'} - (\tilde{p},\nabla \cdot \mathbf{\tilde{w}})= (\mathbf{f},\mathbf{\tilde{w}}) + \tau[(\mathbf{q}\cdot\nabla \mathbf{\tilde{u}},\mathbf{\tilde{w}})_{\Omega'} + (\mathbf{\tilde{u}}\cdot\nabla \mathbf{q},\mathbf{\tilde{w}})_{\Omega'} - \nu(\mathbf{q},\nabla^2{\mathbf{\tilde{w}}})_{\Omega'}] \quad \quad \forall \tilde{\mathbf{w}} \in \tilde{\mathcal{V}_d},
\label{MZ_NS}
\end{equation}

\begin{equation}
\mathbf{q} = (\mathbf{R}({\mathbf{\tilde{u}})}+ \nabla \tilde{p}-\mathbf{f}) - \tilde{\Pi}((\mathbf{R}({\mathbf{\tilde{u}})}+ \nabla \tilde{p}-\mathbf{f})). 
\label{}
\end{equation}
\color{black}Although closure terms were obtained for the momentum equations in Equation \eqref{MZ_NS}, the effect of the velocity sub-scales on the continuity equation should also be accounted for. Hence, an approximate form of the velocity sub-scales is required. To this end, consider Equation \eqref{subscale_find} in a re-arranged form: 
\begin{equation}
\bigg(\frac{\partial \mathbf{\tilde{u}}}{\partial t},\mathbf{\tilde{w}}\bigg)_{\Omega'} + (\mathbf{R}({\mathbf{\tilde{u}}})-\tau \mathbf{R'}(\mathbf{q}),\mathbf{\tilde{w}})_{\Omega'} + (\mathbf{J}(\mathbf{\tilde{u}}-\tau\mathbf{q}),\mathbf{\tilde{w}})_{\Gamma'} = (\mathbf{f},\mathbf{\tilde{w}}) \quad \forall \tilde{\mathbf{w}} \in \tilde{\mathcal{V}_d},
\label{subscale_find2}
\end{equation}
where the operators $\mathbf{R}$ and $\mathbf{J}$ are non-linear and linear respectively and $\mathbf{R'}$ is the linearization of $\mathbf{R}$ about $\mathbf{\tilde{u}}$. For small sub-scale $\mathbf{u'}$ \cite{VMS3} approximation, we have,
\begin{equation}
\mathbf{R}(\mathbf{\tilde{u}}+\mathbf{u'}) \approx \mathbf{R}(\mathbf{\tilde{u}}) +\mathbf{R'}(\mathbf{u'}).
\end{equation}
Consequently, we can express velocity sub-scales approximately as
\begin{equation}
\mathbf{u'} \approx -\tau \mathbf{q}.
\end{equation}
A similar form of sub-scales was also obtained by Wang et al. \cite{VMSE} by writing an asymptotic series in terms of residual \cite{VMS3}\color{black}. Finally, the effect of the sub-scales on the continuity equation is taken into consideration as follows: 
\begin{equation}
(\nabla\cdot(\mathbf{\tilde{u}+u')},\tilde{k})_{\Omega'} = 0 \quad \forall \tilde{k} \in \tilde{\mathcal{K}},
\end{equation}

\begin{equation}
(\nabla\cdot(\mathbf{\tilde{u}-\tau \mathbf{q})},\tilde{k})_{\Omega'} = 0 \quad \forall \tilde{k} \in \tilde{\mathcal{K}}.
\end{equation}
By applying integration by parts and using the fact that sub-scales vanish at the elemental boundaries, we have the following formulation for the continuity equation:
\begin{equation}
(\nabla.\mathbf{\tilde{u}},\tilde{k})_{\Omega'} + \tau(\mathbf{q},\nabla{\tilde{k}})_{\Omega'}= 0 \quad \forall \tilde{k} \in \tilde{\mathcal{K}},
\end{equation}
The next step is to discretize the above equation in time using the $\theta$ family of methods. This resulting variational problem at each time step is to find $\mathbf{\tilde{u}}^{n+\theta} \in \tilde{\mathcal{V}}_d$ and $ \tilde{p}^{n+\theta} \in \tilde{\mathcal{K}}$ such that

\begin{align}
\begin{split}
\bigg(\frac{\mathbf{\tilde{u}}^{n+1}-\mathbf{\tilde{u}}^{n}}{\Delta t},\mathbf{\tilde{w}}\bigg)_{\Omega'} + ({\mathbf{\tilde{u}}^{n+\theta}}\cdot\nabla {\mathbf{\tilde{u}}^{n+\theta}},\mathbf{\tilde{w}})_{\Omega'} + \nu(\nabla {\mathbf{\tilde{u}}^{n+\theta}},\nabla \mathbf{\tilde{w}})_{\Omega'} - (\tilde{p}^{n+\theta},\nabla \cdot \mathbf{\tilde{w}})= \\ (\mathbf{f}^{n+\theta},\mathbf{\tilde{w}}) + \tau[({{\mathbf{q}}^{n+\theta}}\cdot\nabla {{\mathbf{\tilde{u}}}^{n+\theta}},\mathbf{\tilde{w}})_{\Omega'} + ({\mathbf{\tilde{u}}^{n+\theta}}\cdot\nabla {\mathbf{q}^{n+\theta}},\mathbf{\tilde{w}})_{\Omega'} - \nu({\mathbf{q}^{n+\theta}},\nabla^2{\mathbf{\tilde{w}}})_{\Omega'}]  \quad \forall \tilde{\mathbf{w}}, \in \tilde{\mathcal{V}_d}
\end{split}
\end{align}

\begin{equation}
{\mathbf{q}^{n+\theta}} = (\mathbf{R}({{\mathbf{\tilde{u}}^{n+\theta}})}+ \nabla \tilde{p}^{n+\theta}-{\mathbf{f}^{n+\theta}}) - \tilde{\Pi}((\mathbf{R}({{\mathbf{\tilde{u}}^{n+\theta}})}+ \nabla \tilde{p}^{n+\theta}-\mathbf{f}^{n+\theta})),
\end{equation}

\begin{equation}
(\nabla.{\mathbf{\tilde{u}}^{n+\theta}},\tilde{k})_{\Omega'} + \tau({\mathbf{q}^{n+\theta}},\nabla{\tilde{k}})_{\Omega'}= 0 \quad \forall \tilde{k} \in \tilde{\mathcal{K}}.
\label{cweak}
\end{equation}
 One way to linearize the above set of non-linear equations is by using Picard iteration based technique given by

\begin{align}
\begin{split}
\bigg(\frac{{\mathbf{\tilde{u}}^{n+\theta,i+1}}-{\mathbf{\tilde{u}}}^{n}}{\theta \Delta t},\mathbf{\tilde{w}}\bigg)_{\Omega'} + ({\mathbf{\tilde{u}}^{n+\theta,i}}\cdot\nabla {\mathbf{\tilde{u}}^{n+\theta,i+1}},\mathbf{\tilde{w}})_{\Omega'} + \nu(\nabla {\mathbf{\tilde{u}}^{n+\theta,i+1}},\nabla \mathbf{\tilde{w}})_{\Omega'} - (\tilde{p}^{n+\theta,i+1},\nabla \cdot \mathbf{\tilde{w}})= \\ ({\mathbf{f}^{n+\theta,i}},\mathbf{\tilde{w}}) + \tau[({\mathbf{q}^{n+\theta,i}}\cdot\nabla {\mathbf{\tilde{u}}^{n+\theta,i+1}},\mathbf{\tilde{w}})_{\Omega'} + ({\mathbf{\tilde{u}}^{n+\theta,i}}\cdot\nabla {\mathbf{q}^a},\mathbf{\tilde{w}})_{\Omega'} - \nu({\mathbf{q}^a},\nabla^2{\mathbf{\tilde{w}}})_{\Omega'}] \quad \forall \tilde{\mathbf{w}} \in \tilde{\mathcal{V}_d},
\end{split}
\label{nsweak}
\end{align}
where $i+1$ and $i$ denote the present and previous iteration respectively. It can be noted that ${\mathbf{q}^{a}}$ and ${\mathbf{q}^{n+\theta,i}}$ are defined differently. This has been done so that $({\mathbf{\tilde{u}}^{n+\theta,i}}\cdot\nabla {\mathbf{\tilde{u}}^{n+\theta,i+1}},{\mathbf{\tilde{w}}})_{\Omega'}$ and $\tau({\mathbf{q}^{n+\theta,i}}\cdot\nabla {\mathbf{\tilde{u}}^{n+\theta,i+1}},\mathbf{\tilde{w}})_{\Omega'}$ can be merged together. This is possible because ${{\mathbf{q}}^{n+\theta,i}}$ is calculated from previous iteration variables. {\color{black} This is similar to Codina's procedure \cite{OSS} of adding sub-scales to the convective velocity, and a direct consequence of retaining the non-linearity  in the VMS formulation \cite{OSS2}}. Defining ${\mathbf{q}^a}$ in this manner allows for an implicit calculation of the memory terms which is similar to the stabilization term in \cite{OSS,SUPG,GLS,SUPG2} as follows:
\begin{equation}
{\mathbf{q}^a} = (\mathbf{\tilde{u}}^{n+\theta,i}\cdot\nabla {\mathbf{\tilde{u}}^{n+\theta,i+1}} - \nu \nabla^2 {\mathbf{\tilde{u}}^{n+\theta,i+1}} + \nabla \tilde{p}^{n+\theta,i+1}-{\mathbf{f}^{n+\theta,i}}) - \tilde{\Pi}({\mathbf{\tilde{u}}^{n+\theta,i}}\cdot\nabla {\mathbf{\tilde{u}}^{n+\theta,i}}- \nu \nabla^2 {\mathbf{\tilde{u}}^{n+\theta,i}} + \nabla \tilde{p}^{n+\theta,i}-{\mathbf{f}^{n+\theta,i}}), 
\label{nsweak2}
\end{equation}

\begin{equation}
{\mathbf{q}^{n+\theta,i}} = ({\mathbf{\tilde{u}}^{n+\theta,i}}\cdot\nabla {\mathbf{\tilde{u}}^{n+\theta,i}} - \nu \nabla^2 {\mathbf{\tilde{u}}^{n+\theta,i}} + \nabla \tilde{p}^{n+\theta,i}-{\mathbf{f}^{n+\theta,i}}) - \tilde{\Pi}({\mathbf{\tilde{u}}^{n+\theta,i}}\cdot\nabla {\mathbf{\tilde{u}}^{n+\theta,i}}- \nu \nabla^2 {\mathbf{\tilde{u}}^{n+\theta,i}} + \nabla \tilde{p}^{n+\theta,i}-{\mathbf{f}^{n+\theta,i}}),
\label{nsweak3}
\end{equation}
Equations \eqref{cweak}, \eqref{nsweak}, \eqref{nsweak2} and \eqref{nsweak3} are iterated until convergence of the relative norm of the solution vector between two consecutive iterations is achieved.

\subsection{Homogeneous Isotropic Turbulence (HIT)}
\begin{figure}
	\begin{subfigure}[b]{0.5\textwidth}
		\centering
		\includegraphics[width=\textwidth]{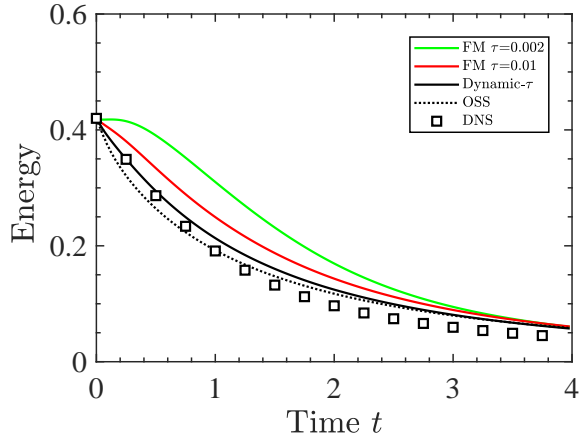}
		\caption{Time evolution of kinetic energy.}
		\label{HIT_1}
	\end{subfigure}
	\begin{subfigure}[b]{0.5\textwidth}
		\centering
		\includegraphics[width=\textwidth]{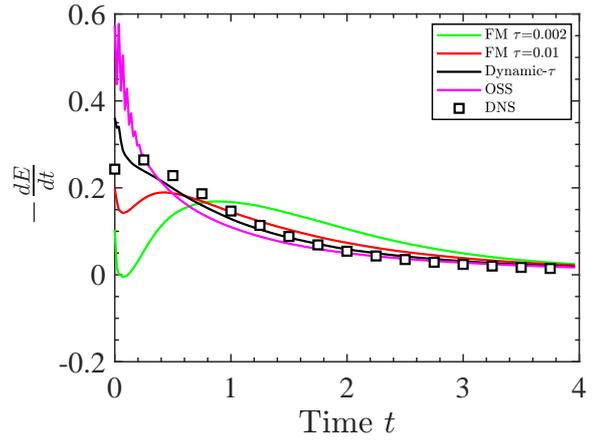}
		\caption{Rate of kinetic energy decay.}
		\label{HIT_2}
	\end{subfigure}
	\begin{subfigure}[b]{0.5\textwidth}
		\centering
		\includegraphics[width=\textwidth]{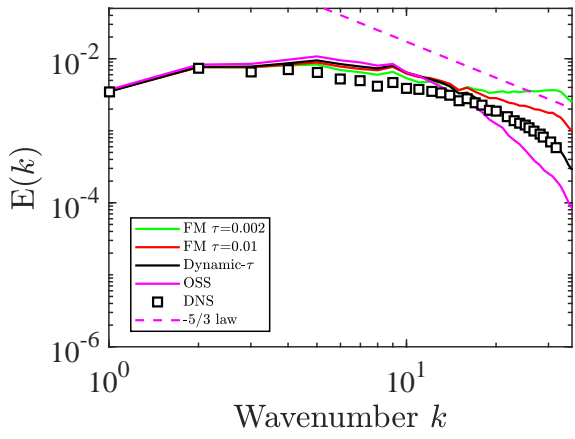}
		\caption{Energy Spectra at T=2.0.}
		\label{HIT_3}
	\end{subfigure}
	\begin{subfigure}[b]{0.5\textwidth}
		\centering
		\includegraphics[width=\textwidth]{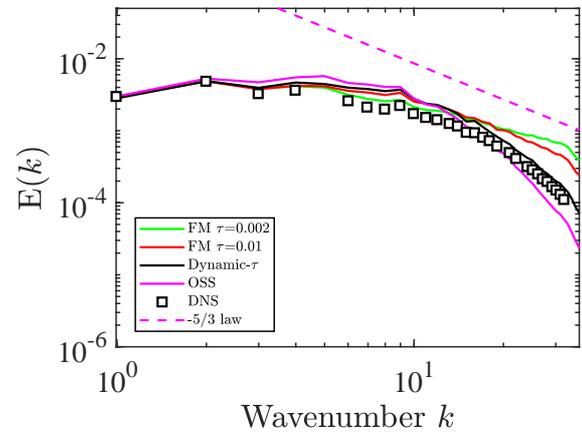}
		\caption{Energy Spectra at T=4.0.}
		\label{HIT_4}
	\end{subfigure}
	\caption{{\color{black} (a) Kinetic energy}, (b) dissipation, (c) energy spectra at T = 2 and (d)  energy spectra at T=4 for homogeneous isotropic turbulence at initial $Re_\lambda \approx 65$.}
	\label{HIT}
\end{figure}

\begin{figure}
	\begin{subfigure}[b]{0.5\textwidth}
		\centering
		\includegraphics[width=\textwidth]{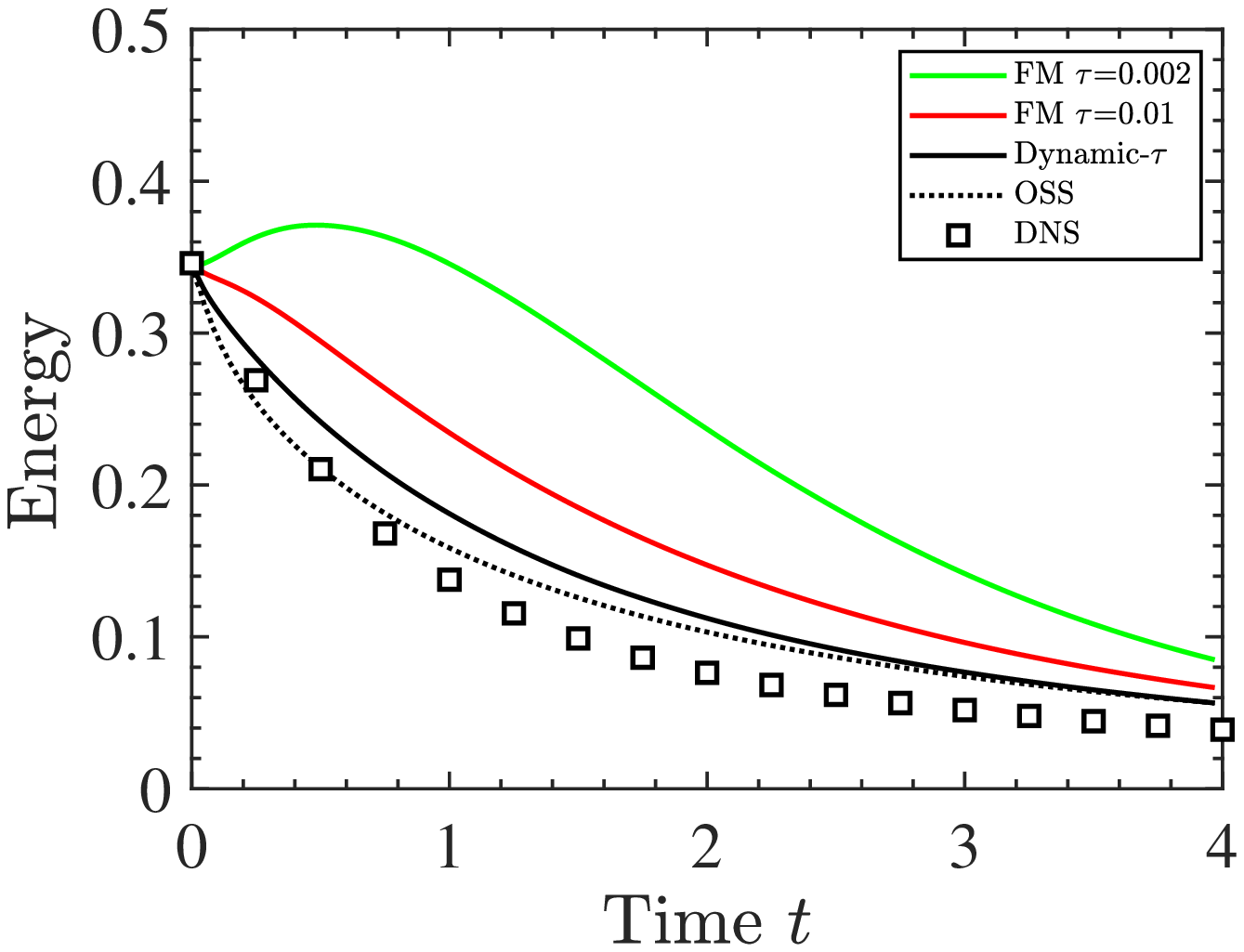}
		\caption{Time evolution of kinetic energy.}
		\label{HIT1_1}
	\end{subfigure}
	\begin{subfigure}[b]{0.5\textwidth}
		\centering
		\includegraphics[width=\textwidth]{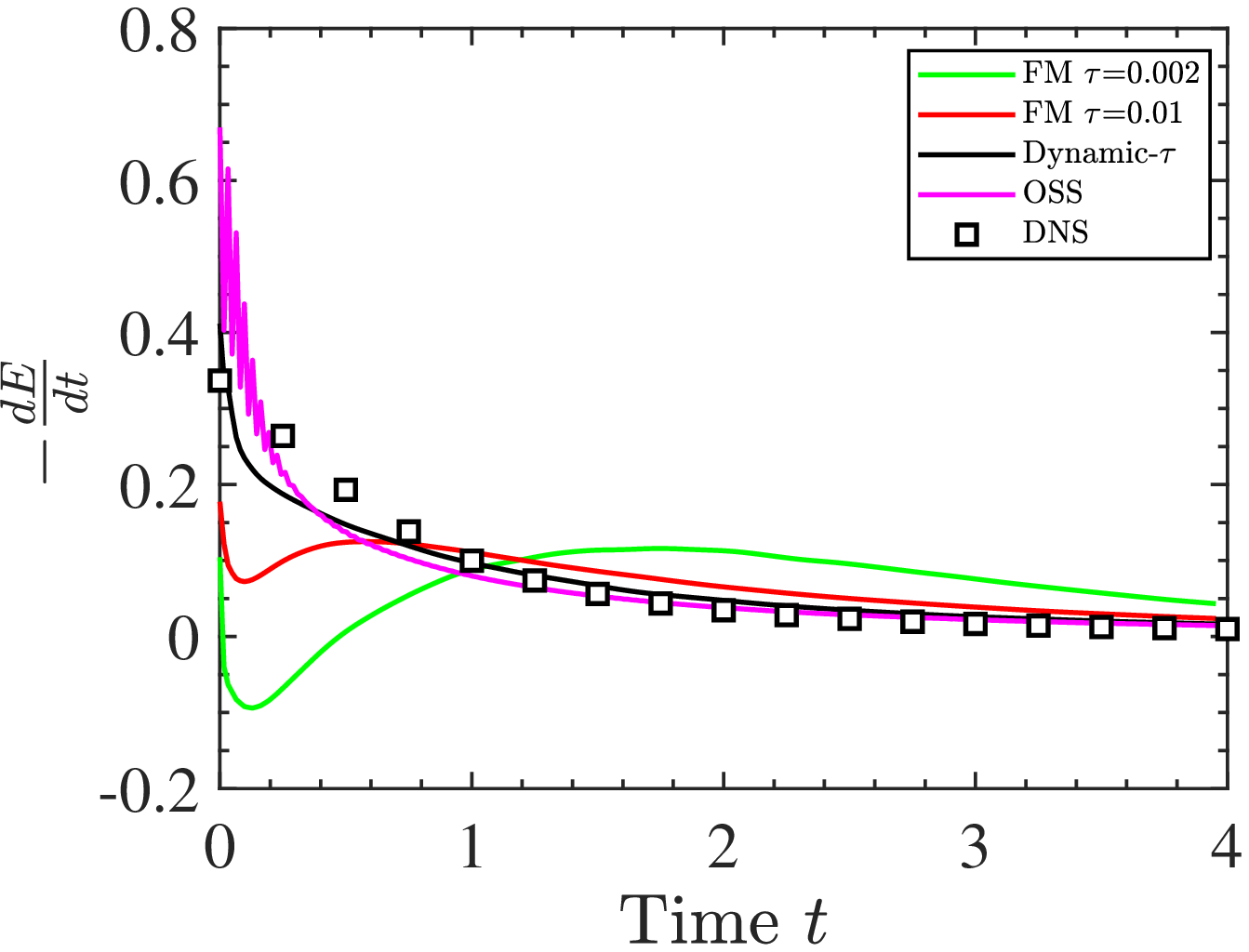}
		\caption{Rate of kinetic energy decay.}
		\label{HIT1_2}
	\end{subfigure}
	\begin{subfigure}[b]{0.5\textwidth}
		\centering
		\includegraphics[width=\textwidth]{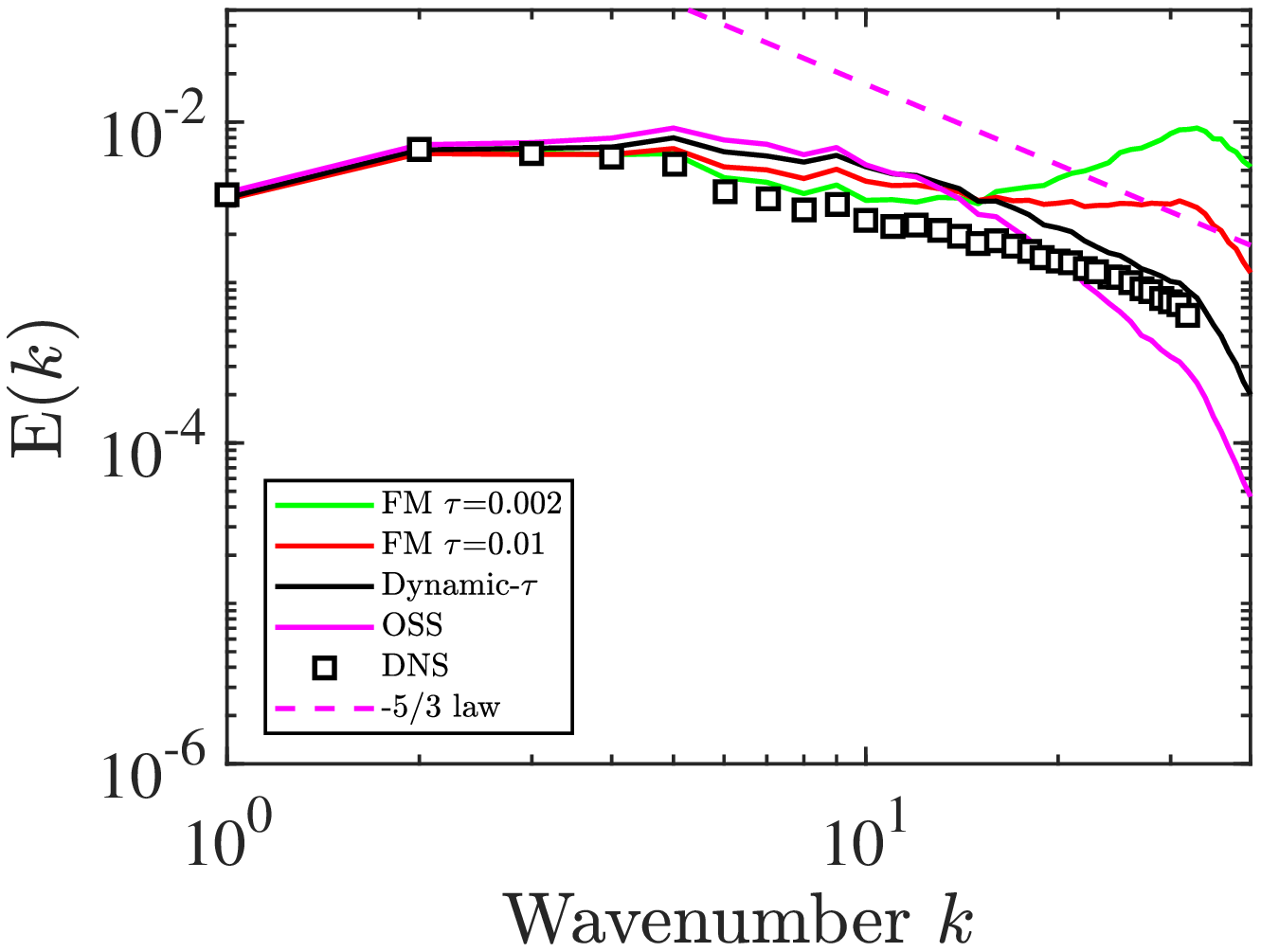}
		\caption{Energy Spectra at T=2.0.}
		\label{HIT1_3}
	\end{subfigure}
	\begin{subfigure}[b]{0.5\textwidth}
		\centering
		\includegraphics[width=\textwidth]{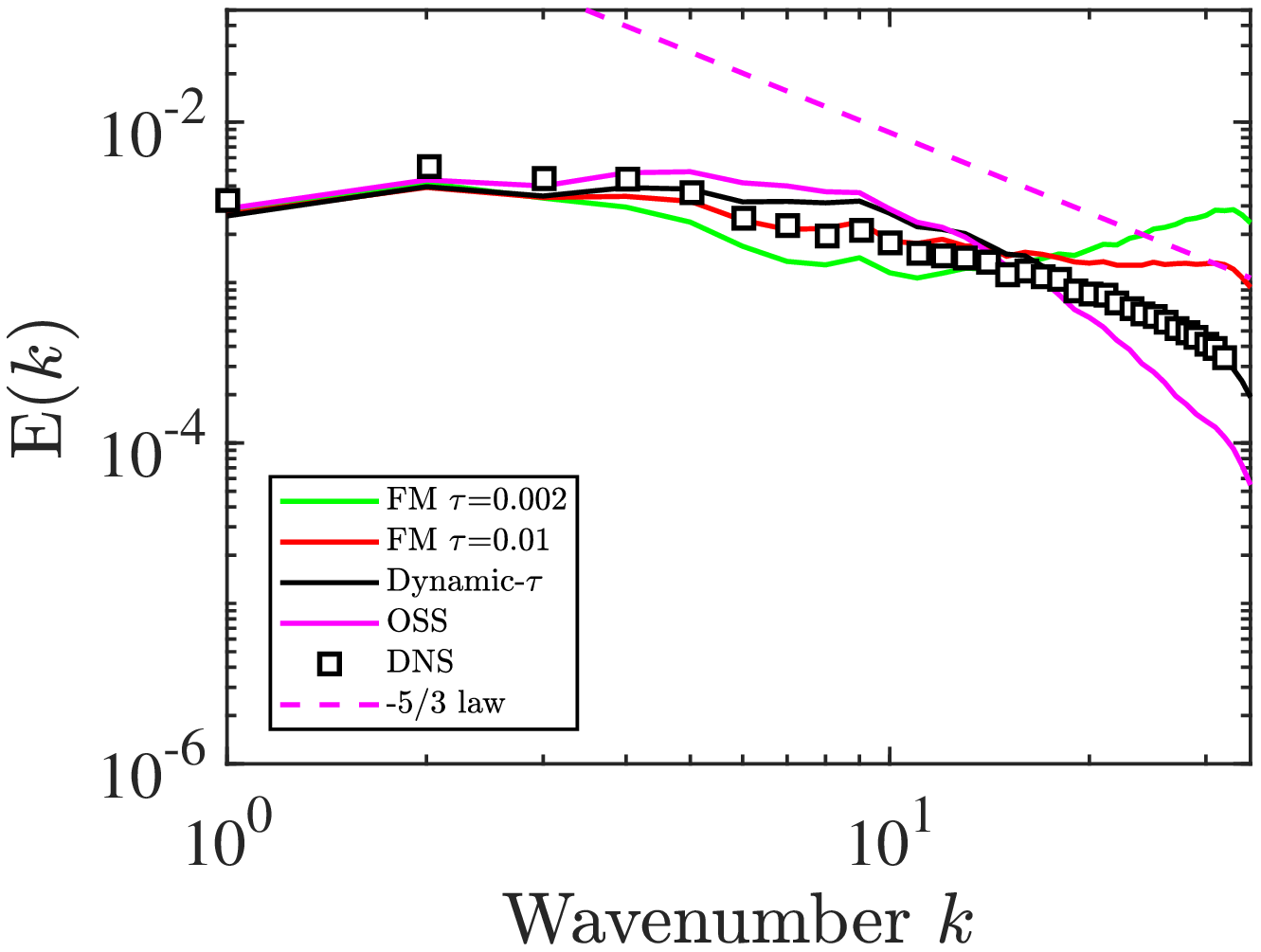}
		\caption{Energy Spectra at T=4.0.}
		\label{HIT1_4}
	\end{subfigure}
	\caption{{\color{black} (a) Kinetic energy}, (b) dissipation, (c) energy spectra at T = 2 and (d)  energy spectra at T=4 for homogeneous isotropic turbulence at initial $Re_\lambda \approx 75$.}
    \label{HIT1}
\end{figure}

\begin{figure}
	\centering
	\begin{subfigure}[b]{0.5\textwidth}
		\centering
		\includegraphics[width=\textwidth]{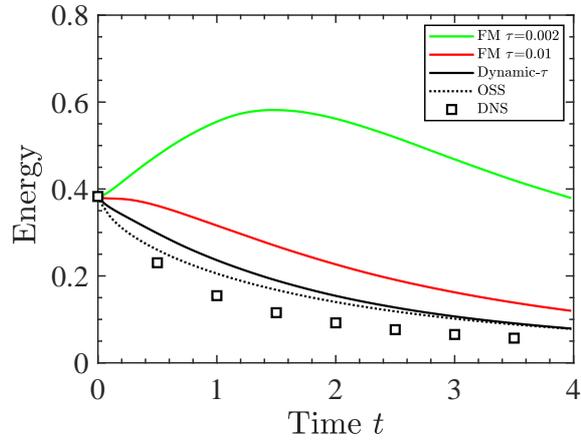}
		\caption{Time evolution of kinetic energy.}
		\label{HIT2_1}
	\end{subfigure}
	\begin{subfigure}[b]{0.5\textwidth}
		\centering
		\includegraphics[width=\textwidth]{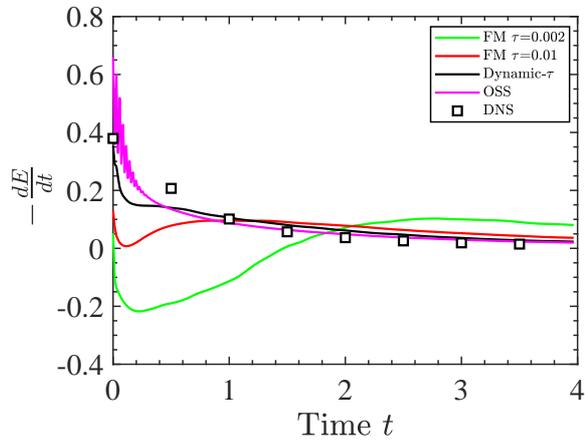}
		\caption{Rate of kinetic energy decay.}
		\label{HIT2_2}
	\end{subfigure}
	\begin{subfigure}[b]{0.5\textwidth}
		\centering
		\includegraphics[width=\textwidth]{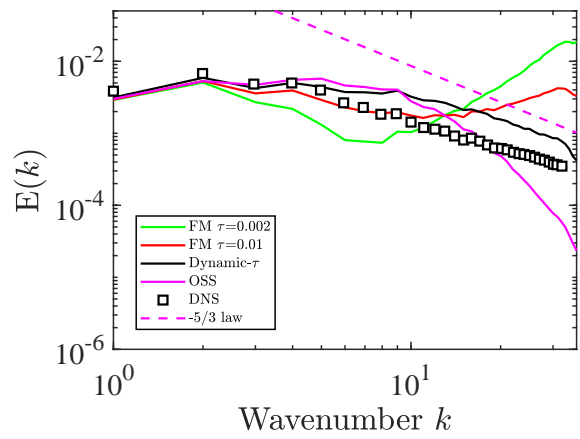}
		\caption{Energy Spectra at T=4.0.}
		\label{HIT2_3}
	\end{subfigure}
	\caption{{\color{black} (a) Kinetic energy}, (b) dissipation and (c)  energy spectra at T=4 for homogeneous isotropic turbulence at initial $Re_\lambda \approx 164$.}
	\label{HIT2}
\end{figure}

In this section, we present results for decaying homogeneous isotropic turbulence (HIT) and compare it to DNS. We choose the OSS model as a reference, as it has been shown to be a good VMS closure for turbulence \cite{OSSLES}.  The HIT problem has been extensively studied in literature both numerically \cite{HIT1,HIT2,HIT3} and experimentally \cite{HIT4}.  This problem is well defined in a 3-D periodic box and the initialization of the initial velocity field for DNS is done using  Rogallo's procedure \cite{ROGALLO} which assumes the following energy spectrum at initial time:

\begin{equation}
E(k,t=0) = {{q^2}\over{2A}}{1\over{k_p^{\sigma+1}}}{k^p}{exp({-{\sigma}\over{2}}({k \over k_p})^2)},
\end{equation}
where $k_p$ is the wavenumber at which the energy spectra peaks and A is defined as $\int_0^{\infty}{k^{\sigma}exp(-\sigma k^2/2)} dk$. The velocity in spectral space is given by

\begin{equation}
\mathbf{a}(\mathbf{k}) = ({{\alpha k k_2 + \beta k k_1} \over {k(k_1^2+k_2^2)^{1/2}}})\hat{i} + ({{\beta k_2 k_3 - \alpha k_1 k_3} \over {k(k_1^2+k_2^2)^{1/2}}})\hat{j} - ({{\beta(k_1^2+k_2^2)^{1/2}} \over {k}})\hat{k},
\end{equation}
where $k$ denotes the magnitude of the wavenumber vector and $\alpha$ and $\beta$ are defined as follows:

\begin{equation}
\alpha = ({E(k) \over {4 \pi k^2}})^{1/2}{e^{i \theta_1}}cos(\phi),
 \ \ 
\beta = ({E(k) \over {4 \pi k^2}})^{1/2}{e^{i \theta_2}}sin(\phi),
\end{equation}
where $\phi$, $\theta_1$, $\theta_2$ are uniformly distributed random numbers from 0 to $2\pi$. In all the simulations, $k_p= 3$,  $q^2=3$ and $\sigma = 4$. Although, the initial velocity field  satisfies the divergence free condition, it does not represent a physical homogeneous isotropic turbulent velocity field. To achieve this state, the field is allowed to decay to a lower  $Re_{\lambda}$ where the field will resemble a more realistic velocity field due to redistribution of energy \cite{HIT1}. Three different initial $Re_{\lambda}$ have been considered here: $Re_{\lambda} \approx$ 65, 75 and 164 where $Re_{\lambda}$ is defined as the Reynolds number based on the Taylor microscale $\lambda$ as follows:

\begin{equation}
Re_{\lambda} = {{u'} \lambda \over \nu},
\end{equation}
where  $u'$ is the velocity fluctuation/root mean square (rms) of the velocity field defined as $\sqrt{2k/3}$. The initial conditions for all these cases are generated from DNS simulations by starting at a higher $Re_{\lambda}$ and allowing it to reach our target $Re_{\lambda}$ of 65, 75 and 164 respectively. The kinematic viscosity for the three different cases are set to $\nu=$ 0.001, 0.0005 and 0.0001 respectively. The LES simulations utilize $64^3$ linear elements for all the three $Re_{\lambda}$ cases, the results of which are presented in Figures \ref{HIT},\ref{HIT1} and \ref{HIT2} respectively. This allows us to study the effects of increasing $Re_\lambda$ by retaining the same resolution.

At a relatively lower Reynolds number of $Re_{\lambda} \approx$ 65, all the models perform fairly well in predicting the time history of the resolved kinetic energy except the fixed memory models where an arbitrary choice of $\tau$ is used. As can be seen from the kinetic energy decay plots in Figures \ref{HIT_2}, \ref{HIT1_2} and \ref{HIT2_2}, for all the three different Reynolds numbers, the OSS model is not stable and can be seen to oscillate initially. This indicates that for the initial time period, the OSS model incorrectly forces the turbulence. We compare the energy spectra at two different times of T=2.0 and T=4.0 in Figure \ref{HIT1_3} and \ref{HIT1_4} respectively and observe that all models perform very well at the lower wavenumber modes. However, the OSS is clearly more dissipative at higher wave-numbers where it predicts a lower energy content compared to the present model and DNS. At the higher $Re_\lambda$ = 75 case, we find that both the Dynamic-$\tau$ and the OSS model predict reasonably the evolution of kinetic energy and rate of kinetic energy decay. If we compare the energy spectra at T=2.0 in Figure \ref{HIT1_3}, all the models predict a higher energy content across different wave-numbers compared to the DNS solution with the dynamic-$\tau$ again performing better at higher wave-numbers again. However, at T=4.0, when it decays to a lower $Re_{\lambda}$, the performance of all the models improve, as can be seen in Figure \ref{HIT1_4}. At the highest $Re_\lambda$ case, $Re_\lambda$ = 174, we find that the performance of all the model becomes worse compared to the DNS results. The energy spectra for this case in presented is Figure \ref{HIT2_3} where only the lower wavenumber modes are resolved accurately in comparison to DNS. One possible reason for the deterioration of performance at high $Re_{\lambda}$ cases is that the current VMS models are efficient in modeling the cross-stress terms and not the Reynolds stress terms \cite{VMSE} which dominate at higher $Re_{\lambda}$ values. 

\begin{figure}
	\centering
	\begin{subfigure}[b]{0.5\textwidth}
		\centering
		\includegraphics[width=\textwidth]{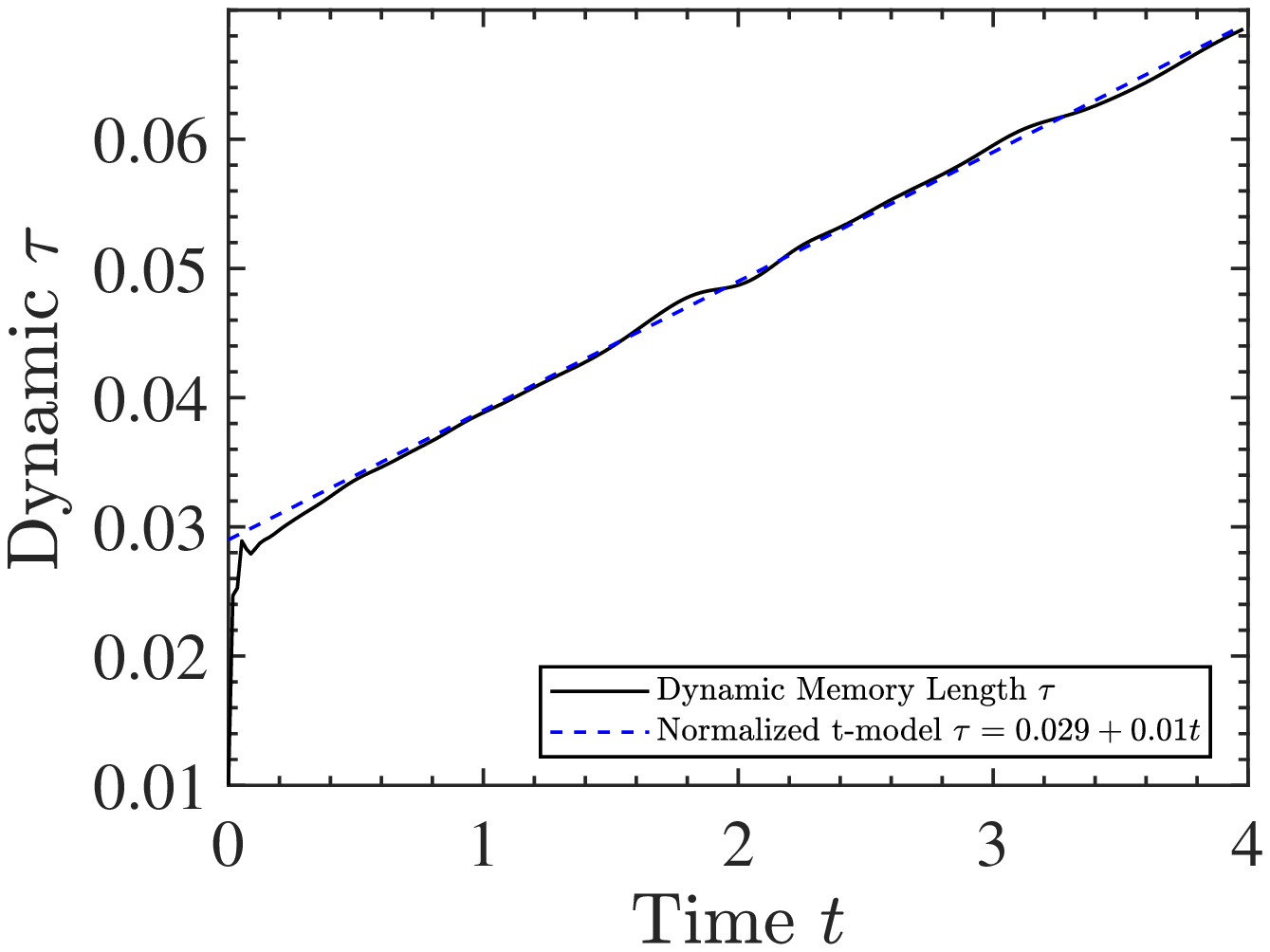}
		\caption{Initial $Re_{\lambda} \approx$ 65.}
		\label{TAU_1}
	\end{subfigure}
	\begin{subfigure}[b]{0.5\textwidth}
		\centering
		\includegraphics[width=\textwidth]{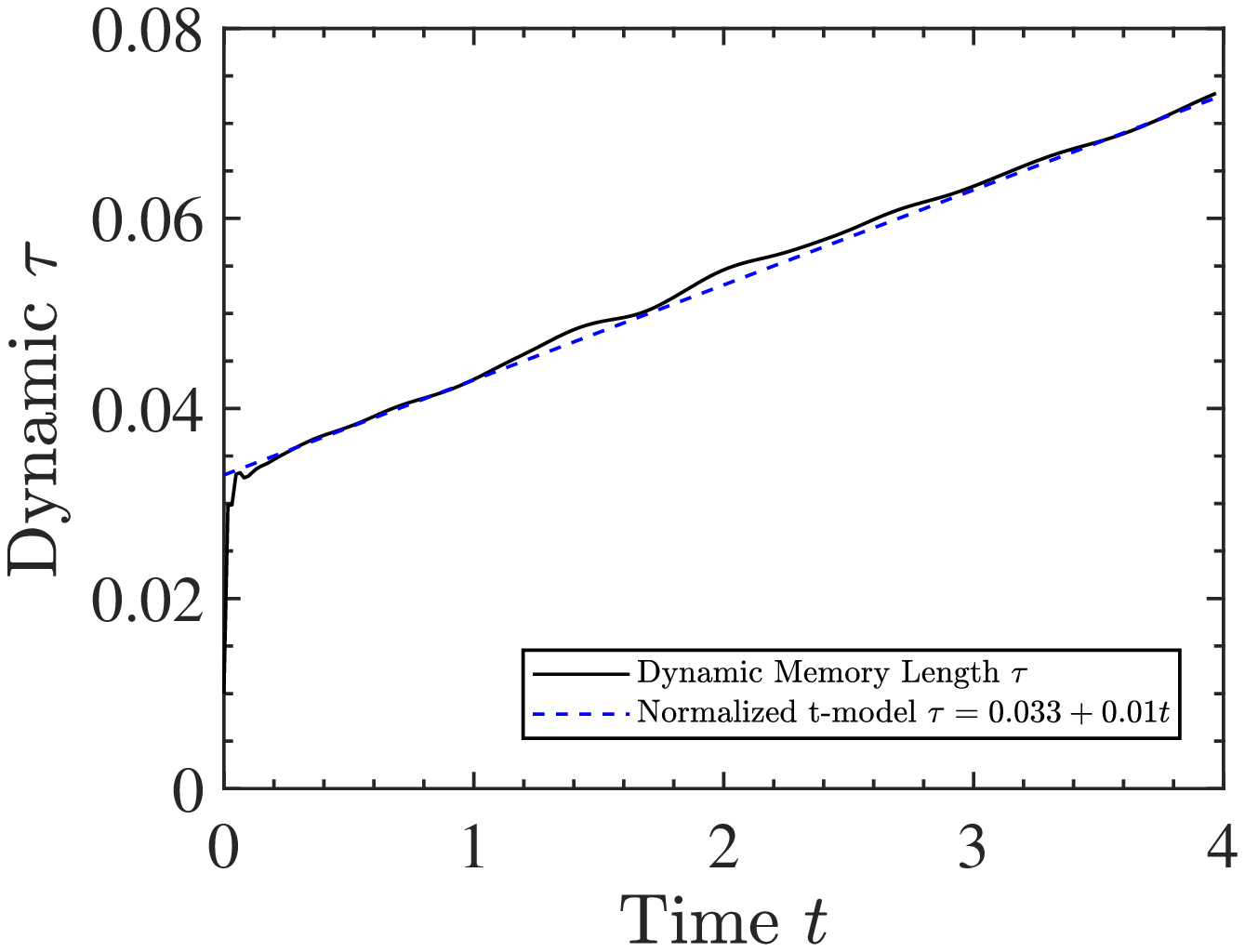}
		\caption{Initial $Re_{\lambda} \approx$ 75.}
		\label{TAU_2}
	\end{subfigure}
	\begin{subfigure}[b]{0.5\textwidth}
		\centering
		\includegraphics[width=\textwidth]{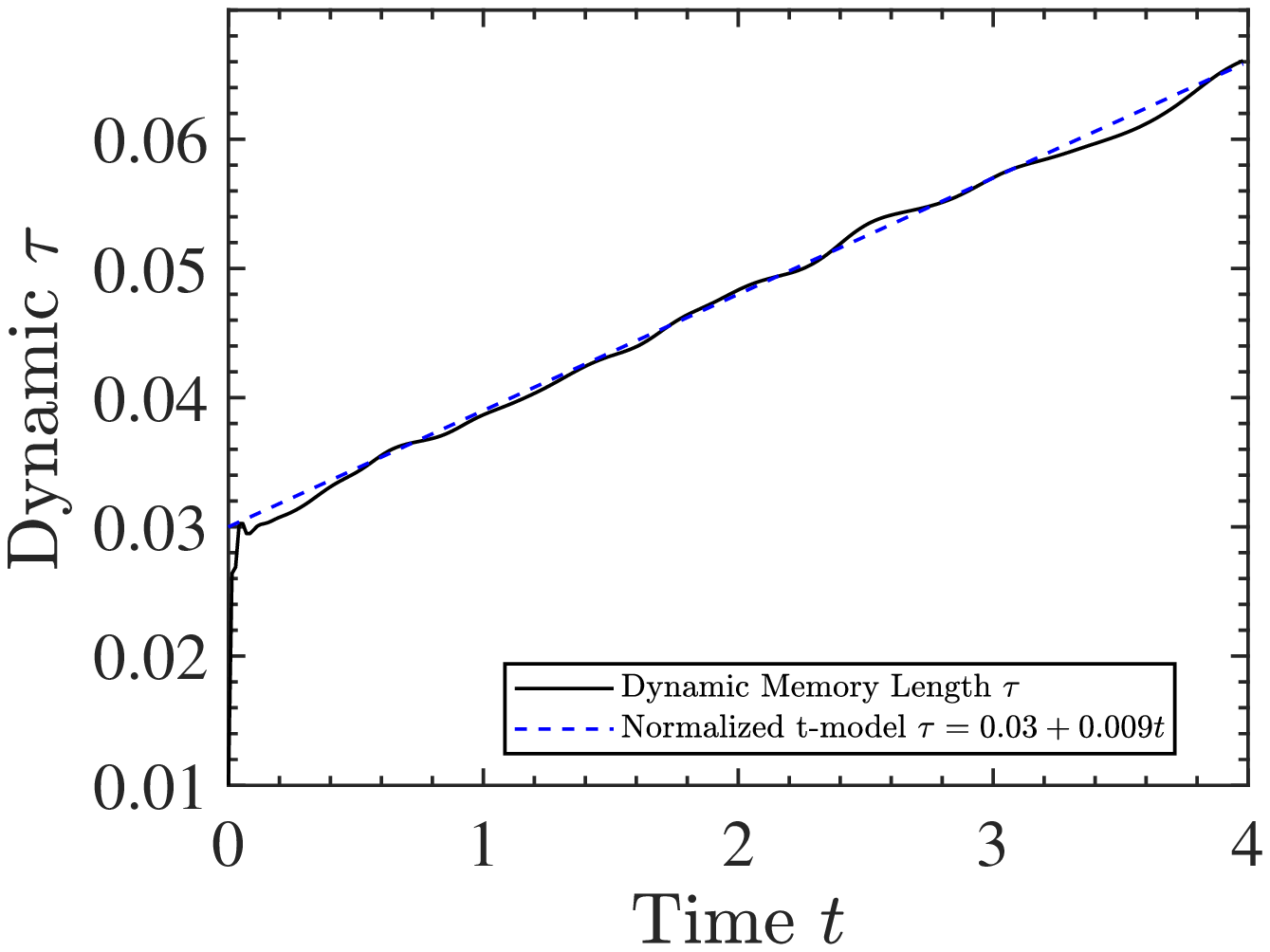}
		\caption{Initial $Re_{\lambda} \approx$ 164.}
		\label{TAU_3}
	\end{subfigure}
	\caption{Evolution of memory length $\tau$ predicted using our dynamic model for homogeneous isotropic turbulence for different initial $Re_\tau$. }
	\label{HITTAU}
\end{figure}

The time variation of the predicted dynamic memory length for all the three cases is shown in Figure \ref{HITTAU}. From the plots, it can be observed that the memory length increases almost linearly with time similar to the viscous Burgers equation. This is consistent with Stinis \cite{STINIST} of re-normalizing the t-model for stability and accuracy. Also, the predicted value for $\tau$ by our dynamic model is higher in comparison to the randomly chosen $\tau$ values for our fixed memory model. The differences in the temporal evolution between the dynamically-selected $\tau$ and the imposed $\tau$ is a further indicate that the dynamic model is necessary for calculating the memory length.

\subsection{Taylor Green Vortex (TGV)}
The next step in understanding the applicability of the proposed method is to employ the model on a turbulent flow that undergoes complex dynamics such the Taylor-Green vortex. This problem involves transition to turbulence-like flow, as well as decay.  Models such as the Smagorinsky which have been derived based on assumptions of homogeneity, isotropy and balance between sub-grid production and dissipation \cite{pope} might not optimally preform in such flows where there is non-homogenity and transition to turbulence.  Similar to HIT, this problem is well-defined on a 3-D periodic box with smooth initial conditions which are given as follows: 

\begin{equation}
u = U_o cos(x)sin(y)cos(z), \ \
v = -U_o sin(x)cos(y)cos(z), \ \
w = 0
\end{equation}
where $u,v, w$ denote the velocity in $x,y$ and $z$ directions respectively and $x,y, z$ $\in [-\pi L,\pi L]$. To study this problem, three different Reynolds numbers are considered: $Re = {U_o L \over \nu}=$ 400, 800 and 1600. The values for $L$ and $U_o$ are unity and the Re is changed solely by varying the kinematic viscosity $\nu$. The initial conditions for the velocity field are kept the same for all the cases. 

\begin{figure}
	\begin{subfigure}[b]{0.5\textwidth}
		\centering
		\includegraphics[width=\textwidth]{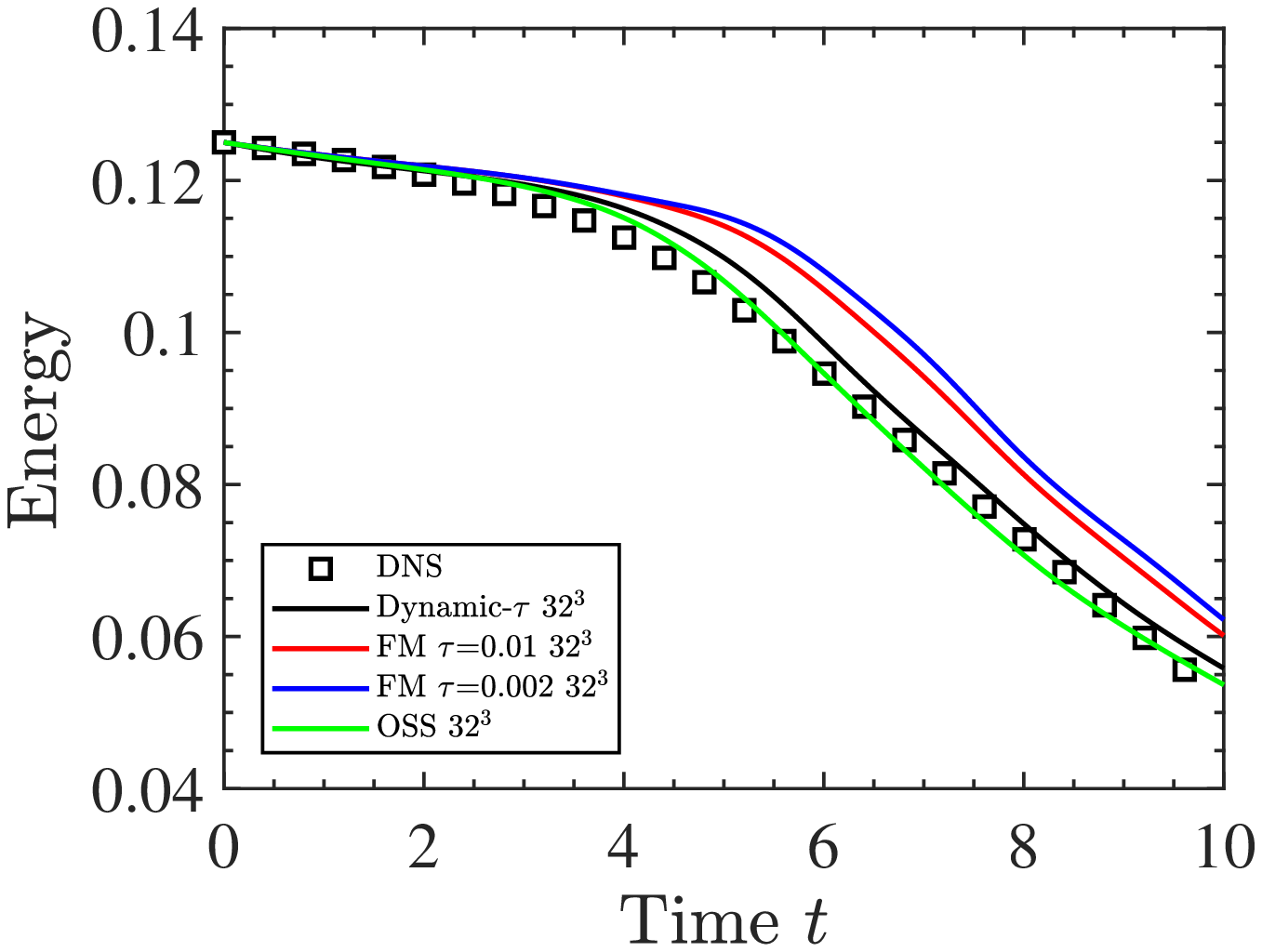}
		\caption{Time evolution of kinetic energy.}
		\label{TGV400_1}
	\end{subfigure}
	\begin{subfigure}[b]{0.5\textwidth}
		\centering
		\includegraphics[width=\textwidth]{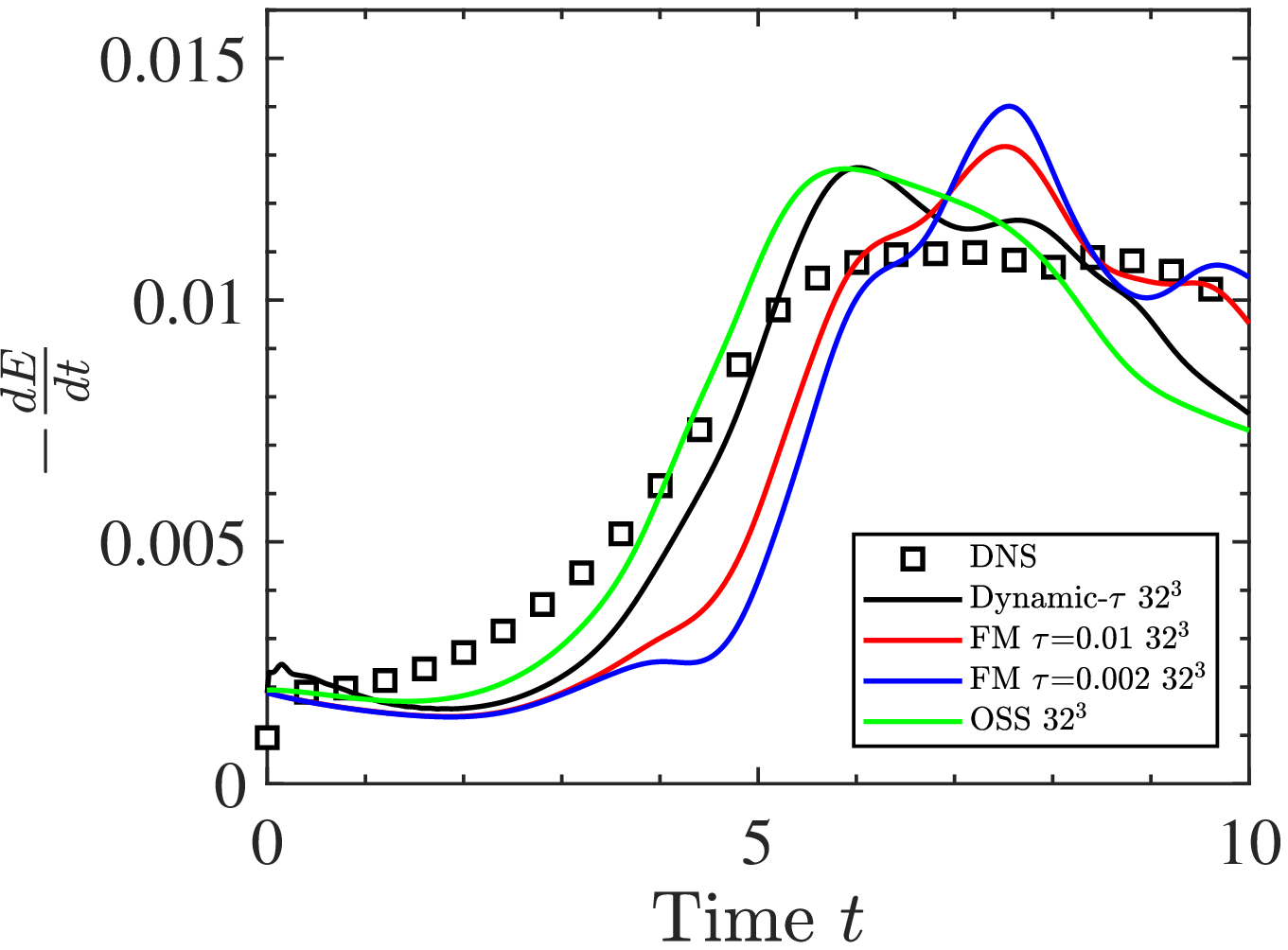}
		\caption{Rate of kinetic energy decay.}
		\label{TGV400_2}
	\end{subfigure}
    \begin{subfigure}[b]{0.5\textwidth}
    	\centering
    	\includegraphics[width=\textwidth]{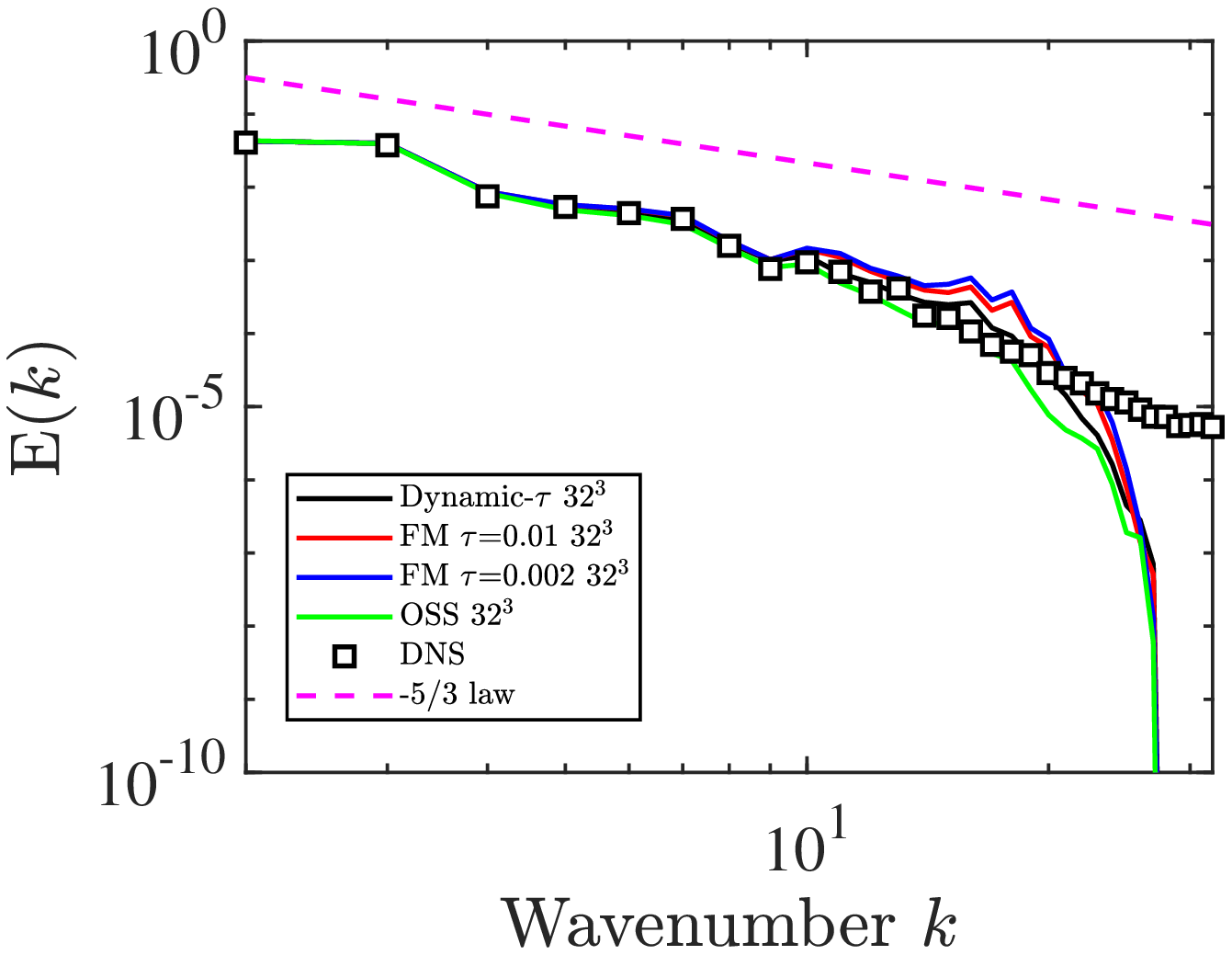}
    	\caption{Energy Spectra at T=5.0.}
    	\label{TGV400_3}
    \end{subfigure}
    \begin{subfigure}[b]{0.5\textwidth}
    	\centering
    	\includegraphics[width=\textwidth]{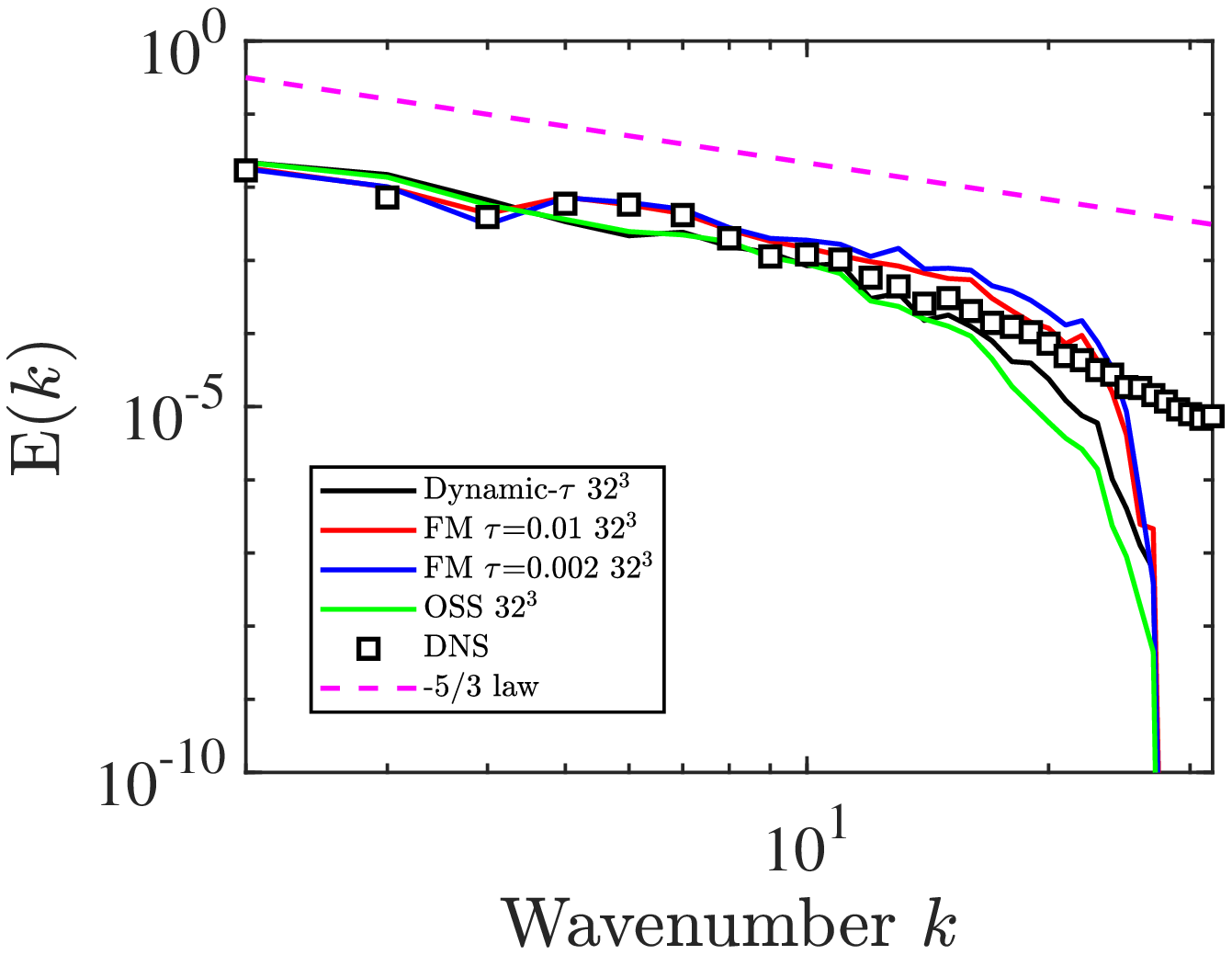}
    	\caption{Energy Spectra at T=10.0.}
    	\label{TGV400_4}
    \end{subfigure}
	\caption{{\color{black} (a) Kinetic energy}, (b) dissipation, (c) energy spectra at T = 5 and (d)  energy spectra at T=10 for Taylor Green vortex at Re=400 using different coarse graining methods.}
	\label{TGV400}
\end{figure}

\begin{figure}
	\begin{subfigure}[b]{0.5\textwidth}
		\centering
		\includegraphics[width=\textwidth]{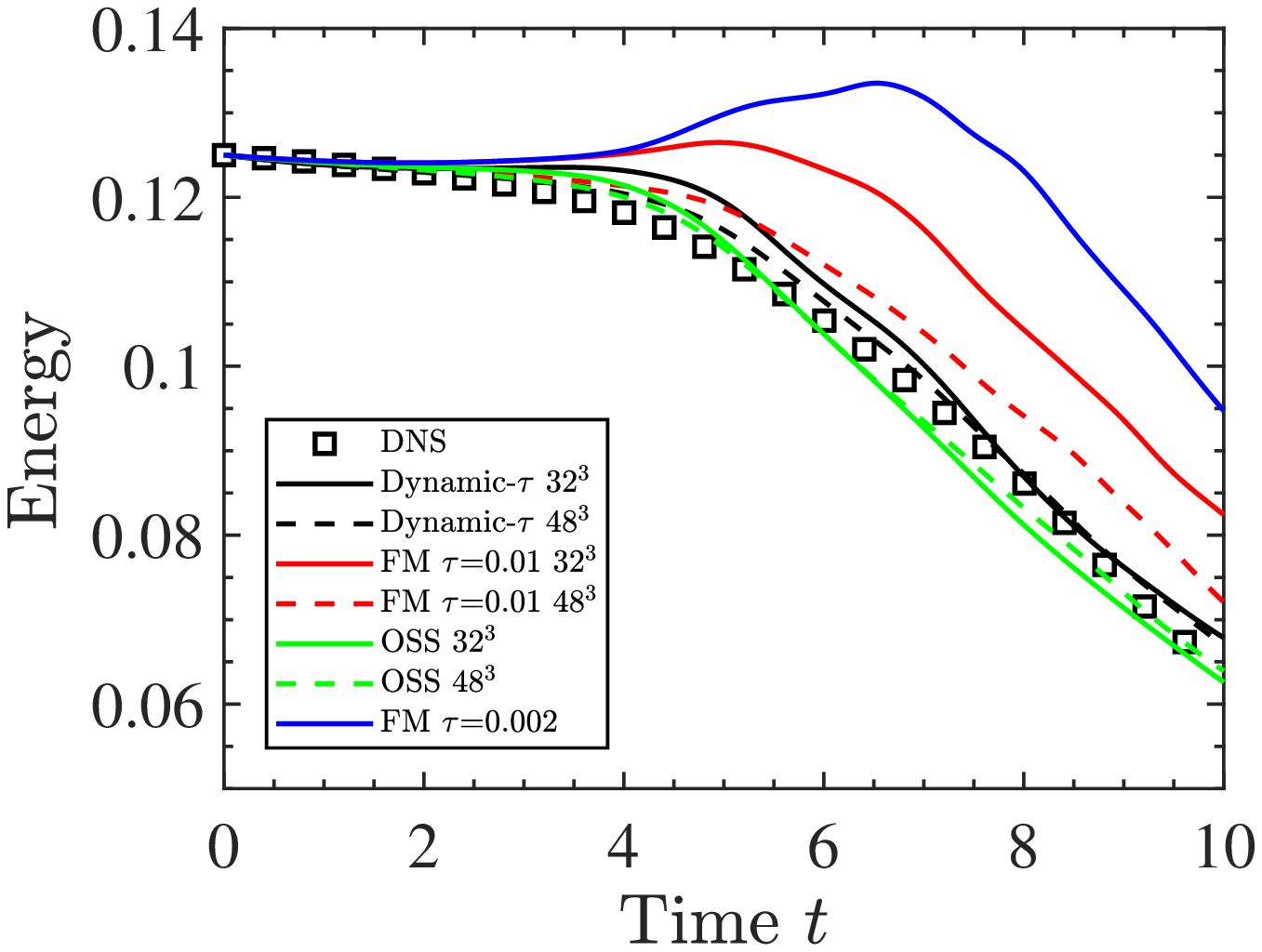}
		\caption{Time evolution of kinetic energy.}
		\label{TGV800_1}
	\end{subfigure}
	\begin{subfigure}[b]{0.5\textwidth}
		\centering
		\includegraphics[width=\textwidth]{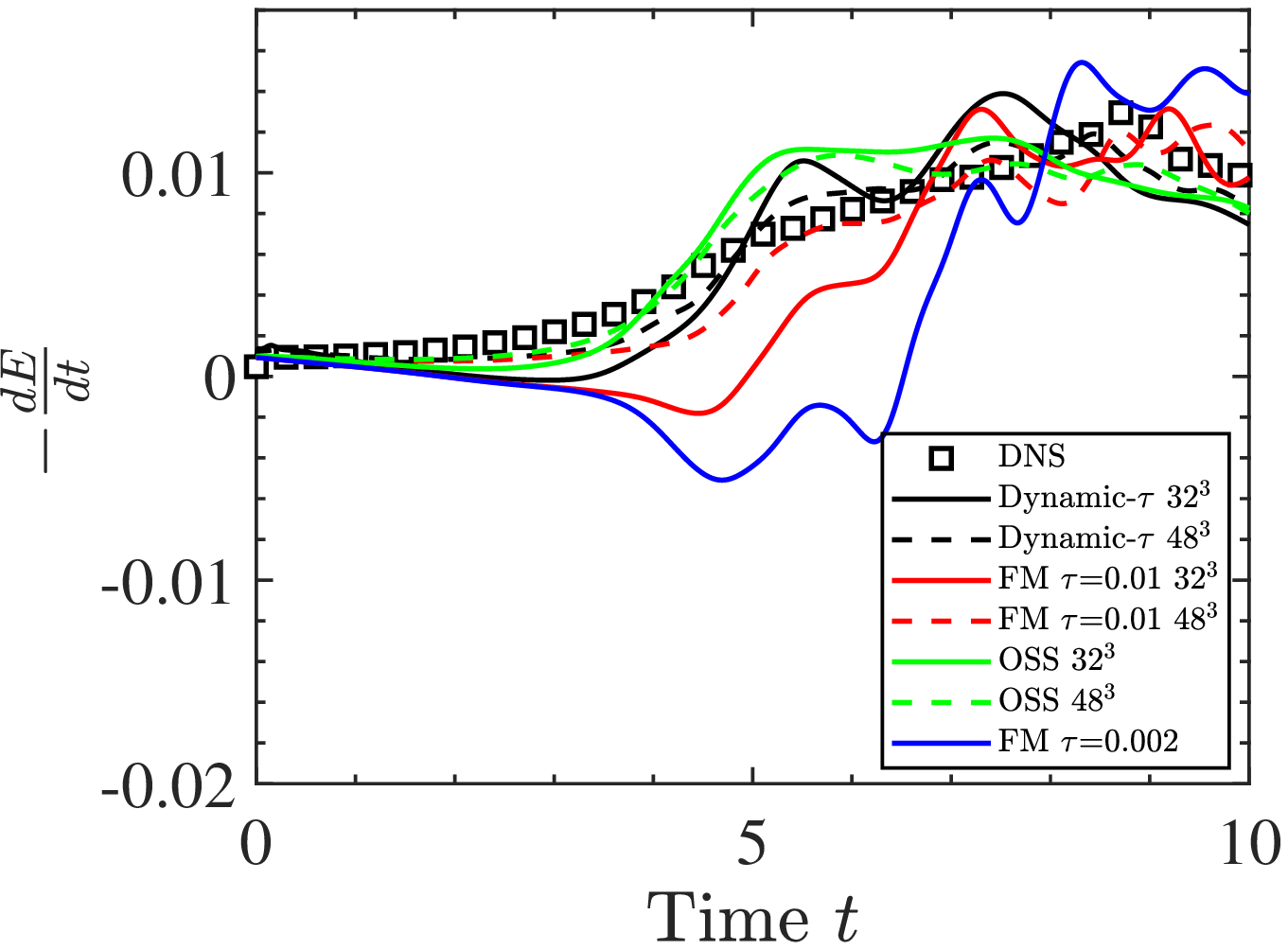}
		\caption{Rate of kinetic energy decay.}
		\label{TGV800_2}
	\end{subfigure}
    \begin{subfigure}[b]{0.5\textwidth}
    	\centering
    	\includegraphics[width=\textwidth]{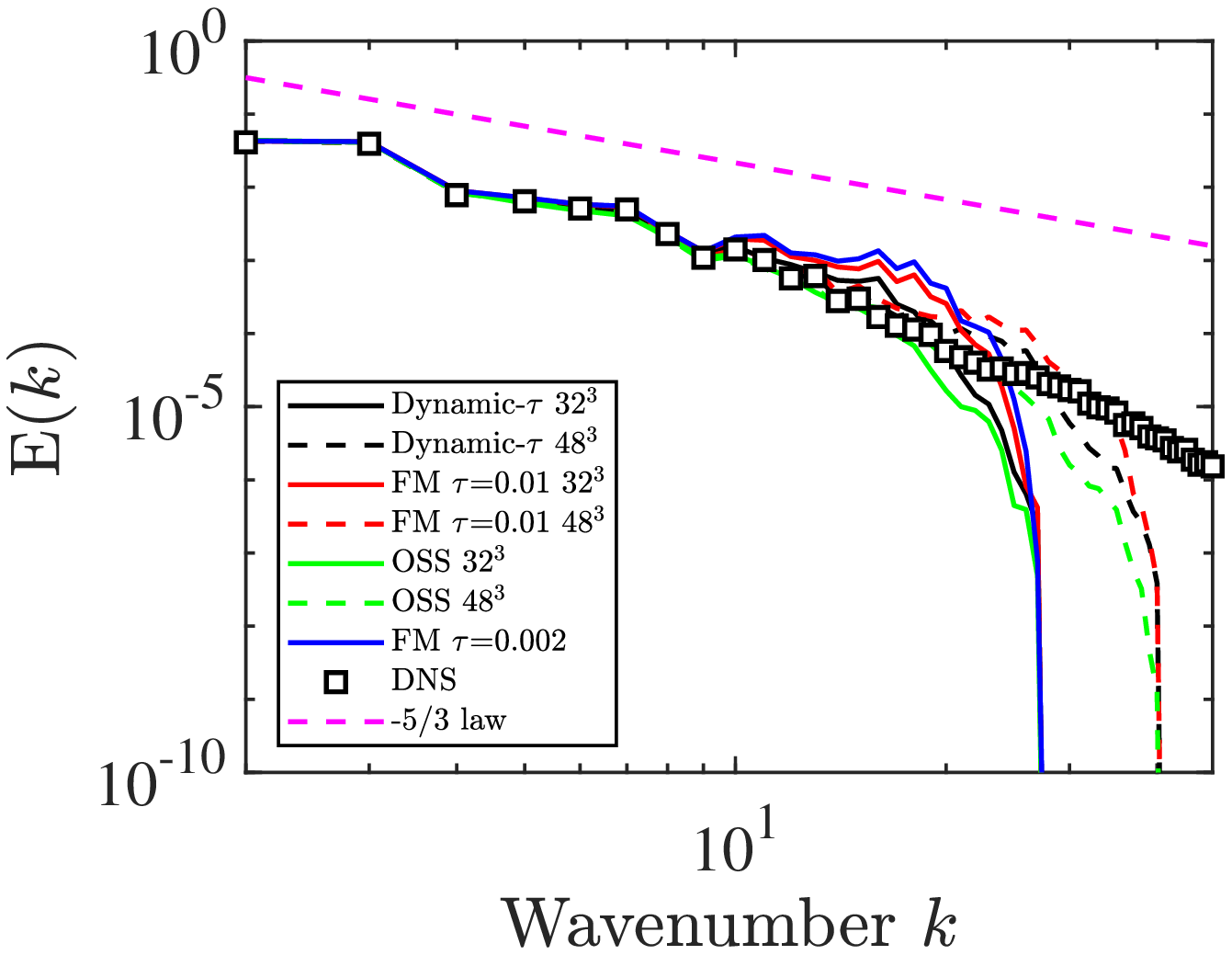}
    	\caption{Energy Spectra at T=5.0.}
    	\label{TGV800_3}
    \end{subfigure}
    \begin{subfigure}[b]{0.5\textwidth}
    	\centering
    	\includegraphics[width=\textwidth]{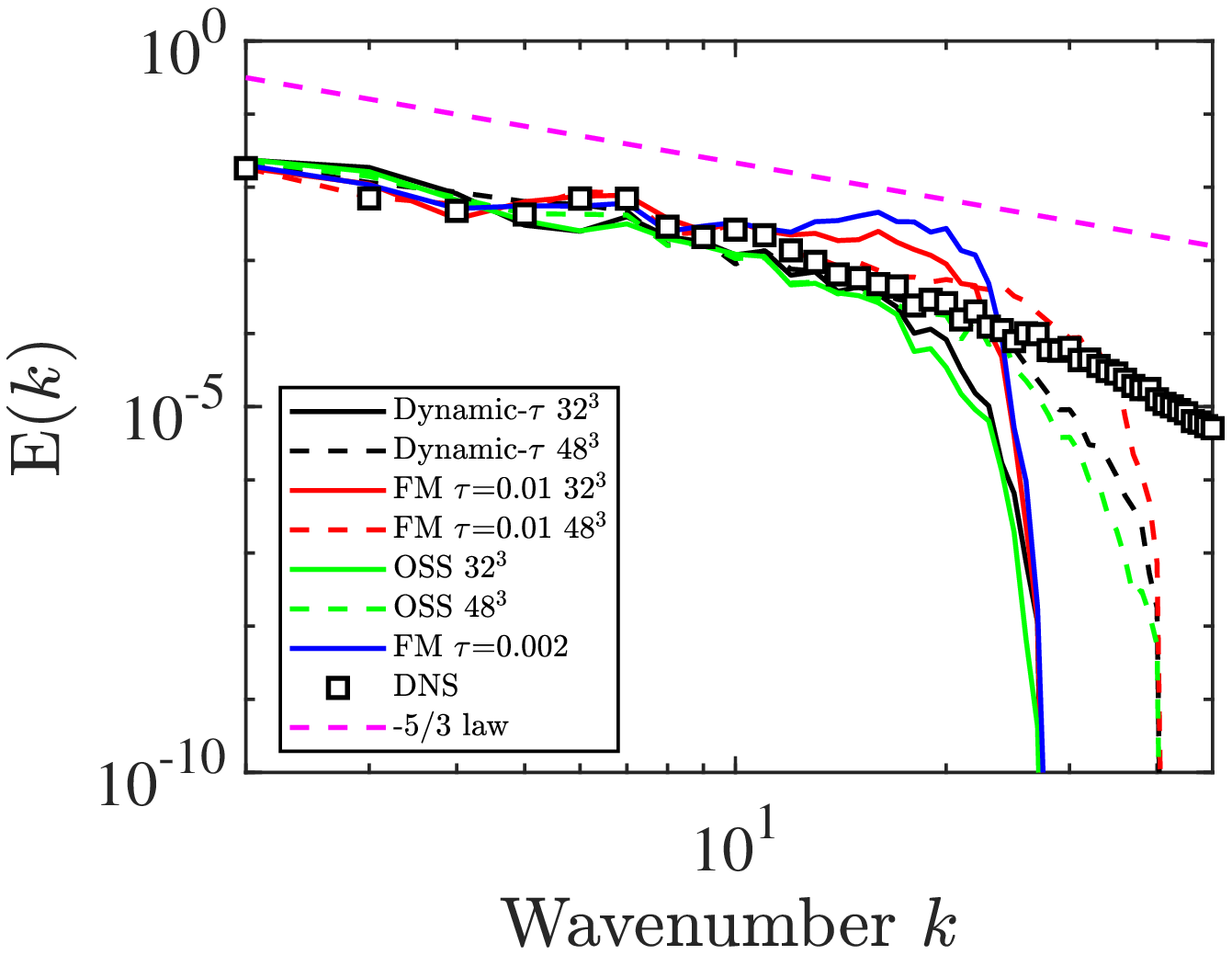}
    	\caption{Energy Spectra at T=10.0.}
    	\label{TGV800_4}
    \end{subfigure}

	\caption{{\color{black} (a) Kinetic energy}, (b) dissipation, (c) energy spectra at T = 5 and (d)  energy spectra at T=10 for Taylor Green vortex at Re=800 using different coarse graining methods.}
	\label{TGV800}
\end{figure}

\begin{figure}
	\begin{subfigure}[b]{0.5\textwidth}
		\centering
		\includegraphics[width=\textwidth]{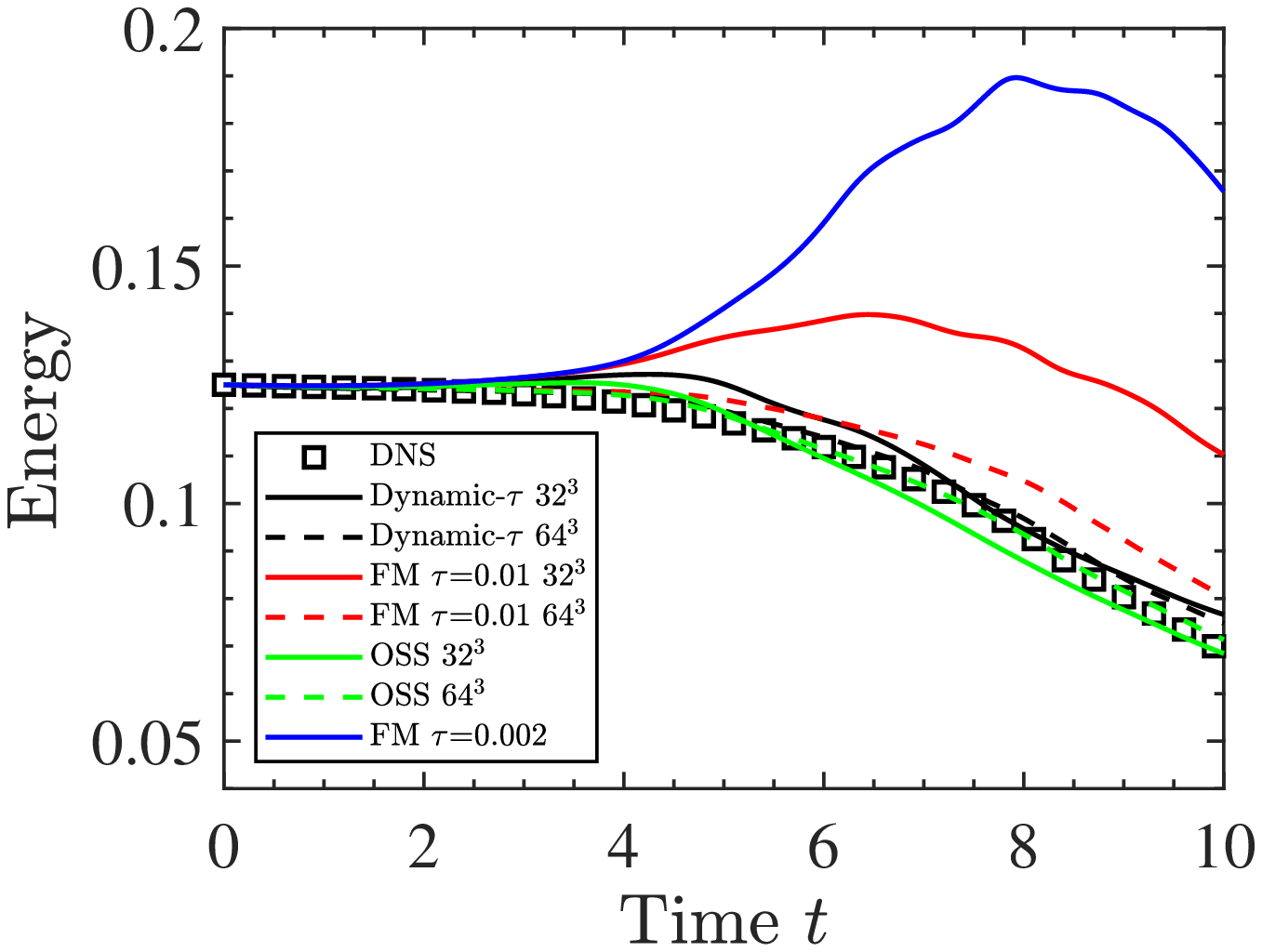}
		\caption{Time evolution of kinetic energy.}
		\label{TGV1600_1}
	\end{subfigure}
	\begin{subfigure}[b]{0.5\textwidth}
		\centering
		\includegraphics[width=\textwidth]{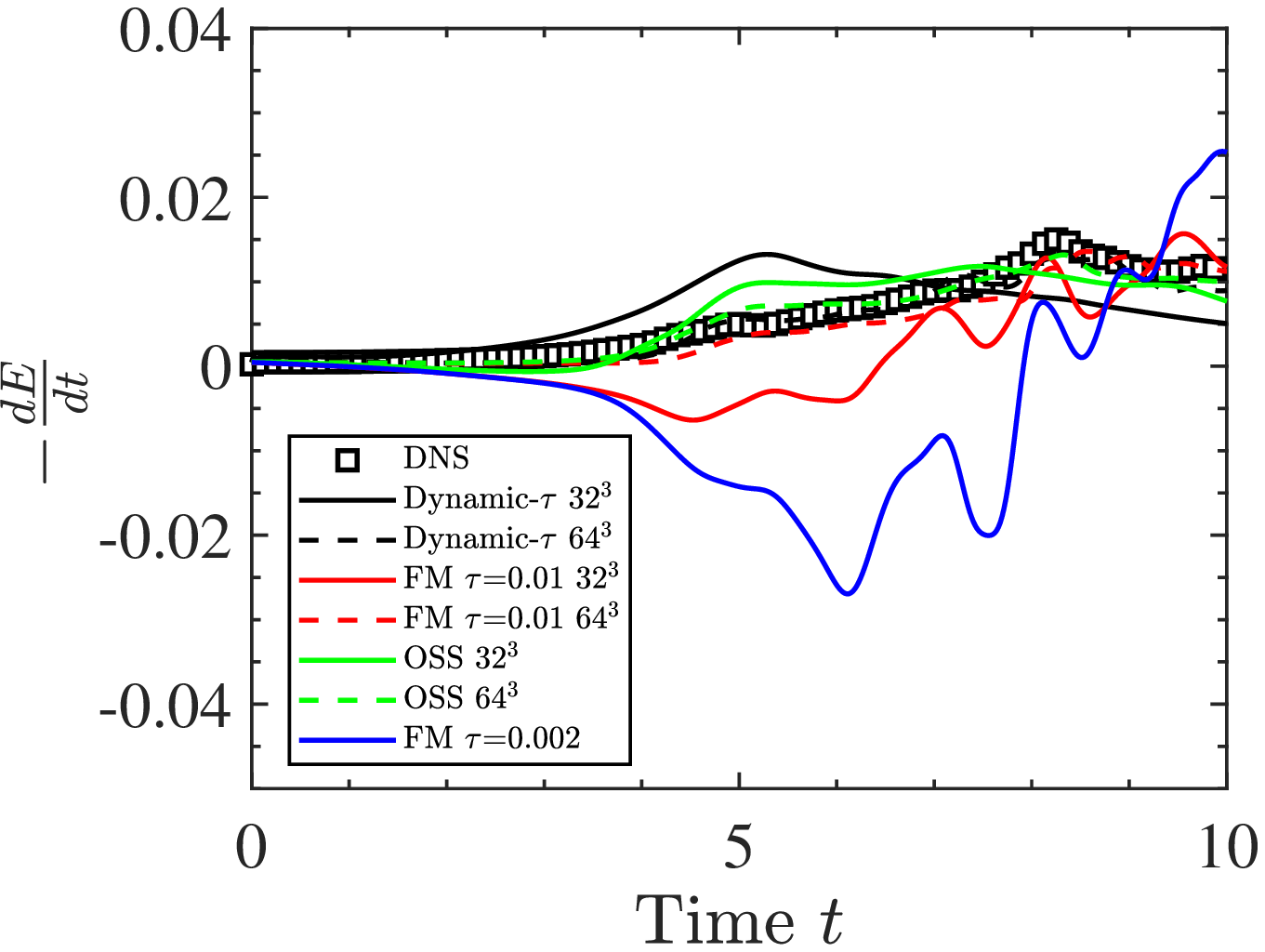}
		\caption{Rate of kinetic energy decay.}
		\label{TGV1600_2}
	\end{subfigure}
    \begin{subfigure}[b]{0.5\textwidth}
    	\centering
    	\includegraphics[width=\textwidth]{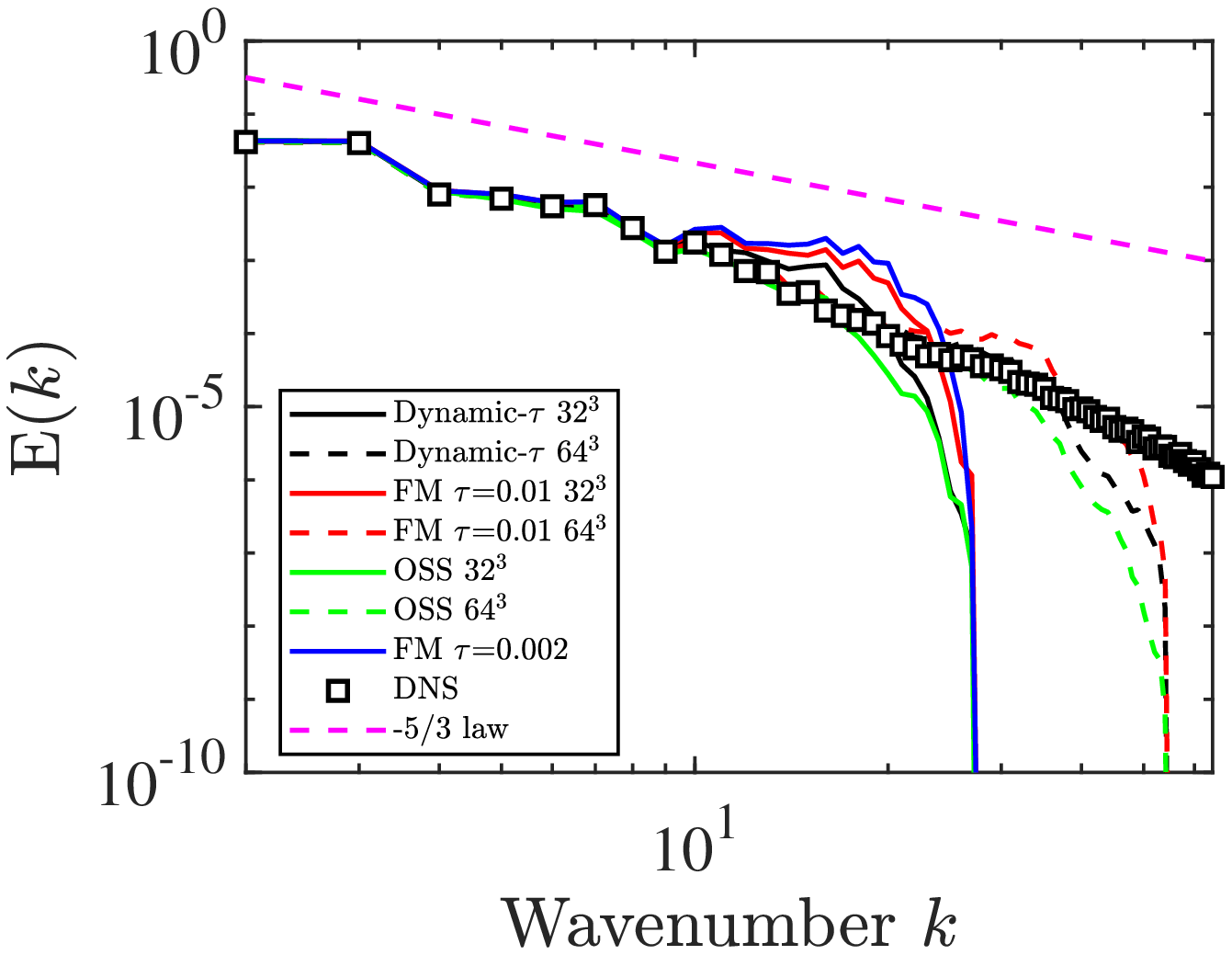}
    	\caption{Energy Spectra at T=5.0.}
    	\label{TGV1600_3}
    \end{subfigure}
    \begin{subfigure}[b]{0.5\textwidth}
    	\centering
    	\includegraphics[width=\textwidth]{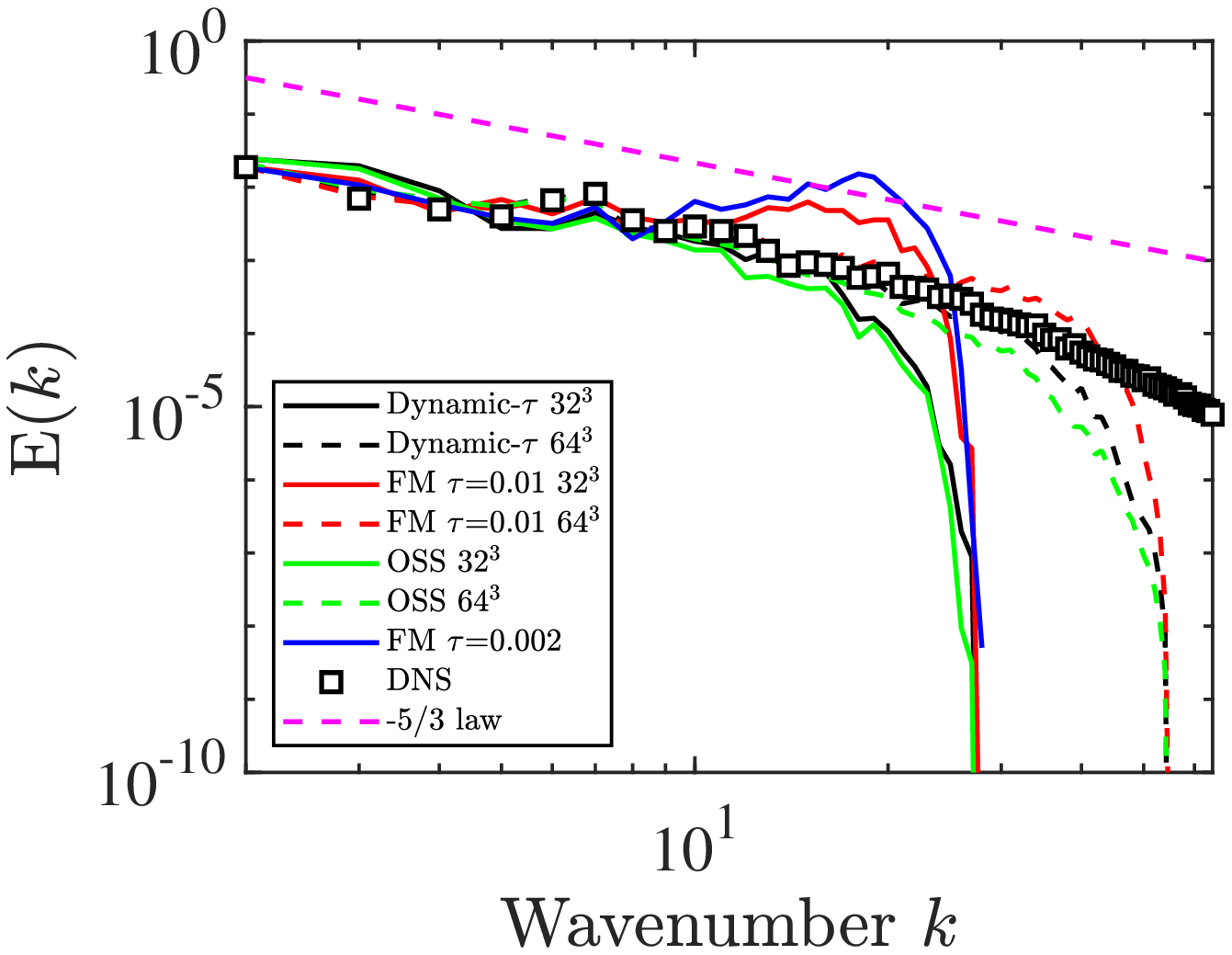}
    	\caption{Energy Spectra at T=10.0.}
    	\label{TGV1600_4}
    \end{subfigure}
	\caption{{\color{black}(a) Kinetic energy}, (b) dissipation, (c) energy spectra at T = 5 and (d)  energy spectra at T=10 for Taylor Green vortex at Re=1600 using different coarse graining methods.}
	\label{TGV1600}
\end{figure}

The profiles for the resolved kinetic energy for $Re =$ 400, 800 and 1600 are shown in Figures \ref{TGV400_1}, \ref{TGV800_1} and \ref{TGV1600_1} respectively. The evolution of the kinetic energy indicates that both the OSS and Dynamic-$\tau$ perform well with only $32^3$ degrees of freedom. Similar trends are also observed for the rate of KE energy decay for $Re =$ 400, 800 and 1600 in Figures \ref{TGV400_2},\ref{TGV800_2} and \ref{TGV1600_2} respectively, where the fixed $\tau$ models fail to accurately predict the correct results in comparison to DNS. When $48^3$ and $64^3$ degrees of freedoms are used for $Re =$ 800 and 1600 respectively, there is an overall improvement in the results for the fixed memory model. The present dynamic model and the OSS model perform well at finer resolutions.  

Figures \ref{TGV400_3}, \ref{TGV800_3} and \ref{TGV1600_3}, and Figures \ref{TGV400_4}, \ref{TGV800_4} and \ref{TGV1600_4} show the energy spectra of the resolved velocity fields at two time instants $T=5.0$ and $T=10.0$ respectively. At $T=5.0$, all the models are in agreement with DNS at the low wavenumber modes even with just $32^3$ degrees of freedom. However, the constant $\tau$ models produces a build-up of energy which grows with $Re$ at high-wavenumber modes. This suggests that either an incorrect value of $\tau$ is used for the FM models or the assumption  of constant memory length throughout the simulation is not very accurate. As a result, model with constant memory length is not capable of producing enough dissipation and the energy increases at the high wavenumber modes.  At a later time $T=10.0$, a similar trend is also observed with the constant $\tau$ model where energy increases at high-wave numbers. On the other hand, the dynamic $\tau$-model and OSS do not result in energy increase at high wavenumbers. Although our dynamic model and OSS model perform closely for the $32^3$ cases, at higher resolutions OSS is clearly more dissipative wherein lower energy is present at high wavenumber especially for the high Reynolds number case. In spite of the OSS model and the dynamic-$\tau$ model performing closely, the stabilization parameter in OSS and the memory length in dynamic-$\tau$ is computed differently. 

\begin{table}[h!]
	\centering
	\begin{tabular}{ccccccccc}
		\toprule
		Case       & $DOFs$      &  $dx$    & $dt$     & $\nu$   & $U_0$& $L$   & FM $\tau$'s   \\ \midrule
		DNS-400    & $64^3$    & $9.81\times10^{-2}$ & $2\times10^{-2}$ & $2.5\times10^{-3}$  &     1& 1     &-                     \\ 
		LES-FM-400    & $32^3$     & $1.96\times10^{-1}$  & $1.96\times10^{-2}$ & $2.5\times10^{-3}$  &     1& 1     &0.01 and 0.002 \\ 
		LES-DY-400    & $32^3$    & $1.96\times10^{-1}$ & $1.96\times10^{-2}$ & $2.5\times10^{-3}$  &     1& 1     &Dynamic
	    \\ 
		DNS-800    & $128^3$    &  $4.90\times10^{-2}$ & $2\times10^{-2}$ & $1.25\times10^{-3}$  &     1& 1     &-                     \\ 
		LES-FM-800    & $32^3, 48^3$    & $1.96\times10^{-1}$, $1.31\times10^{-1}$ & $1.96\times10^{-2}$ &  $1.25\times10^{-3}$ &     1& 1     &0.01 and 0.002 \\ 
		LES-DY-800    & $32^3, 48^3$    & $1.96\times10^{-1}$, $1.31\times10^{-1}$ & $1.96\times10^{-2}$ &  $1.25\times10^{-3}$ &     1& 1     &Dynamic        \\ 
		DNS-1600   &  $256^3$   &  $2.45\times10^{-2}$ & $5\times10^{-3}$      &  $6.25\times10^{-4}$ &     1& 1     &-                     \\ 
		LES-FM-1600   & $32^3, 64^3$    & $1.96\times10^{-1}$, $9.81\times10^{-2}$ & $1.96\times10^{-2}$, $10^{-2}$  & $6.25\times10^{-4}$ &     1& 1     &0.01 and 0.002 \\ 
		LES-DY-1600   & $32^3, 64^3$    & $1.96\times10^{-1}$, $9.81\times10^{-2}$ & $10^{-2}$& $6.25\times10^{-4}$ &     1& 1     &Dynamic        \\ 
		\bottomrule
	\end{tabular}
	\caption{Simulation parameters for DNS and LES of the Taylor Green Vortex problem.}
	\label{table_TGV}
\end{table}   

\section{Conclusion}
The Variational Multiscale method and the Mori-Zwanzig formalism are combined within the Continuous Galerkin method to develop coarse grained models for multiscale PDEs. This approach utilizes the Variational Multiscale method to separate scales with the capability of the Mori-Zwanzig formalism to represent the impact of unresolved dynamics on the resolved dynamics. {\color{black} This  approach - in a similar spirit to  existing non-linear VMS models \cite{OSS2,OSS,VMS3,VMSE,NLVMS,NLVMS2} - is developed to provide sub-grid scale models without phenomenological assumptions}. This procedure is generalizable and can potentially be applied to arbitrarily complex non-linear multiscale PDEs.
 In context of turbulent flows, this approach provides a general framework for large-eddy simulation that eschews assumptions such as those based on energy balance between scale. Sub-grid scale models are developed for the Burgers equation and the Navier-Stokes using the proposed approach. The sub-grid scale models include a parameter called the memory length, $\tau$, which represents the the time correlation of unresolved dynamics, and controls the stabilization. We impose different memory lengths $\tau$ and observe that there is an optimum memory length $\tau$ which provides results comparable to the full order solution \cite{ETHESIS}. {\color{black}To avoid the imposition of an adhoc memory length, and recognizing that the model should adapt to the instantaneous level of resolution, we derived a dynamic-$\tau$ model and found that it can accurately predict the temporal evolution of $\tau$}. The predicted value of $\tau$ was observed to be generally linear in time, as conceptualized by the renormalized t-model \cite{STINIST}. In general, for the range of problems that were investigated, the proposed technique performs favorably in comparison to existing counterparts.

 This work was focused on fixed memory type models leading to Markovian  closures, however, other approaches leading to non-Markovian type closures can be implemented \cite{ETHESIS}. Alternatively, different approximations to the orthogonal dynamics \cite{FABER,MZ3} can be used to construct models.  The single memory length model was - in part - successful because of the nature  of the problems investigated herein. Future extensions to highly anisotropic and inhomogeneous problems will require the development of local definitions for the memory length. 
 
 \section*{Acknowledgement}
This research was funded by the AFOSR under the project LES Modeling of \textit{Non-local effects using Statistical Coarse-graining}, grant number FA9550-16-1-0309. The authors also thank Dr. Eric Parish for providing DNS results for comparison for the HIT and TGV cases, and Dr. Daniel Foti for his valuable suggestions.

\bibliography{mybibfile}

\begin{thebibliography}{51}
\expandafter\ifx\csname natexlab\endcsname\relax\def\natexlab#1{#1}\fi
\providecommand{\url}[1]{\texttt{#1}}
\providecommand{\href}[2]{#2}
\providecommand{\path}[1]{#1}
\providecommand{\DOIprefix}{doi:}
\providecommand{\ArXivprefix}{arXiv:}
\providecommand{\URLprefix}{URL: }
\providecommand{\Pubmedprefix}{pmid:}
\providecommand{\doi}[1]{\href{http://dx.doi.org/#1}{\path{#1}}}
\providecommand{\Pubmed}[1]{\href{pmid:#1}{\path{#1}}}
\providecommand{\bibinfo}[2]{#2}
\ifx\xfnm\relax \def\xfnm[#1]{\unskip,\space#1}\fi
%Type = Article
\bibitem[{Smagorinsky(1963)}]{SMAG}
\bibinfo{author}{J.~Smagorinsky},
\newblock \bibinfo{title}{General circulation experiments with the primitive
  equations: I. the basic experiment},
\newblock \bibinfo{journal}{Monthly weather review} \bibinfo{volume}{91}
  (\bibinfo{year}{1963}) \bibinfo{pages}{99--164}.
%Type = Article
\bibitem[{Vreman(2004)}]{VREMEN}
\bibinfo{author}{A.~Vreman},
\newblock \bibinfo{title}{An eddy-viscosity subgrid-scale model for turbulent
  shear flow: Algebraic theory and applications},
\newblock \bibinfo{journal}{Physics of fluids} \bibinfo{volume}{16}
  (\bibinfo{year}{2004}) \bibinfo{pages}{3670--3681}.
%Type = Article
\bibitem[{Nicoud and Ducros(1999)}]{WALE}
\bibinfo{author}{F.~Nicoud}, \bibinfo{author}{F.~Ducros},
\newblock \bibinfo{title}{Subgrid-scale stress modelling based on the square of
  the velocity gradient tensor},
\newblock \bibinfo{journal}{Flow, turbulence and Combustion}
  \bibinfo{volume}{62} (\bibinfo{year}{1999}) \bibinfo{pages}{183--200}.
%Type = Article
\bibitem[{Germano et~al.(1991)Germano, Piomelli, Moin, and Cabot}]{DSM}
\bibinfo{author}{M.~Germano}, \bibinfo{author}{U.~Piomelli},
  \bibinfo{author}{P.~Moin}, \bibinfo{author}{W.~H. Cabot},
\newblock \bibinfo{title}{A dynamic subgrid-scale eddy viscosity model},
\newblock \bibinfo{journal}{Physics of Fluids A: Fluid Dynamics}
  \bibinfo{volume}{3} (\bibinfo{year}{1991}) \bibinfo{pages}{1760--1765}.
%Type = Article
\bibitem[{Meneveau et~al.(1996)Meneveau, Lund, and Cabot}]{DSM2}
\bibinfo{author}{C.~Meneveau}, \bibinfo{author}{T.~S. Lund},
  \bibinfo{author}{W.~H. Cabot},
\newblock \bibinfo{title}{A lagrangian dynamic subgrid-scale model of
  turbulence},
\newblock \bibinfo{journal}{Journal of fluid mechanics} \bibinfo{volume}{319}
  (\bibinfo{year}{1996}) \bibinfo{pages}{353--385}.
%Type = Article
\bibitem[{You and Moin(2007)}]{GDSM}
\bibinfo{author}{D.~You}, \bibinfo{author}{P.~Moin},
\newblock \bibinfo{title}{A dynamic global-coefficient subgrid-scale
  eddy-viscosity model for large-eddy simulation in complex geometries},
\newblock \bibinfo{journal}{Physics of Fluids} \bibinfo{volume}{19}
  (\bibinfo{year}{2007}) \bibinfo{pages}{065110}.
%Type = Book
\bibitem[{Pope and Pope(2000)}]{pope}
\bibinfo{author}{S.~B. Pope}, \bibinfo{author}{S.~B. Pope},
  \bibinfo{title}{Turbulent flows}, \bibinfo{publisher}{Cambridge university
  press}, \bibinfo{year}{2000}.
%Type = Article
\bibitem[{Nicoud et~al.(2011)Nicoud, Toda, Cabrit, Bose, and Lee}]{SIGMA}
\bibinfo{author}{F.~Nicoud}, \bibinfo{author}{H.~B. Toda},
  \bibinfo{author}{O.~Cabrit}, \bibinfo{author}{S.~Bose},
  \bibinfo{author}{J.~Lee},
\newblock \bibinfo{title}{Using singular values to build a subgrid-scale model
  for large eddy simulations},
\newblock \bibinfo{journal}{Physics of Fluids} \bibinfo{volume}{23}
  (\bibinfo{year}{2011}) \bibinfo{pages}{085106}.
%Type = Article
\bibitem[{Hughes et~al.(1998)Hughes, Feij{\'o}o, Mazzei, and Quincy}]{VMS}
\bibinfo{author}{T.~J. Hughes}, \bibinfo{author}{G.~R. Feij{\'o}o},
  \bibinfo{author}{L.~Mazzei}, \bibinfo{author}{J.-B. Quincy},
\newblock \bibinfo{title}{The variational multiscale method—a paradigm for
  computational mechanics},
\newblock \bibinfo{journal}{Computer methods in applied mechanics and
  engineering} \bibinfo{volume}{166} (\bibinfo{year}{1998})
  \bibinfo{pages}{3--24}.
%Type = Article
\bibitem[{Hughes et~al.(1989)Hughes, Franca, and Hulbert}]{GLS}
\bibinfo{author}{T.~J. Hughes}, \bibinfo{author}{L.~P. Franca},
  \bibinfo{author}{G.~M. Hulbert},
\newblock \bibinfo{title}{A new finite element formulation for computational
  fluid dynamics: Viii. the galerkin/least-squares method for
  advective-diffusive equations},
\newblock \bibinfo{journal}{Computer methods in applied mechanics and
  engineering} \bibinfo{volume}{73} (\bibinfo{year}{1989})
  \bibinfo{pages}{173--189}.
%Type = Article
\bibitem[{Brooks and Hughes(1982)}]{SUPG}
\bibinfo{author}{A.~N. Brooks}, \bibinfo{author}{T.~J. Hughes},
\newblock \bibinfo{title}{Streamline upwind/petrov-galerkin formulations for
  convection dominated flows with particular emphasis on the incompressible
  navier-stokes equations},
\newblock \bibinfo{journal}{Computer methods in applied mechanics and
  engineering} \bibinfo{volume}{32} (\bibinfo{year}{1982})
  \bibinfo{pages}{199--259}.
%Type = Article
\bibitem[{Codina(2000)}]{ADJ}
\bibinfo{author}{R.~Codina},
\newblock \bibinfo{title}{On stabilized finite element methods for linear
  systems of convection--diffusion-reaction equations},
\newblock \bibinfo{journal}{Computer Methods in Applied Mechanics and
  Engineering} \bibinfo{volume}{188} (\bibinfo{year}{2000})
  \bibinfo{pages}{61--82}.
%Type = Article
\bibitem[{Hughes et~al.(1986)Hughes, Franca, and Balestra}]{SUPG2}
\bibinfo{author}{T.~J. Hughes}, \bibinfo{author}{L.~P. Franca},
  \bibinfo{author}{M.~Balestra},
\newblock \bibinfo{title}{A new finite element formulation for computational
  fluid dynamics: V. circumventing the babu{\v{s}}ka-brezzi condition: a stable
  petrov-galerkin formulation of the stokes problem accommodating equal-order
  interpolations},
\newblock \bibinfo{journal}{Computer Methods in Applied Mechanics and
  Engineering} \bibinfo{volume}{59} (\bibinfo{year}{1986})
  \bibinfo{pages}{85--99}.
%Type = Article
\bibitem[{Codina et~al.(2007)Codina, Principe, Guasch, and Badia}]{OSS2}
\bibinfo{author}{R.~Codina}, \bibinfo{author}{J.~Principe},
  \bibinfo{author}{O.~Guasch}, \bibinfo{author}{S.~Badia},
\newblock \bibinfo{title}{Time dependent subscales in the stabilized finite
  element approximation of incompressible flow problems},
\newblock \bibinfo{journal}{Computer Methods in Applied Mechanics and
  Engineering} \bibinfo{volume}{196} (\bibinfo{year}{2007})
  \bibinfo{pages}{2413--2430}.
%Type = Article
\bibitem[{Codina(2002)}]{OSS}
\bibinfo{author}{R.~Codina},
\newblock \bibinfo{title}{Stabilized finite element approximation of transient
  incompressible flows using orthogonal subscales},
\newblock \bibinfo{journal}{Computer Methods in Applied Mechanics and
  Engineering} \bibinfo{volume}{191} (\bibinfo{year}{2002})
  \bibinfo{pages}{4295--4321}.
%Type = Article
\bibitem[{Bazilevs et~al.(2007)Bazilevs, Calo, Cottrell, Hughes, Reali, and
  Scovazzi}]{VMS3}
\bibinfo{author}{Y.~Bazilevs}, \bibinfo{author}{V.~Calo},
  \bibinfo{author}{J.~Cottrell}, \bibinfo{author}{T.~Hughes},
  \bibinfo{author}{A.~Reali}, \bibinfo{author}{G.~Scovazzi},
\newblock \bibinfo{title}{Variational multiscale residual-based turbulence
  modeling for large eddy simulation of incompressible flows},
\newblock \bibinfo{journal}{Computer Methods in Applied Mechanics and
  Engineering} \bibinfo{volume}{197} (\bibinfo{year}{2007})
  \bibinfo{pages}{173--201}.
%Type = Article
\bibitem[{Wang and Oberai(2010)}]{VMSE}
\bibinfo{author}{Z.~Wang}, \bibinfo{author}{A.~Oberai},
\newblock \bibinfo{title}{Spectral analysis of the dissipation of the
  residual-based variational multiscale method},
\newblock \bibinfo{journal}{Computer Methods in Applied Mechanics and
  Engineering} \bibinfo{volume}{199} (\bibinfo{year}{2010})
  \bibinfo{pages}{810--818}.
%Type = Article
\bibitem[{Gravemeier et~al.(2010)Gravemeier, Gee, Kronbichler, and
  Wall}]{NLVMS}
\bibinfo{author}{V.~Gravemeier}, \bibinfo{author}{M.~W. Gee},
  \bibinfo{author}{M.~Kronbichler}, \bibinfo{author}{W.~A. Wall},
\newblock \bibinfo{title}{An algebraic variational multiscale--multigrid method
  for large eddy simulation of turbulent flow},
\newblock \bibinfo{journal}{Computer Methods in Applied Mechanics and
  Engineering} \bibinfo{volume}{199} (\bibinfo{year}{2010})
  \bibinfo{pages}{853--864}.
%Type = Article
\bibitem[{Masud and Calderer(2011)}]{NLVMS2}
\bibinfo{author}{A.~Masud}, \bibinfo{author}{R.~Calderer},
\newblock \bibinfo{title}{A variational multiscale method for incompressible
  turbulent flows: Bubble functions and fine scale fields},
\newblock \bibinfo{journal}{Computer Methods in Applied Mechanics and
  Engineering} \bibinfo{volume}{200} (\bibinfo{year}{2011})
  \bibinfo{pages}{2577--2593}.
%Type = Article
\bibitem[{Franca et~al.(1992)Franca, Frey, and Hughes}]{TAU}
\bibinfo{author}{L.~P. Franca}, \bibinfo{author}{S.~L. Frey},
  \bibinfo{author}{T.~J. Hughes},
\newblock \bibinfo{title}{Stabilized finite element methods: I. application to
  the advective-diffusive model},
\newblock \bibinfo{journal}{Computer Methods in Applied Mechanics and
  Engineering} \bibinfo{volume}{95} (\bibinfo{year}{1992})
  \bibinfo{pages}{253--276}.
%Type = Article
\bibitem[{Chorin et~al.(2002)Chorin, Hald, and Kupferman}]{CHORIN}
\bibinfo{author}{A.~J. Chorin}, \bibinfo{author}{O.~H. Hald},
  \bibinfo{author}{R.~Kupferman},
\newblock \bibinfo{title}{Optimal prediction with memory},
\newblock \bibinfo{journal}{Physica D: Nonlinear Phenomena}
  \bibinfo{volume}{166} (\bibinfo{year}{2002}) \bibinfo{pages}{239--257}.
%Type = Book
\bibitem[{Chorin and Hald(2009)}]{CHORIN2}
\bibinfo{author}{A.~J. Chorin}, \bibinfo{author}{O.~H. Hald},
  \bibinfo{title}{Stochastic tools in mathematics and science},
  volume~\bibinfo{volume}{3}, \bibinfo{publisher}{Springer},
  \bibinfo{year}{2009}.
%Type = Article
\bibitem[{Parish and Duraisamy(2017{\natexlab{a}})}]{MZ1}
\bibinfo{author}{E.~J. Parish}, \bibinfo{author}{K.~Duraisamy},
\newblock \bibinfo{title}{Non-markovian closure models for large eddy
  simulations using the mori-zwanzig formalism},
\newblock \bibinfo{journal}{Physical Review Fluids} \bibinfo{volume}{2}
  (\bibinfo{year}{2017}{\natexlab{a}}) \bibinfo{pages}{014604}.
%Type = Article
\bibitem[{Parish and Duraisamy(2017{\natexlab{b}})}]{MZ2}
\bibinfo{author}{E.~J. Parish}, \bibinfo{author}{K.~Duraisamy},
\newblock \bibinfo{title}{A dynamic subgrid scale model for large eddy
  simulations based on the mori--zwanzig formalism},
\newblock \bibinfo{journal}{Journal of Computational Physics}
  \bibinfo{volume}{349} (\bibinfo{year}{2017}{\natexlab{b}})
  \bibinfo{pages}{154--175}.
%Type = Article
\bibitem[{Gouasmi et~al.(2017)Gouasmi, Parish, and Duraisamy}]{MZ3}
\bibinfo{author}{A.~Gouasmi}, \bibinfo{author}{E.~Parish},
  \bibinfo{author}{K.~Duraisamy},
\newblock \bibinfo{title}{A priori estimation of memory effects in
  coarse-grained nonlinear systems using the mori-zwanzig formalism}
  (\bibinfo{year}{2017}).
%Type = Article
\bibitem[{Stinis(2007)}]{STINIS}
\bibinfo{author}{P.~Stinis},
\newblock \bibinfo{title}{Higher order mori--zwanzig models for the euler
  equations},
\newblock \bibinfo{journal}{Multiscale Modeling \& Simulation}
  \bibinfo{volume}{6} (\bibinfo{year}{2007}) \bibinfo{pages}{741--760}.
%Type = Article
\bibitem[{Parish and Duraisamy(2017)}]{MZVMS}
\bibinfo{author}{E.~J. Parish}, \bibinfo{author}{K.~Duraisamy},
\newblock \bibinfo{title}{A unified framework for multiscale modeling using the
  mori-zwanzig formalism and the variational multiscale method},
\newblock \bibinfo{journal}{arXiv preprint arXiv:1712.09669}
  (\bibinfo{year}{2017}).
%Type = Article
\bibitem[{Parish(2018)}]{ETHESIS}
\bibinfo{author}{E.~Parish},
\newblock \bibinfo{title}{Variational multiscale modeling and memory effects in
  turbulent flow simulations}  (\bibinfo{year}{2018}).
%Type = Article
\bibitem[{Mori(1965)}]{MORI}
\bibinfo{author}{H.~Mori},
\newblock \bibinfo{title}{Transport, collective motion, and brownian motion},
\newblock \bibinfo{journal}{Progress of theoretical physics}
  \bibinfo{volume}{33} (\bibinfo{year}{1965}) \bibinfo{pages}{423--455}.
%Type = Incollection
\bibitem[{Zwanzig(1980)}]{ZWANZIG}
\bibinfo{author}{R.~Zwanzig},
\newblock \bibinfo{title}{Problems in nonlinear transport theory},
\newblock in: \bibinfo{booktitle}{Systems far from equilibrium},
  \bibinfo{publisher}{Springer}, \bibinfo{year}{1980}, pp.
  \bibinfo{pages}{198--225}.
%Type = Article
\bibitem[{Parish et~al.(2018)Parish, Wentland, and Duraisamy}]{CHRIS}
\bibinfo{author}{E.~J. Parish}, \bibinfo{author}{C.~Wentland},
  \bibinfo{author}{K.~Duraisamy},
\newblock \bibinfo{title}{A residual-based petrov-galerkin reduced-order model
  with memory effects},
\newblock \bibinfo{journal}{arXiv preprint arXiv:1810.03455}
  (\bibinfo{year}{2018}).
%Type = Article
\bibitem[{Chorin et~al.(2000)Chorin, Hald, and Kupferman}]{CHORIN3}
\bibinfo{author}{A.~J. Chorin}, \bibinfo{author}{O.~H. Hald},
  \bibinfo{author}{R.~Kupferman},
\newblock \bibinfo{title}{Optimal prediction and the mori--zwanzig
  representation of irreversible processes},
\newblock \bibinfo{journal}{Proceedings of the National Academy of Sciences}
  \bibinfo{volume}{97} (\bibinfo{year}{2000}) \bibinfo{pages}{2968--2973}.
%Type = Article
\bibitem[{Zhu and Venturi(2018)}]{FABER}
\bibinfo{author}{Y.~Zhu}, \bibinfo{author}{D.~Venturi},
\newblock \bibinfo{title}{Faber approximation of the mori-zwanzig equation},
\newblock \bibinfo{journal}{Journal of Computational Physics}
  (\bibinfo{year}{2018}).
%Type = Article
\bibitem[{Stinis(2015)}]{STINIS2}
\bibinfo{author}{P.~Stinis},
\newblock \bibinfo{title}{Renormalized mori--zwanzig-reduced models for systems
  without scale separation},
\newblock \bibinfo{journal}{Proceedings of the Royal Society A: Mathematical,
  Physical and Engineering Sciences} \bibinfo{volume}{471}
  (\bibinfo{year}{2015}) \bibinfo{pages}{20140446}.
%Type = Article
\bibitem[{Stinis(2012)}]{STINIS3}
\bibinfo{author}{P.~Stinis},
\newblock \bibinfo{title}{Mori-zwanzig reduced models for uncertainty
  quantification i: Parametric uncertainty},
\newblock \bibinfo{journal}{arXiv preprint arXiv:1211.4285}
  (\bibinfo{year}{2012}).
%Type = Article
\bibitem[{Masud and Calderer(2011)}]{BUBBLE}
\bibinfo{author}{A.~Masud}, \bibinfo{author}{R.~Calderer},
\newblock \bibinfo{title}{A variational multiscale method for incompressible
  turbulent flows: Bubble functions and fine scale fields},
\newblock \bibinfo{journal}{Computer Methods in Applied Mechanics and
  Engineering} \bibinfo{volume}{200} (\bibinfo{year}{2011})
  \bibinfo{pages}{2577--2593}.
%Type = Article
\bibitem[{Franca and Farhat(1995)}]{BUBBLE1}
\bibinfo{author}{L.~P. Franca}, \bibinfo{author}{C.~Farhat},
\newblock \bibinfo{title}{Bubble functions prompt unusual stabilized finite
  element methods},
\newblock \bibinfo{journal}{Computer Methods in Applied Mechanics and
  Engineering} \bibinfo{volume}{123} (\bibinfo{year}{1995})
  \bibinfo{pages}{299--308}.
%Type = Article
\bibitem[{Brezzi et~al.(1992)Brezzi, Bristeau, Franca, Mallet, and
  Rog{\'e}}]{BUBBLE2}
\bibinfo{author}{F.~Brezzi}, \bibinfo{author}{M.-O. Bristeau},
  \bibinfo{author}{L.~P. Franca}, \bibinfo{author}{M.~Mallet},
  \bibinfo{author}{G.~Rog{\'e}},
\newblock \bibinfo{title}{A relationship between stabilized finite element
  methods and the galerkin method with bubble functions},
\newblock \bibinfo{journal}{Computer Methods in Applied Mechanics and
  Engineering} \bibinfo{volume}{96} (\bibinfo{year}{1992})
  \bibinfo{pages}{117--129}.
%Type = Article
\bibitem[{Hughes(1995)}]{BUBBLE3}
\bibinfo{author}{T.~J. Hughes},
\newblock \bibinfo{title}{Multiscale phenomena: Green's functions, the
  dirichlet-to-neumann formulation, subgrid scale models, bubbles and the
  origins of stabilized methods},
\newblock \bibinfo{journal}{Computer methods in applied mechanics and
  engineering} \bibinfo{volume}{127} (\bibinfo{year}{1995})
  \bibinfo{pages}{387--401}.
%Type = Article
\bibitem[{Germano(1992)}]{GDSM2}
\bibinfo{author}{M.~Germano},
\newblock \bibinfo{title}{Turbulence: the filtering approach},
\newblock \bibinfo{journal}{Journal of Fluid Mechanics} \bibinfo{volume}{238}
  (\bibinfo{year}{1992}) \bibinfo{pages}{325--336}.
%Type = Article
\bibitem[{Oberai and Wanderer(2005)}]{OGERMANO}
\bibinfo{author}{A.~Oberai}, \bibinfo{author}{J.~Wanderer},
\newblock \bibinfo{title}{Variational formulation of the germano identity for
  the navier--stokes equations},
\newblock \bibinfo{journal}{Journal of Turbulence}  (\bibinfo{year}{2005})
  \bibinfo{pages}{N7}.
%Type = Article
\bibitem[{Akkerman et~al.(2010)Akkerman, van~der Zee, and Hulshoff}]{OGERMANO2}
\bibinfo{author}{I.~Akkerman}, \bibinfo{author}{K.~van~der Zee},
  \bibinfo{author}{S.~Hulshoff},
\newblock \bibinfo{title}{A variational germano approach for stabilized finite
  element methods},
\newblock \bibinfo{journal}{Computer methods in applied mechanics and
  engineering} \bibinfo{volume}{199} (\bibinfo{year}{2010})
  \bibinfo{pages}{502--513}.
%Type = Inproceedings
\bibitem[{Tezduyar(2001)}]{JUMP}
\bibinfo{author}{T.~E. Tezduyar},
\newblock \bibinfo{title}{Adaptive determination of the finite element
  stabilization parameters},
\newblock in: \bibinfo{booktitle}{Proceedings of the ECCOMAS computational
  fluid dynamics conference}, \bibinfo{year}{2001}, pp. \bibinfo{pages}{1--17}.
%Type = Book
\bibitem[{Donea and Huerta(2003)}]{DONEA}
\bibinfo{author}{J.~Donea}, \bibinfo{author}{A.~Huerta}, \bibinfo{title}{Finite
  element methods for flow problems}, \bibinfo{publisher}{John Wiley \& Sons},
  \bibinfo{year}{2003}.
%Type = Article
\bibitem[{Stinis(2013)}]{STINIST}
\bibinfo{author}{P.~Stinis},
\newblock \bibinfo{title}{Renormalized reduced models for singular pdes},
\newblock \bibinfo{journal}{Communications in Applied Mathematics and
  Computational Science} \bibinfo{volume}{8} (\bibinfo{year}{2013})
  \bibinfo{pages}{39--66}.
%Type = Article
\bibitem[{Colom{\'e}s et~al.(2015)Colom{\'e}s, Badia, Codina, and
  Principe}]{OSSLES}
\bibinfo{author}{O.~Colom{\'e}s}, \bibinfo{author}{S.~Badia},
  \bibinfo{author}{R.~Codina}, \bibinfo{author}{J.~Principe},
\newblock \bibinfo{title}{Assessment of variational multiscale models for the
  large eddy simulation of turbulent incompressible flows},
\newblock \bibinfo{journal}{Computer Methods in Applied Mechanics and
  Engineering} \bibinfo{volume}{285} (\bibinfo{year}{2015})
  \bibinfo{pages}{32--63}.
%Type = Article
\bibitem[{Mansour and Wray(1994)}]{HIT1}
\bibinfo{author}{N.~Mansour}, \bibinfo{author}{A.~Wray},
\newblock \bibinfo{title}{Decay of isotropic turbulence at low reynolds
  number},
\newblock \bibinfo{journal}{Physics of Fluids} \bibinfo{volume}{6}
  (\bibinfo{year}{1994}) \bibinfo{pages}{808--814}.
%Type = Article
\bibitem[{Orszag and Patterson~Jr(1972)}]{HIT2}
\bibinfo{author}{S.~A. Orszag}, \bibinfo{author}{G.~Patterson~Jr},
\newblock \bibinfo{title}{Numerical simulation of three-dimensional homogeneous
  isotropic turbulence},
\newblock \bibinfo{journal}{Physical Review Letters} \bibinfo{volume}{28}
  (\bibinfo{year}{1972}) \bibinfo{pages}{76}.
%Type = Article
\bibitem[{Ishida et~al.(2006)Ishida, Davidson, and Kaneda}]{HIT3}
\bibinfo{author}{T.~Ishida}, \bibinfo{author}{P.~Davidson},
  \bibinfo{author}{Y.~Kaneda},
\newblock \bibinfo{title}{On the decay of isotropic turbulence},
\newblock \bibinfo{journal}{Journal of Fluid Mechanics} \bibinfo{volume}{564}
  (\bibinfo{year}{2006}) \bibinfo{pages}{455--475}.
%Type = Article
\bibitem[{Comte-Bellot and Corrsin(1971)}]{HIT4}
\bibinfo{author}{G.~Comte-Bellot}, \bibinfo{author}{S.~Corrsin},
\newblock \bibinfo{title}{Simple eulerian time correlation of full-and
  narrow-band velocity signals in grid-generated,‘isotropic’turbulence},
\newblock \bibinfo{journal}{Journal of Fluid Mechanics} \bibinfo{volume}{48}
  (\bibinfo{year}{1971}) \bibinfo{pages}{273--337}.
%Type = Article
\bibitem[{Rogallo(1981)}]{ROGALLO}
\bibinfo{author}{R.~S. Rogallo},
\newblock \bibinfo{title}{Numerical experiments in homogeneous turbulence}
  (\bibinfo{year}{1981}).

\end{thebibliography}

\end{document}